\documentclass[fleqn,usenatbib]{mnras}

\usepackage[T1]{fontenc}
\usepackage{ae,aecompl}

\usepackage{graphicx}

\usepackage{amsmath,amssymb}

\newcommand{\beq}	{\begin{equation}}
\newcommand{\eeq}	{\end{equation}}
\newcommand{\beqa}{\begin{eqnarray}}
\newcommand{\eeqa}{\end{eqnarray}}
\newcommand{\beqs}	{\begin{displaymath}}
\newcommand{\eeqs}	{\end{displaymath}}
\newcommand{\beqas}	{\begin{eqnarray*}}
\newcommand{\eeqas}	{\end{eqnarray*}}
\newcommand{\avg}[1]  {{\langle #1 \rangle}} 
\newcommand{\dis}{\displaystyle}
\def\bit{\begin{itemize}}
\def\eit{\end{itemize}}
\newcommand{\e}	{$^{-1}$}
\newcommand{\ee}	{$^{-2}$}
\newcommand{\eee}	{$^{-3}$}
\def\simlt{\lower.5ex\hbox{$\; \buildrel < \over \sim \;$}}
\def\simgt{\lower.5ex\hbox{$\; \buildrel > \over \sim \;$}}
\def\la{\simlt}
\def\ga{\simgt}

\font\tenbi=cmmib10 
\newfam\bifam  \textfont\bifam=\tenbi

\font\tenbr=cmbx10
\newfam\brfam  \textfont\brfam=\tenbr

\font\squinttenbi=cmbx10 at 9pt
\scriptfont\brfam=\squinttenbi

\def\vecnabla{
              \setbox1=\hbox{$\bigtriangledown$}
                           \raise.45ex\hbox{$\bigtriangledown$\hskip-.97\wd1
                           $\bigtriangledown$\hskip-.97\wd1
                           $\bigtriangledown$\hskip-.97\wd1}
                           \raise.47ex\hbox{$\bigtriangledown$}}
\def\grad{\vecnabla}

\def\div{\vecnabla\mathbf{\cdot}}

\def\vecr{{\textbfit{r}}}

\def\vecv{{\textbfit{v}}}

\def\vecx{{\textbfit{x}}}

\def\vecB{{\textbfit{B}}}

\def\symbol#1{\ifmmode#1\else$#1$\fi}
\def\cala{\symbol{{\cal A}}}

\def\cale{\symbol{{\cal E}}}

\def\calm{\symbol{{\cal M}}}
\def\caln{\symbol{{\cal N}}}

\newcommand\cs		{c_{\rm s}}
\newcommand\eff		{{\rm eff}}

\newcommand\mbe     {M_{\rm BE}}
\newcommand{\msun}   {M$_\odot$}
\newcommand{\msunm}     {\mbox{M}_\odot}

\newcommand{\nh}		{n_{\rm H}}

\newcommand{\nhm}		{n_{\rm H,\,max}}
\newcommand{\nhn}		{n_{{\rm H},\,\nu}}
\newcommand{\nho}		{n_{{\rm H},0}}

\newcommand{\rms}       	{{\rm rms}}
\newcommand{\tff}		{t_{\rm ff}}
\newcommand{\tffo}		{t_{\rm ff,0}}

\usepackage{color}

\newcommand\ead		{\eta_{\rm AD}}
\newcommand{\eb}		{{\cal E}_B}

\newcommand{\ebn}		{{\cal E}_{B_\nu}}

\newcommand{\ebone}		{{\cal E}_{B1}}
\newcommand{\eq}		{{\rm eq}}

\newcommand\gad		{\gamma_{\rm AD}}

\newcommand\he		{{\rm He}}
\newcommand\hhp		{{\rm H\,H^+}}
\newcommand\kin		{{\rm kin}}

\newcommand\lj		{{\lambda_{\rm J}}}

\newcommand\nuni		{\nu_{ni}}
\newcommand\odon		{\omega_{d0,\nu}}
\newcommand\phiff		{\phi_{\rm ff}}

\newcommand\rad		{R_{\rm AD}}

\newcommand\sat		{{\rm sat}}
\newcommand\sigv		{\avg{\sigma v}}

\newcommand\sph		{{\rm sph}}

\newcommand\visc		{{\rm visc}}

\newcommand\vtf		{v_{t,5}}
\newcommand\xif		{x_{i,-4}}

\newcommand\akin    {\cala_{\kin}}

\newcommand{\crit}  {{\rm crit}}

\newcommand\refer     {{\rm ref}}
\newcommand\tcoll	{{t_{\rm coll}}}

\title[Magnetic fields in first star formation--II.]
      {Magnetic fields in the formation of the first stars.--II Results}

\author[A. Stacy et al.]
       {Athena Stacy$^{1}$\thanks{E-mail: athena.stacy@gmail.com},  Christopher F. McKee$^{1,2}$,  Aaron T. Lee$^{3}$,  Richard I. Klein$^{1,4}$,
        \newauthor
        Pak Shing Li$^{1}$\\
        $^{1}$Department of Astronomy, University of California, Berkeley, CA 94720, USA \\
        $^{2}$Department of Physics, University of California, Berkeley, CA 94720, USA \\
        $^{3}$Department of Physics and Astronomy, Saint Mary’s College of California, Moraga, CA 94575\\
        $^{4}$Lawrence Livermore National Laboratory, Livermore, CA 94550
        }

\pagerange{\pageref{firstpage}--\pageref{lastpage}}
\pubyear{2015}

\begin{document}

\maketitle
\topmargin-1cm

\label{firstpage}

\begin{abstract}
Beginning with cosmological initial conditions at $z=100$, we simulate the effects of magnetic fields on the formation of Population III stars and compare our results with the predictions of Paper I.  We use {\sc gadget-2} to follow the evolution of the system while the field is weak. We introduce a new method for treating kinematic fields by tracking the evolution of the deformation tensor. The growth rate in this stage of the simulation is lower than expected for diffuse astrophysical plasmas, which have a very low resistivity (high magnetic Prandtl number); we attribute this to the large numerical resistivity in simulations, corresponding to a magnetic Prandtl number of order unity. When the magnetic field begins to be dynamically significant in the core of the minihalo at $z=27$, we map it onto a uniform grid and follow the evolution in an adaptive mesh refinement, MHD simulation in {\sc orion2}.  The nonlinear evolution of the field in the {\sc orion2} simulation violates flux-freezing and is consistent with the theory proposed by Xu \& Lazarian.  The fields approach equipartition with kinetic energy at densities $\sim 10^{10}-10^{12}$~cm$^{-3}$.  When the same calculation is carried out in {\sc orion2} with no magnetic fields, several protostars form, ranging in mass from $\sim$ 1 to 30 $\msunm$; with magnetic fields, only a single $\sim$ 30 $\msunm$ protostar forms by the end of the simulation. Magnetic fields thus suppress the formation of low-mass Pop III stars, yielding a top-heavy Pop III IMF and contributing to the absence of observed Pop III stars.

\end{abstract}


\section{Introduction}

Magnetic fields affect the formation of stars today by reducing the rate of star formation, suppressing fragmentation, and creating outflows \citep{krum19}. What were the effects of magnetic fields on the formation of the first stars? If magnetic fields suppress fragmentation in primordial stars, that would increase the mass of the first stars and act to prevent the formation of stars small enough ($\la 0.8\,\msunm$) to survive until today. In Paper I \citep{mcke20}, we reviewed the creation of magnetic fields via the Biermann battery \citep{bier50,bier51} and their amplification in a small-scale dynamo (basic theory: \citealp{bat50,kazantsev1968,kuls92,sche02b,sche02a}; theory and astrophysical application: \citealp{schleicheretal2010,schoberetal2012a,xu&lazarian2016,xu20}). Because there is no direct observational evidence on how magnetic fields affect the formation of the first stars, this issue must be addressed through theory and simulation. As discussed in Paper I, the central challenge in simulating magnetic fields in the formation of the first stars is that the numerical viscosity and resistivity available with current computational resources are several orders of magnitude greater than the actual values. As a result, simulated dynamos amplify fields much more slowly than real dynamos. This has several consequences: the initial field in the simulation must be chosen to be much larger than in reality, the subsequent growth of the field is often due more to compression than to dynamo action, and the field becomes dynamically significant in a much smaller fraction of the mass.

In this work we simulate the formation of Pop III stars within 
a minihalo environment, beginning with cosmological initial conditions at $z=100$ and eventually resolving down to scales of several au. In contrast to \cite{machida&doi2013} and \cite{petersetal2014}, we self-consistently follow the evolution of the magnetic field over cosmological time scales,
beginning at a redshift $z\simeq 50$. 
In our simulations we do not include the streaming of the baryons relative to the dark matter
(\citealt{tse&hirata2010}), which may increase the turbulent velocity dispersion of the collapsing gas \citep{stacyetal2011,greifetal2011a} and thereby increase the rate at which the dynamo enhances minihalo magnetic fields. The streaming also
delays the star formation and increases the minimum halo mass in which it can occur (\citealp{scha19} and references therein).
We follow the protostellar growth for 2000 yr (longer than \citealp{machida&doi2013} and \citealp{petersetal2014}), which is
when the largest protostar reaches $\sim$ 30 $\msunm$.

Recently, \citet{shar20} have also studied the effect of magnetic fields on Population III star formation.  They performed a suite of AMR-MHD simulations of the collapse of primordial clouds of mass $1000\,\msunm$ with initial magnetic fields ranging from $10^{-15}\,$G to $30\,\mu$G. They found that strong magnetic fields have a moderate effect in suppressing fragmentation in primordial clouds and a significant effect in reducing the number of low-mass stars.  In a follow-up work, \citet{shar21} studied how numerical resolution affects magnetic field growth within primordial protostellar disks.  They increased the Jeans length resolution of a subset of the \citet{shar20} simulations from 32 to 64 cells, and they found that even a small magnetic field ($10^{-15}\,$G) can rapidly grow to dynamically significant levels through the small-scale turbulent dynamo; at later times, the field is amplified by a large-scale mean-field dynamo in the disk.  They concluded that magnetic fields will alter the Pop III IMF regardless of initial field strength, which is consistent with the theoretical results of Paper I.
Our study has key differences from those of \citet{shar20} and \citet{shar21}:  First, unlike their idealized initial conditions, we initialize our simulation on cosmological scales 
and then zoom to a region of $\sim 1000$~\msun\ (similar to their cores) 
that has a non-uniform turbulent Mach number, a more realistic cloud geometry, and a self-consistently initialized magnetic field. The zoom-in region has a minimum magnetic field value of roughly $10^{-5}$ G, similar to their two largest initial magnetic fields, $B_\rms = 9\times 10^{-6}$~G  and $B_\rms = 2.8\times 10^{-5}$~G.  Our study thus provides important evidence that simulations with more realistic initializations will also show suppressed fragmentation.  Our study provides more detail on a single example of primordial star formation instead of the general statistical overview provided in their work. Finally, we compare our the results of our simulation to the ones predicted theoretically in Paper I.

The content of this paper is somewhat complex in that we use two different numerical codes, the {\sc gadget-2} SPH cosmological code and the {\sc orion2} adaptive mesh refinement (AMR) MHD code, to track three stages in the formation of a Pop III star in the presence of a magnetic field, and then compare the results of these simulations with the predictions from Paper I. The outline of the paper is:
\\
\\
\noindent
SIMULATION

2. Numerical Methodology\\
        \hspace*{0.5 cm} 2.1 Cosmological Simulation ({\sc gadget-2} SPH)\\
        \hspace*{0.5cm} 2.2 Pop III Star Formation Simulation ({\sc orion2} AMR\\\hspace*{1cm}MHD)
        
3. Initial Collapse in a Cosmic Minihalo with Kinematic $B$\\
\hspace*{1.0cm}({\sc gadget-2})\\
\hspace*{0.5cm} 3.1 Hydrodynamic Collapse\\
\hspace*{0.5cm} 3.2 Kinematic Evolution of the Magnetic Field

4. Final Stage Collapse ({\sc orion2})

5. Evolution of the Protostar-Disk System ({\sc orion2})\\
\hspace*{0.5cm} 5.1 Disk Fragmentation\\
\hspace*{0.5cm} 5.2 Sink Accretion and Merging: The IMF\\

\vspace*{-0.2cm}
\noindent
THEORY

6. Theory vs Simulation: Growth of the Magnetic Field\\
\hspace*{0.5cm} 6.1 Prediction for {\sc gadget-2}: The Kinematic Dynamo\\
\hspace*{0.5cm} 6.2 Prediction for {\sc orion2}: The nonlinear dynamo
        
We then discuss the implications for the detection of Pop III stars and list the caveats in our treatment (Section \ref{sec:discussion}), and we wrap up with a summary and our conclusions in Section \ref{sec:summ}.
In the appendixes we describe and test a new method for following the evolution of kinematic magnetic fields in SPH (Appendix \ref{app:mag}), the
mapping from SPH to an AMR grid (Appendix \ref{app:mapping}), refinement and sink particles in {\sc orion2} (Appendix \ref{app:refine}), chemistry and cooling in {\sc orion2} (Appendix \ref{app:chem}), and the growth rate of kinematic magnetic fields in simulations (Appendix {\ref{app:growth}).

\section{Numerical Methodology}
\label{sec:num}

We performed high-resolution cosmological simulations of Pop III star formation in a minihalo environment in two main steps -- the {\sc gadget-2} cosmological simulation and the {\sc orion2} primordial star-forming simulation.  We first use the {\sc gadget-2} SPH code to follow cosmological-scale evolution of the density from $z=100$ to $z=27.5$ 
and of kinematic magnetic fields starting at $z\sim 50$.  
After following the initial minihalo collapse in {\sc gadget-2}, the subsequent evolution of the primordial star-forming clump was continued in {\sc orion2} with increased resolution and MHD physics. 

Since our goal is to follow the development of the magnetic field by a small scale dynamo, resolution is a crucial issue. \citet{fede11b} showed that the properties of turbulence in a gravitationally collapsing cloud are governed by the number of resolution elements per Jeans length, $\lj=(\pi\cs^2/G\rho)^{1/2}=1.19\times 10^{21}(T_3/\nh)^{1/2}$ cm, where $T_3=T/(10^3$~K). They inferred that a minimum of 16-32 cells per Jeans length were needed for the dynamo to operate.
In Paper I, we showed that this resolution requirement corresponds to the minimum magnetic Reynolds number found by \citet{haug04}, provided the numerical magnetic Prandtl number (the ratio of the viscosity to the resistivity) is $P_m\simeq 1-2$, as found by \citet{les07}.
However, even at 128 cells per Jeans length, the simulations were not converged. As \citet{fede11b} pointed out, the growth rate of the dynamo increases with Reynolds number (see Section 4 in Paper I), and since it is presently not possible to simulate the very large Reynolds numbers in astrophysics, one cannot expect to resolve the dynamo. \cite{turketal2012} studied the growth of the magnetic field during the formation of the first stars and found that a minimum of 64 cells per Jeans length was required for their somewhat more dissipative code to obtain dynamo action. Furthermore, they found that if the dynamo is insufficiently resolved, the simulated gas within a minihalo exhibits slower collapse and less magnetic field amplification as well as a more disk-like central gas morphology.

\subsection{Cosmological Simulation ({\sc gadget-2} SPH)}
\label{sec:sph}

The cosmological simulation  employed  {\sc gadget-2,} a widely-tested three-dimensional $N$-body and SPH code (\citealt{springel2005}). It was initialized as described in \cite{stacy&bromm2013},
with a 1.4~Mpc (comoving) box containing 512$^3$ SPH gas particles and the same number of DM particles at $z=100$.   
Positions and velocities were assigned to the particles in accordance with a 
$\Lambda$CDM cosmology with $\Omega_{\Lambda}=0.7$, $\Omega_{\rm M}=0.3$, $\Omega_{\rm B}=0.04$, $\sigma_8=0.9$, and $h=0.7$.   
Each gas particle had a mass of $m_{\rm sph} = 120$ $\msunm$, while DM particles had a mass of  $m_{\rm DM} = 770$ $\msunm$.   
The  \cite{stacy&bromm2013} simulation followed the collapse and subsequent star formation of the first $\sim$ 10 minihalos that formed within the cosmological box from $z \sim 15-30$.

Once the simulation was evolved to the point that the location of the first ten minihalos was ascertained, the cosmological box was reinitialized at $z=100$ for each individual minihalo, but with 64 `child' particles added around a 100-140 kpc (physical) region  where the target halo will form. Larger regions of refinement were used for minihalos whose mass originated from a larger area of the cosmological box.  Particles at progressively larger distances from the minihalo were given increasingly large masses, such that 
in the refined initial conditions there was a total of $\sim$ 10$^7$ particles.
The most resolved particles were of mass $m_{\rm sph} = $1.85~$\msunm$ and $m_{\rm DM} = 12$~$\msunm$.

For this current work, we use a slightly modified technique once this refined simulation reached $z \sim 50$.  
At this point we cut out the central 800 pc of SPH and DM particles around the target minihalo, 
where we chose the minihalo that had reached the highest maximum gas density. The cut-out region contained
a total mass $\sim 6\times10^6$ $\msunm$, which is ten times the mass of the virializing minihalo. We
carried out two simulations, one at high resolution and one at low resolution. For the high resolution run, we split the SPH particles into 64 child particles with a mass $m_{\rm sph} = 0.029\,\msunm$.
Since the mass of gas in this stage of the simulation is $8\times 10^5\,\msunm$, 
the high-resolution simulation had $2.7\times 10^7$ gas particles. For the low resolution run, we split the SPH into 2 child particles, each of mass $0.92\,\msunm$.

Our goal for the SPH simulation was to follow the evolution of the kinematic dynamo, in which the field is so weak that it has no dynamical effects. As a result,
we could follow the magnetic field evolution by tracking the evolution of the deformation tensor (see Appendix \ref{app:mag}) without invoking the full MHD equations.
Tests of the accuracy of this formulation are described in Appendixes \ref{appsub:whirl} and \ref{appsub:div} and shown in Fig. \ref{divB}.
Since our code treats only kinematic fields, the presence of magnetic monopoles associated with finite values of $\div\vecB$ has no dynamical effects. The field is divergence-free in the second part of our simulation, which used the {\sc orion2} MHD code.

As noted above, high resolution is important in simulating a dynamo.
  \citet{fede11b} recommended that simulations use at least 64 cells per Jeans length for dynamo simulations, and \citet{turketal2012} adopted this recommendation. It is not clear how to implement this recommendation in an SPH code, and furthermore our treatment of the field is unique. 
 In our simulation, the smoothing length for the high-resolution run was 
\beq
h_{\rm sm}=h_f\left(\frac{m_\sph}{\rho}\right)^{1/3}=9.62\times 10^{18}\left(\frac{h_f m_\sph'^{1/3}}{\nh^{1/3}}\right)~~~~\mbox{cm},
\label{eq:hsm}
\eeq
where $m_\sph'=m_\sph/(1\,\msunm)$ and $\nh$ is the density of H nuclei. The factor $h_f$ depends on the number of neighbour particles in a kernel; we set $h_f=3.63$ so that at high resolution $h_f m_\sph'^{1/3}=1.12$. 
The criterion for adequate resolution scales as 
\beq
\frac{\lj}{h_{\rm sm}}=111\,\frac{T_3^{1/2}}{\nh^{1/6}}.
\label{eq:lj1}
\eeq
We anticipate that the dynamo will be well resolved at low densities, but not at high densities. Fig. \ref{lres} shows the relative behavior of the resolution in the \citet{turketal2012} simulation vs. that in the SPH stage of our simulation.

\begin{figure}
\includegraphics[width=.48\textwidth]{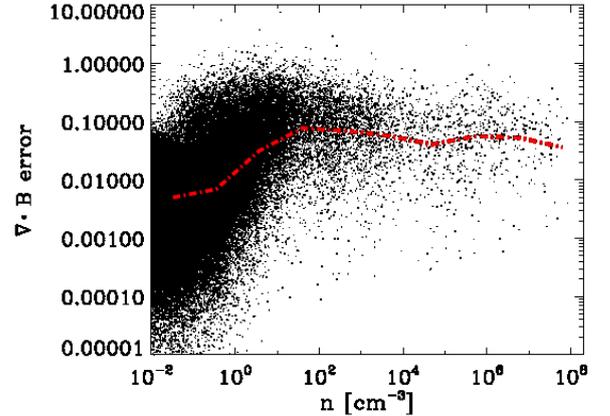}
 \caption
 {
 Divergence error of the magnetic field for the high-resolution {\sc gadget-2} simulation.  The values of $\mbox{div}\,B$ are normalized to $|B|/h_{\rm sm}$. Dashed red line follows the average divergence of over the range of density bins.
 }
\label{divB}
\end{figure}

We continued the simulation until the 
the maximum density was $\nh\simeq 10^8$~cm\eee, corresponding to $\rho\simeq 2\times 10^{-16}$~g~cm\eee. We assigned a value to the initial field such that the magnetic energy was 10 percent of the kinetic energy at that time (see Section \ref{sec:kinesim}), 
We then
mapped the particles and magnetic field onto the grid of the {\sc orion2} MHD code.  
This occurred at a redshift $z=27.5$; 
the first star formed 9000 years later.
We give more detail of this mapping procedure in Appendix \ref{app:mapping}.
Before the magnetic field values were mapped onto the {\sc orion2} grid, we performed an additional divergence cleaning as described in Appendix \ref{appsub:add}.  

\begin{figure}
\includegraphics[width=.45\textwidth]{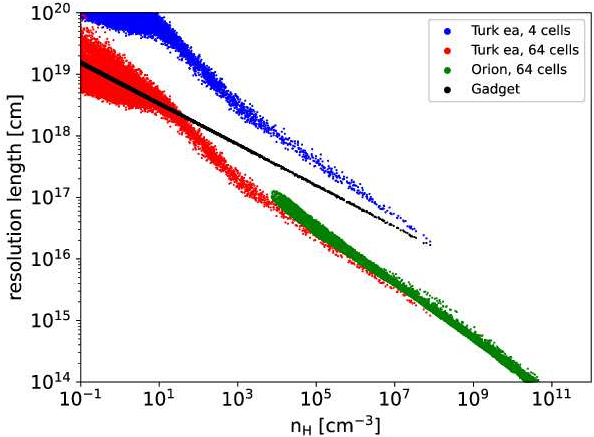}
 \caption
 {
 Comparison of SPH smoothing lengths to typical AMR grid cell sizes.
 Black points represent SPH particles from our {\sc gadget-2} cosmological simulation.  Blue points depict the hypothetical case in which gas of the same density and temperature is followed with an adaptive mesh simulation where $\lj$ is resolved with 4 grid cells, and red points are for a resolution of 64 grid cells.  Similar to \citet{turketal2012}, our cosmological simulation typically resolves $\lj$ with between 10 and 100 smoothing lengths.
 }
\label{lres}
\end{figure}

\subsection{Pop III Star Formation Simulation ({\sc orion2} AMR MHD)}
\label{sec:amr}

To treat the dynamical effects of the magnetic field, we use
the {\sc orion2} adaptive-mesh refinement (AMR), ideal MHD code \citep{lietal2012}. For an accurate treatment of ideal MHD, the code uses constrained transport (CT) \citep{ston08}
coupled with a dimensionally unsplit corner transport upwind (CTU) scheme 
\citep{cole90,Mign07} so that the solenoidal constraint $\vecnabla{\mathbf{\cdot}}{\vecB} = 0$ is maintained to machine accuracy.  
For AMR, a variant of the face-centered projection described in \citet{mart00} is used to ensure the interpolated face-centered magnetic fields of the newly refined regions are discretely divergence free.  We use the Harten-Lax-van Leer discontinuities (HLLD) approximate Riemann solver to obtain fluxes at the cell interfaces based on the reconstructed cell interface states \citep{miyo05}. The result is an accurate and robust AMR MHD scheme that can handle  MHD turbulence simulations driven at Mach numbers up to 30. The inertial range for supersonically driven turbulence on a $512^3$ uniform grid extends up to wavenumbers $k$ corresponding to $kL/(2\pi) \sim 30$ (Paper I).

The {\sc orion2} phase of this study employed a base grid of 128$^3$ cells spanning a length of 0.5 pc.  This is small enough that
we can ignore the DM, since the total DM mass within the central 1 pc is generally $\la$10\% that of the gas.  
In mapping the {\sc gadget-2} results into {\sc orion2}, we did not include SPH particles outside the 0.5 pc box; since the SPH particles have a finite size, this means that the {\sc orion2} density is accurate only within about $6\times 10^{17}$ cm of the center.
The initial mass of gas in the {\sc orion2} simulation was $1300\,\msunm$.
We employed outflow boundary conditions at the edge of the box, such that the gradients of the hydrodynamic quantities were set to zero when the system advances in time (e.g. \citealt{myersetal2013,rosenetal2016}).  
Once the simulation began, up to 8 additional levels of refinement were allowed.  
We ensured that the Jeans length was always resolved by at least 64 cells, as recommended by \citet{fede11b} and \citet{turketal2012}.
On the finest level, one grid cell has a length of $4.7 \times 10^{13}$ cm (3.1 au).  
We describe the criteria for refinement to higher levels in Appendix \ref{app:refine}.  Cells on the highest level of refinement can additionally form mass-accreting sink particles, as also described in Appendix \ref{app:refine}.  
These sinks serve as numerical representations of Pop III protostars, and they accrete mass from within a radius of 4 grid cells (i.e. 12.5 au).
We employed a merging criterion such that sinks are always merged if they come within the an accretion radius of each other, regardless of their mass (see Appendix \ref{app:refine} for further detail).
To complete the simulation in a timely manner, we did not include radiation in our simulations. We therefore stopped the simulations 2000 yr after the first sink forms, when the most massive sink reached $\sim$ 30 $\msunm$ and radiative effects became important (e.g. \citealt{stacyetal2016}).

The chemothermal evolution of each cell was updated at every time step. 
The adiabatic index was determined from the relative proportions of atomic and molecular gas.
Here we mention in particular the uncertainty of the three-body H$_2$ formation rate; we used the rate published by \cite{forrey2013}.
We describe the chemistry update procedure in more detail in Appendix \ref{app:chem}.

In sum, we followed the hydrodynamic evolution of the minihalo from $z=100$ to 9000 yr before the formation of the first
star using  {\sc gadget-2} SPH code; the evolution of the weak magnetic field during this time was tracked by following the evolution of the deformation tensor.  When the field became dynamically important, we switched to the {\sc orion2} AMR MHD simulation and followed the evolution of the field until 2000 years after the first star formed. 
We set $t=0$ at the time of the formation of the first sink, so the transition from {\sc gadget-2} to {\sc orion2} occurred at $t=-9000$~yr.
In addition to these two main simulations, we carried out a hydrodynamic
run with {\sc orion2} in order to determine the effects of magnetic fields on star formation and also a low-resolution simulation with {\sc gadget-2}.

\section{Initial Collapse in a Cosmic Minihalo with Kinematic $\bf B$
({\sc gadget-2})}
\label{sec:initc}

\subsection{Hydrodynamic Collapse}

The initial stages of the collapse,
beginning at $z=100$ and
leading to the first star, are covered by the {\sc gadget-2} SPH simulation. Because the magnetic field has negligible strength in this state, the collapse is hydrodynamic.
Fig. \ref{nHmax} shows the growth of the maximum density with time, while Fig.
 \ref{morph_cosmoA} is a snapshot of 
the density  of the gas 
at the end of the SPH simulation.
The roughly spherical shape with increasing density toward the center is consistent with previous cosmological simulations by various authors (\citealt{yoshidaetal2006,hiranoetal2014}).   The temperature at the end of the {\sc gadget-2} simulation is shown in Fig. \ref{nprof_temp_gadget}.

\begin{figure}
   \includegraphics[width=.48\textwidth]{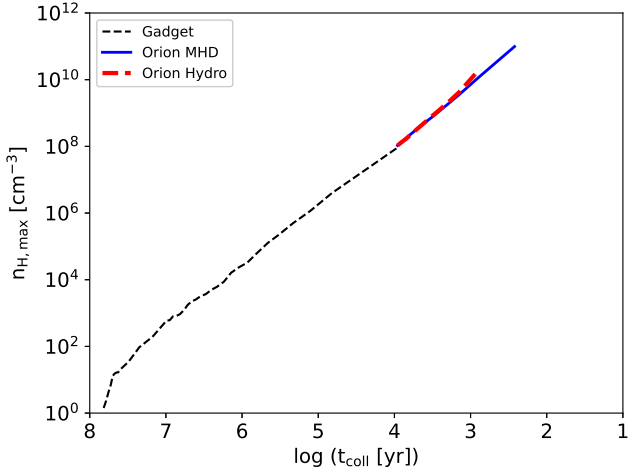}
    \caption{ 
    Maximum density versus $t_{\rm coll}$, the time until initial sink formation. 
    The {\sc gadget-2} simulation begins at $z=100$. That simulation ends, and the {\sc orion2} simulation begins, 9000 years before sink formation.}
    \label{nHmax}
\end{figure}

\begin{figure*}
\includegraphics[width=.47\textwidth]{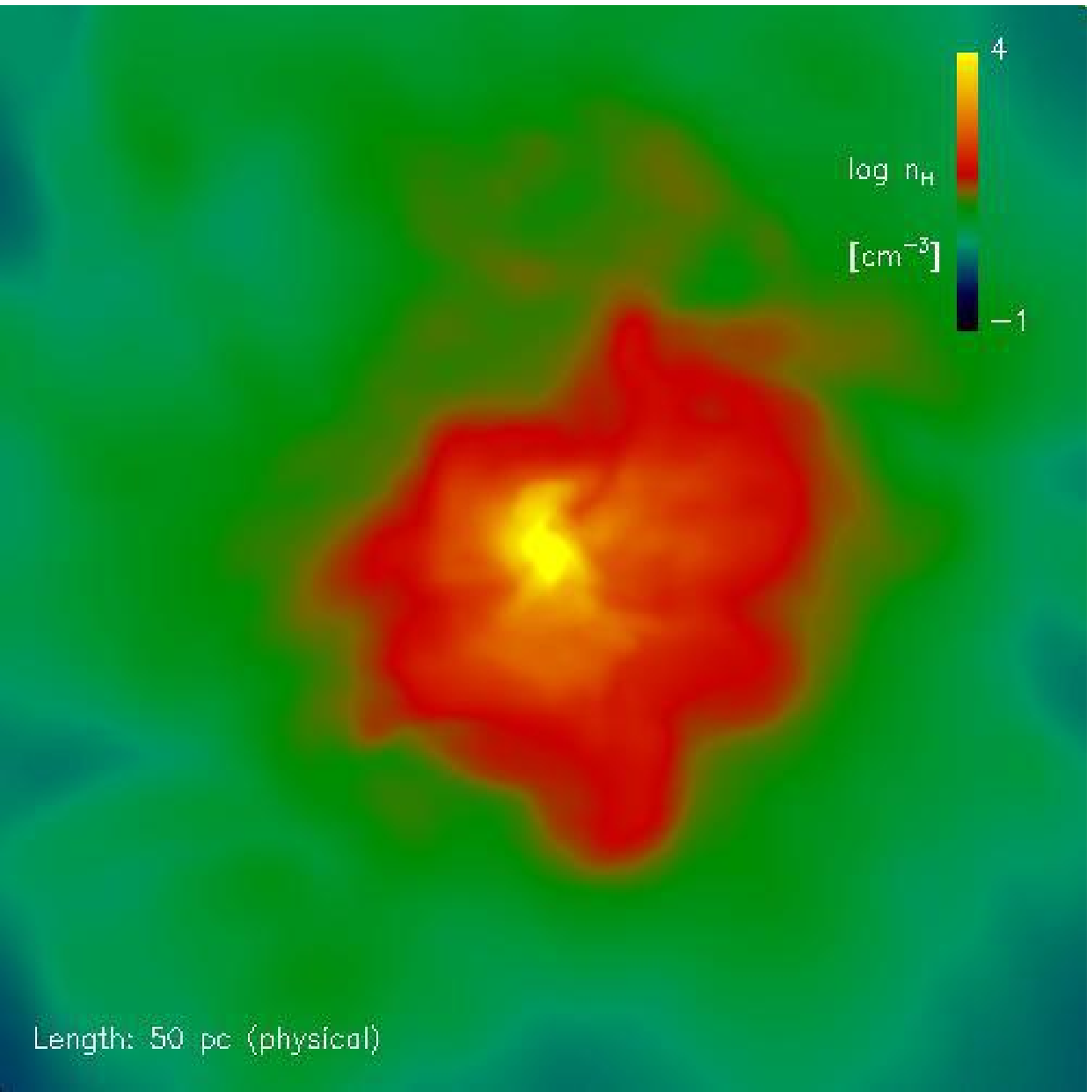}
\includegraphics[width=.47\textwidth]{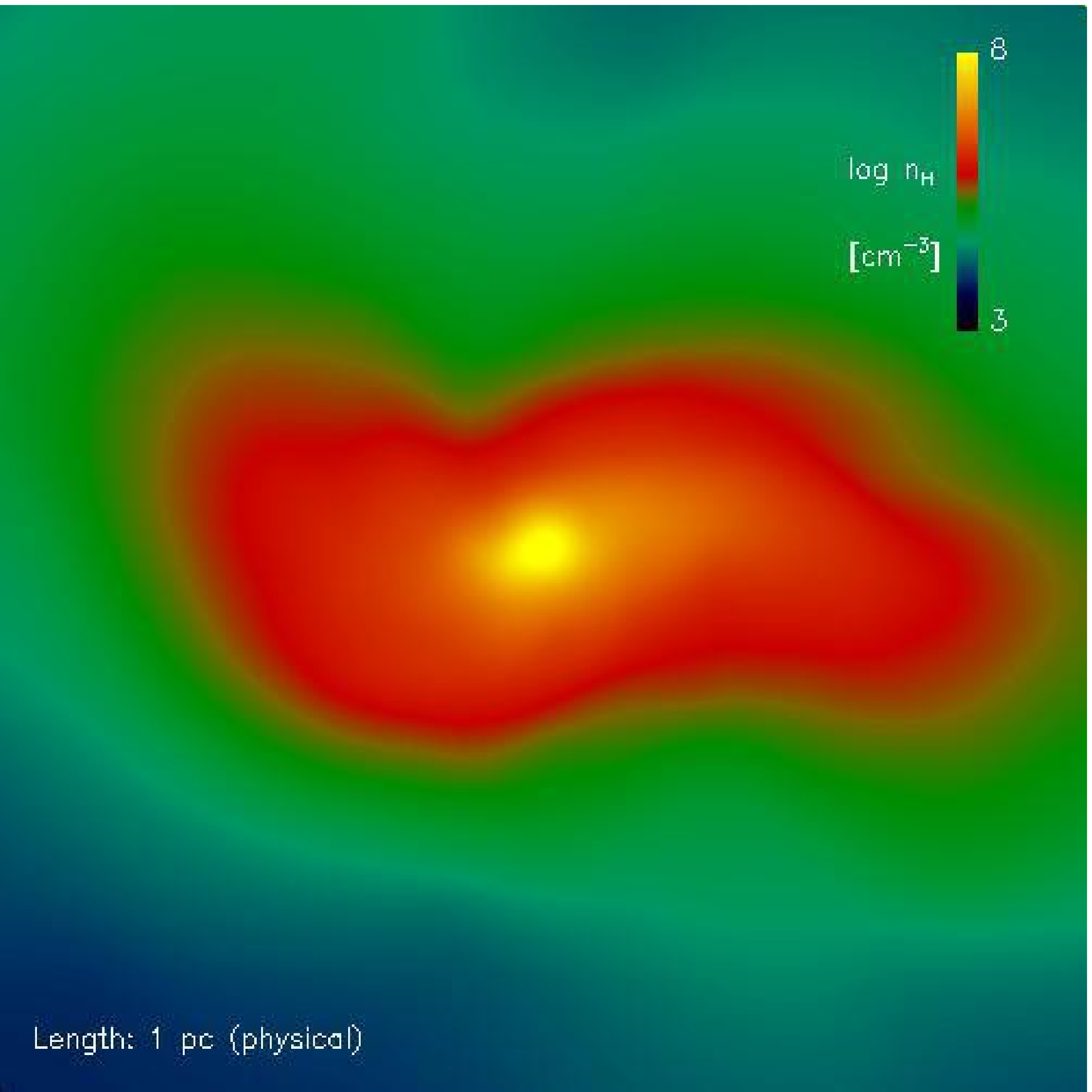}
\includegraphics[width=.47\textwidth]{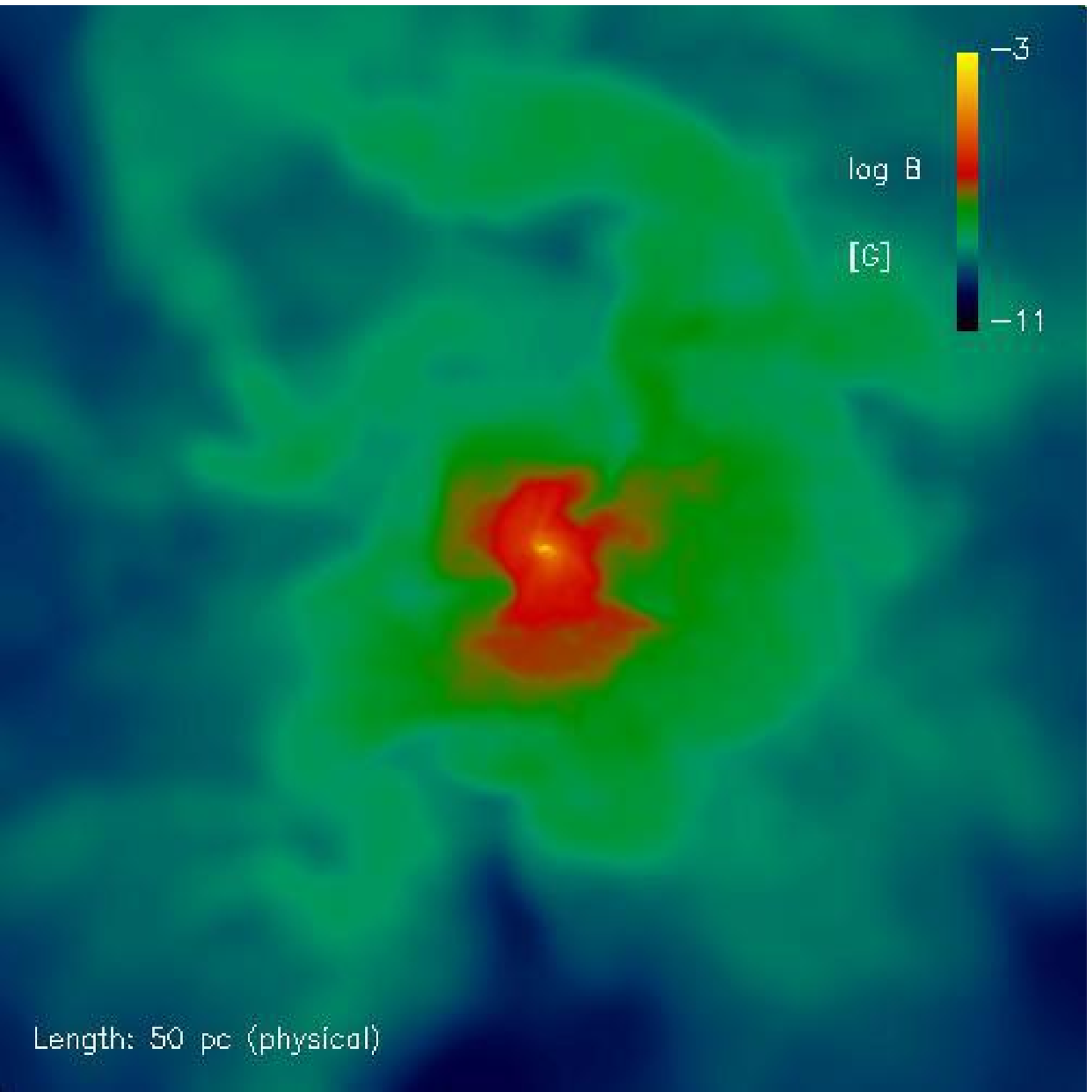}
\includegraphics[width=.47\textwidth]{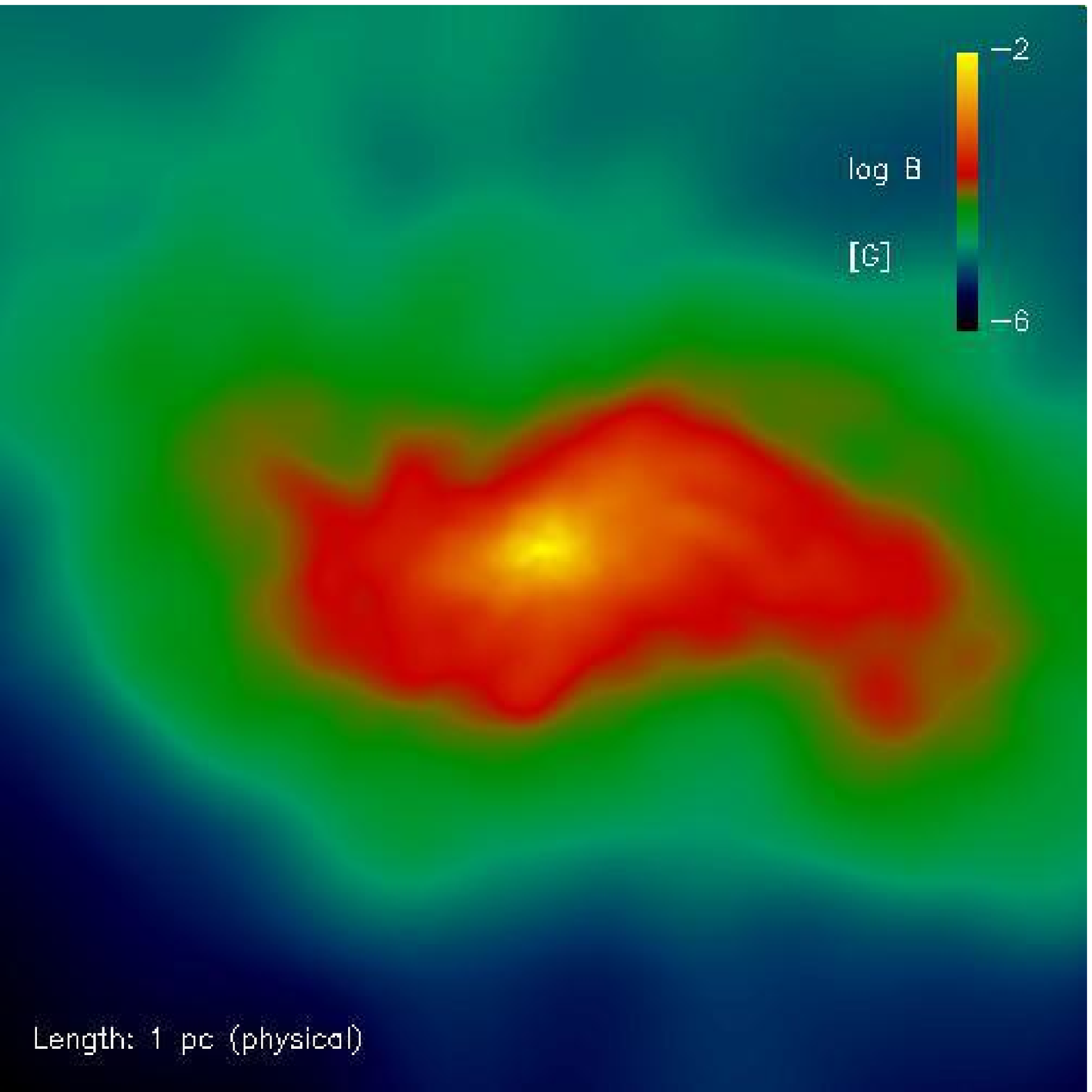}
 \caption
 {
Projection of the density (top) and magnetic field (bottom) of the minihalo
at the end of the {\sc gadget-2} simulation ($t=-9000$~yr).  
Box sizes are 50 and 1 pc (left and right, respectively).  Image is centered on the densest SPH particle.  Both density and magnetic fields grow gradually larger towards the center of the minihalo, while the density structure is smoother than the magnetic field structure.
 }
\label{morph_cosmoA}
\end{figure*}

\begin{figure}
   \includegraphics[width=.48\textwidth]{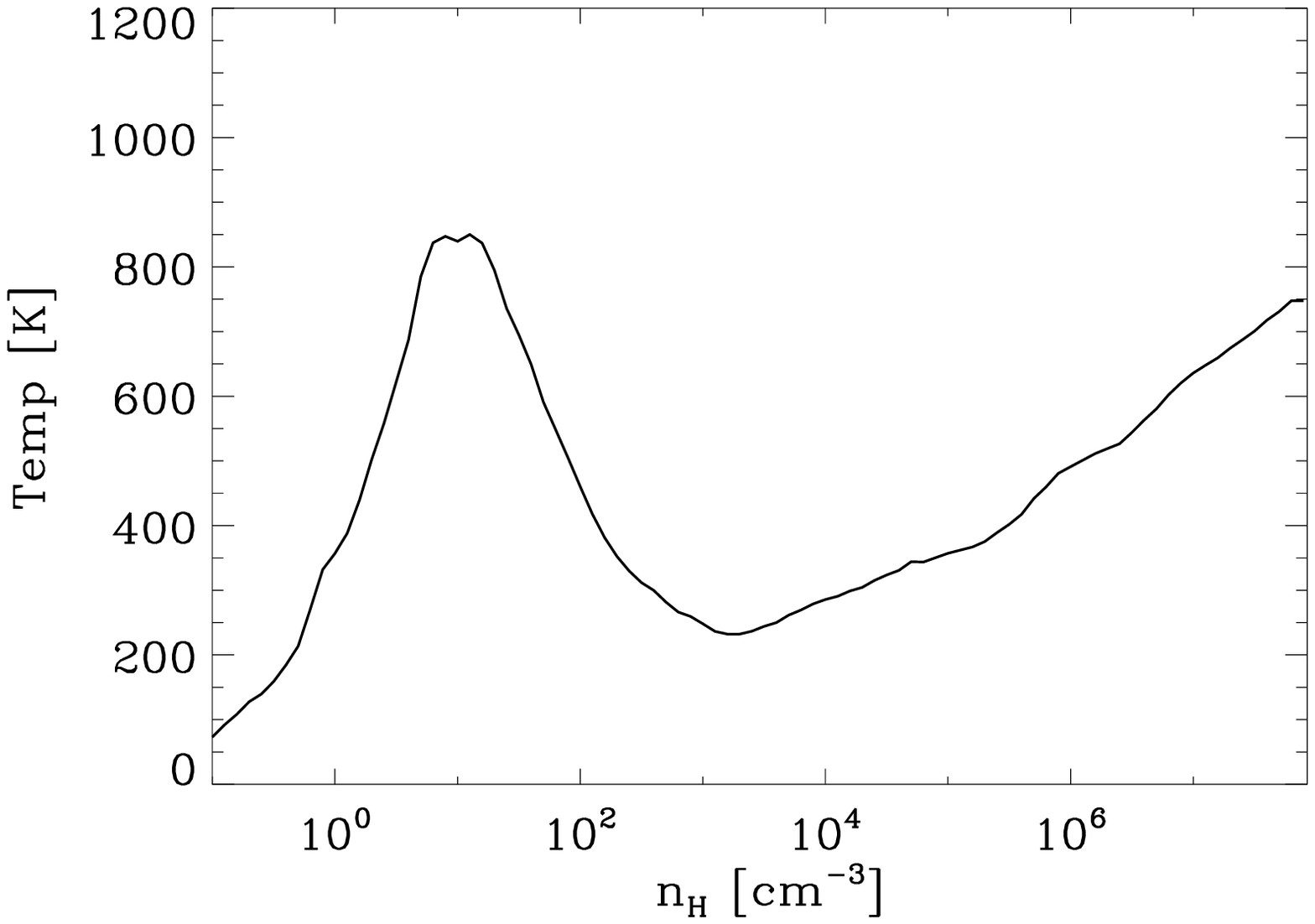}
    \caption{
    Temperature versus density at the end of the {\sc gadget-2} simulation.}
    \label{nprof_temp_gadget}
\end{figure}

In Paper I, we adopted the simple model of an impeded pressureless collapse for the the infall, in which the
velocity is reduced by a factor $\phiff$ so that an initially uniform, static sphere in the absence of dark matter
collapses in a time
\beq
\phiff\tffo=\phiff\left(\frac{3\pi}{32G\rho_0}\right)^{1/2},
\label{eq:phiff}
\eeq
where $\tffo$ is the free-fall time at the initial density, $\rho_0$.
Since the actual collapse is not pressureless, we expect that $\phiff>1$ so that
the collapse is slowed.  
To infer the value of $\phiff$ from the simulation, we consider the collapse at late times, when the gravitational field is dominated by the gas. 
The infall velocity is then given by
\beq
v_r=-\frac{1}{\phiff}\left[\frac{2GM(r)}{r}\right]^{1/2}~~~~(\rho\gg\rho_0),
\label{eq:vr}
\eeq
where $M(r)$ is the gas mass interior to $r$. 
Integration of this equation under the assumption that $\phiff$ is constant implies that the time for the gas with a maximum density $\rho$ to collapse to 
a star is
\beq
t_{\rm coll}(\rho)=\left(\frac{4}{3\pi}\right)\phiff \tff(\rho),
\label{eq:deltat}
\eeq 
where $\tff(\rho)$ is the free-fall time at a density $\rho$. This time is shorter than the initial collapse time in equation (\ref{eq:phiff}) by a factor $4/3\pi\simeq 0.42$ because here the gas is initially moving at the free-fall velocity.
Fig. \ref{phi_ff} shows the ratio of the collapse time to the free-fall time.
\citet{turketal2012} obtained a similar result: For their run with 64 cells per Jeans length, they found a peak value $t_{\rm coll}/\tff(\rho)\simeq 7$ at $\nh\sim 500$~cm\eee, similar to the peak value $\sim 5$ that we find
at $\nh\sim 600$~cm\eee.
Drawing upon the {\sc orion2} results discussed in Section \ref{sec:final} below, we note that the gas collapsed from an initial peak density of $1.0\times 10^8$ cm\eee\ to stellar densities in 9000 years, corresponding to $t_{\rm coll}\simeq 2\tff$; \citet{turketal2012} found the same ratio of $t_{\rm coll}/\tff$
for $10^7\mbox{ cm\eee}\la\nh\la 10^{13}\mbox{ cm\eee}$. Our results show that $\phiff=(3\pi/4)t_{\rm coll}/\tff=2.36t_{\rm coll}/\tff$
is in the range 4-12 over the entire range of densities in our simulations. In the density range covered by the {\sc gadget-2} simulation ($1\mbox{ cm\eee}\la \nh\la 10^8$~cm\eee), the average value of $\phiff$ is about 7,
whereas in the {\sc orion2} density range, we have $t_{\rm coll}/\tff\simeq 2$ so that
$\phiff\simeq 4.7$.

\begin{figure}
\includegraphics[width=.48\textwidth]{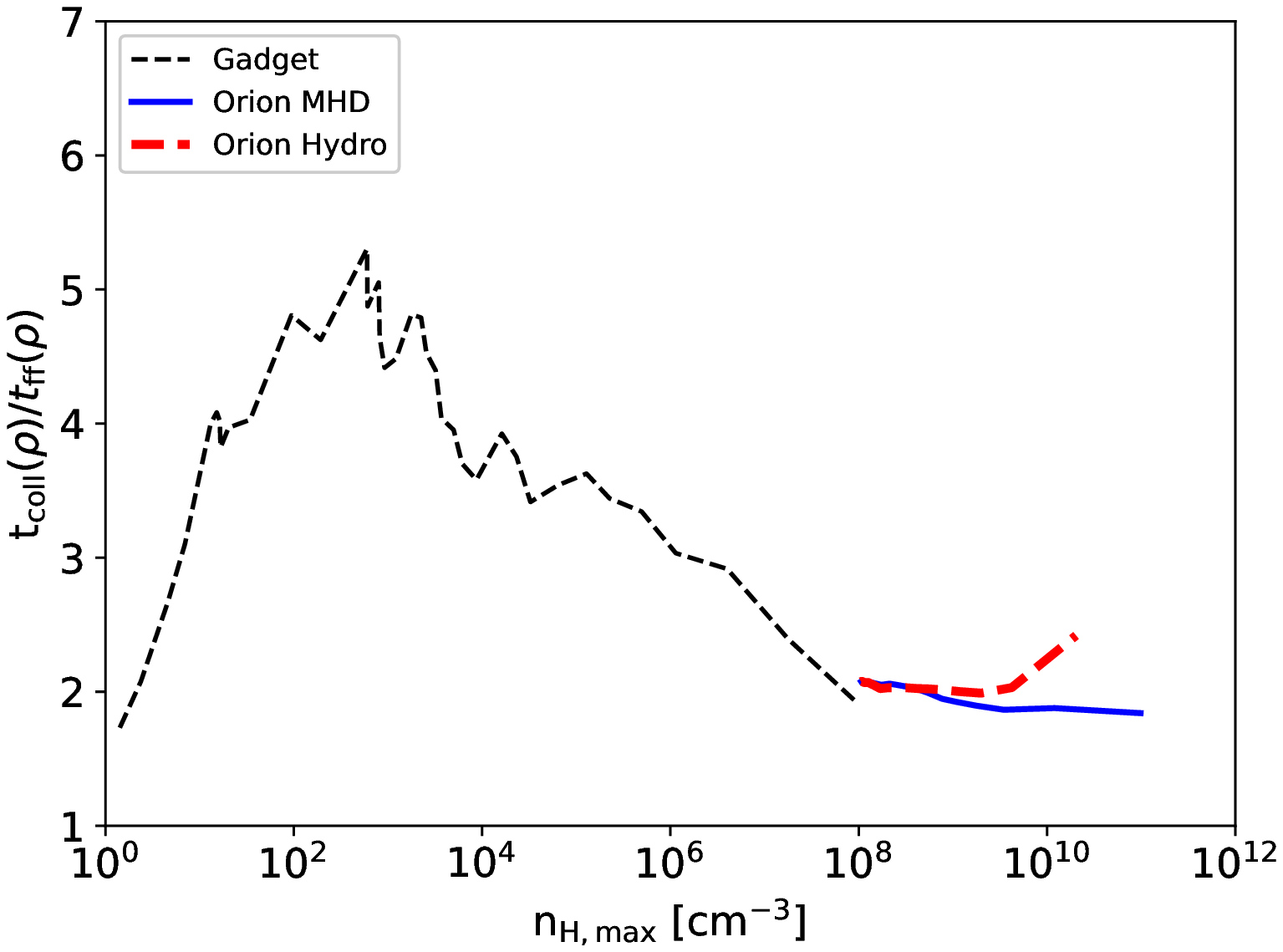}
 \caption
 {
The ratio of the time to collapse to a star, $t_{\rm coll}$, to the free-fall time  as a function of the maximum density in the simulation.
  }
\label{phi_ff}
\end{figure}

The collapse of the gas generates turbulence. Previous simulations  \citep{greifetal2012}  of the collapse of minihalos
of mass $M_m\sim (2-3)\times 10^5\,\msunm$ found turbulent velocities
$\sim 2$ ~km~s\e, about half the virial velocity.
In Fig. \ref{velprof} we show the radial velocity relative to the density maximum and the turbulent velocity and Mach number, all measured at the end of the cosmological simulation.  
All velocities are measured 
relative to the velocity of the gas density peak and
across 100 density bins, evenly spaced on a logarithmic scale.
We define $\vecv_{\theta,i}$ as the velocity in the direction normal to both $ \hat\vecr$ and  
the angular momentum of the shell.
The turbulent Mach number in the $i$th shell, ${\cal M}_{{\rm turb},i}$, 
is then defined as  
\begin{equation}
{\cal M}_{{\rm turb},i}^2 c_{s,i}^2  = 3 \, |\vecv_{\theta,i}|^2 \mbox{.}
\end{equation}
The decrease in the velocities at high densities is most likely due to limited resolution: For example, in the {\sc gadget-2} simulation, the gas at $\nh=10^7$~cm\eee\ was at a radius $r=1.5\times 10^{17}$~cm and that at $\nh=10^8$ was at $r=6\times 10^{16}$~cm, whereas the resolution was $2\times 10^{16}$~cm.

\begin{figure}
\includegraphics[width=.48\textwidth]{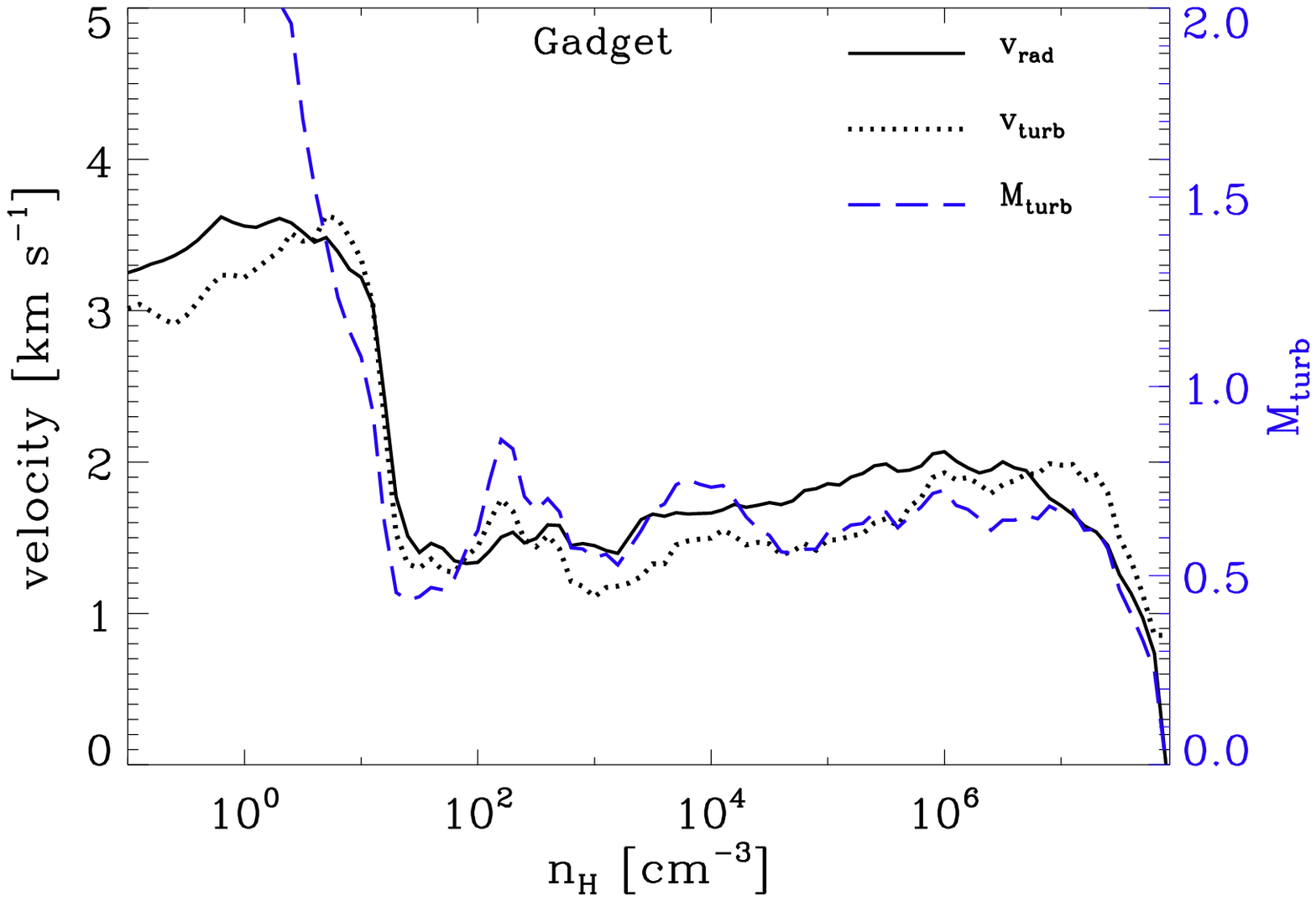}
\includegraphics[width=.48\textwidth]{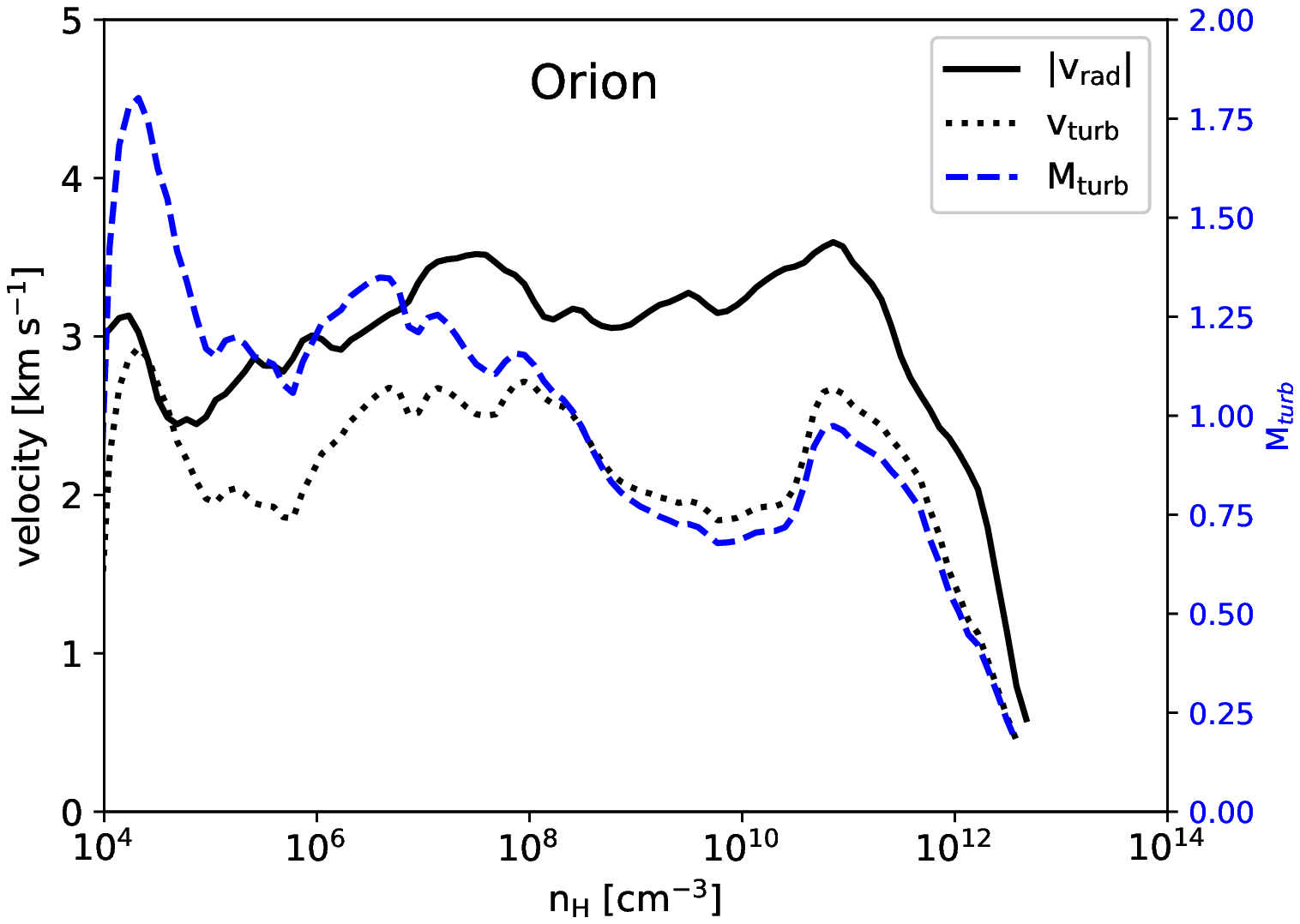}
 \caption
 {
{\it Upper}: Gas velocity with respect to density in the {\sc gadget-2} cosmological simulation at the final snapshot
($t=-9000$~yr).
Dashed blue line shows turbulent Mach number, ${\cal M}_{\rm turb}$.
Dotted line is turbulent velocity $v_{\rm turb}$, while solid line depicts radial velocity $v_{\rm rad}$.  
Initial infall of the gas into the minihalo at densities below 10 cm$^{-3}$ is slightly supersonic, while it becomes sonic to mildly subsonic at higher densities.
{\it Lower}: Radial and turbulent gas velocity and Mach number as a function of density in {\sc orion2} simulation at snapshot just prior to initial sink formation
($t=0$).
  }
\label{velprof}
\end{figure}

\subsection{Kinematic Evolution of the Magnetic Field}
\label{sec:kinesim}
 
As discussed in Section \ref{sec:num}, our simulation of the evolution of the magnetic field occurred in two stages. In the first stage, we used the deformation tensor to track the evolution of a kinematic magnetic field with the
{\sc gadget-2} SPH code, starting at a redshift $z= 54$ 
and ending when the maximum density was about $10^8$~cm\eee. In the second stage, we 
mapped the data in a box of size 0.5 pc centered on the density maximum from the {\sc gadget-2} code to the {\sc orion2} code and followed 
the subsequent evolution of the field using ideal MHD.
The absolute value of the field is irrelevant during the {\sc gadget-2} simulation since it is purely kinematic, but a value must be chosen for the {\sc orion2} MHD simulation. 
We chose an initial field 
$B_0=4.5\times 10^{-12}$ G (physical) at $z=54$,
which corresponded to a field energy that was 10 percent of the kinetic energy at the onset of the {\sc orion2} simulation. 
This initial field is much larger than the $\sim 10^{-16}\,$G field expected from the Biermann battery operating in the turbulent minihalo (Paper I), but this is necessitated by the reduction in the dynamo growth rate due to the large numerical viscosity.

\begin{figure}
\includegraphics[width=.48\textwidth]{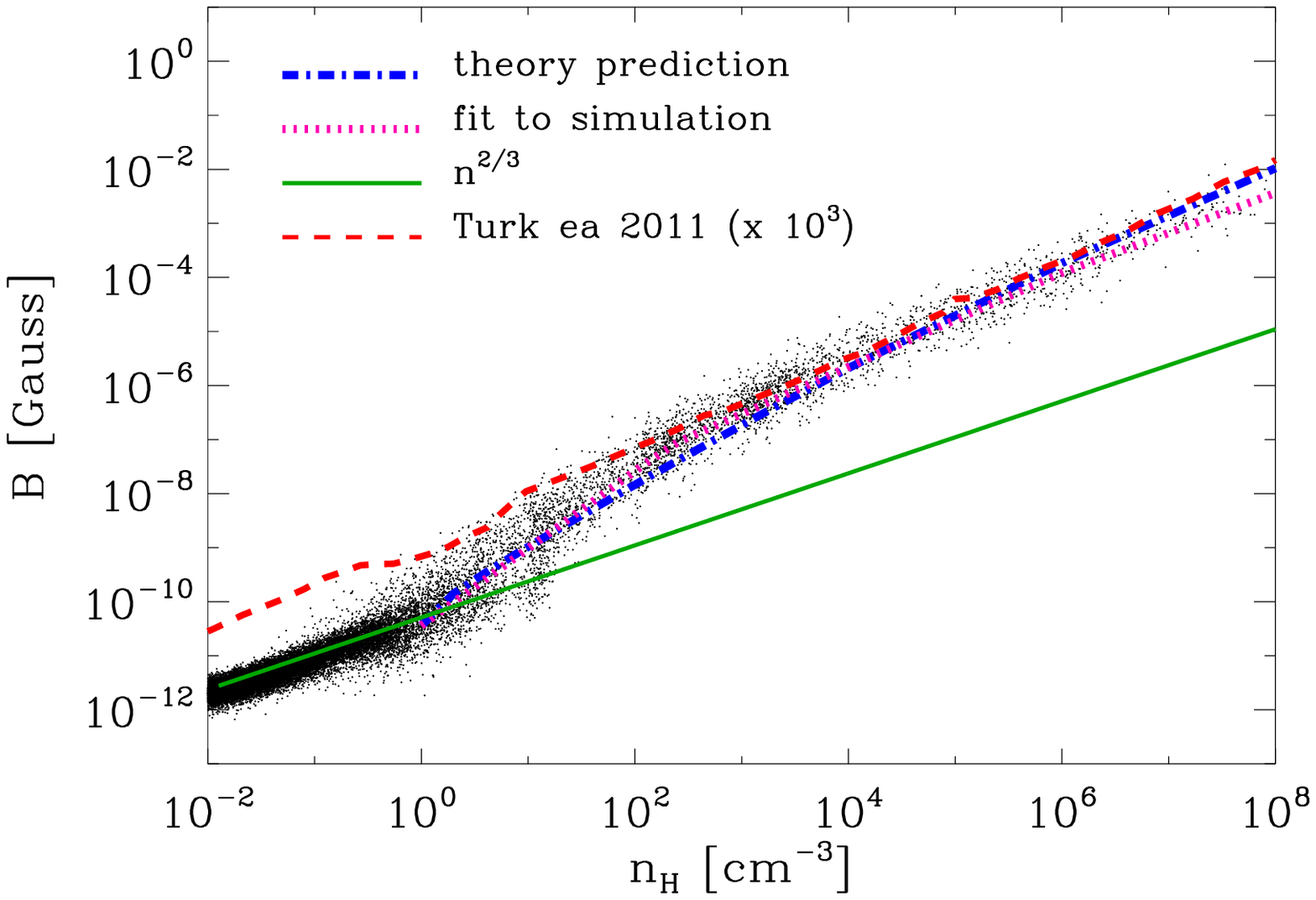}
 \caption
 {
 Magnetic field of SPH particles with respect to density, as determined at the end of the {\sc gadget-2} cosmological simulation.  Dots show $B$ versus $\nh$ for a subset of SPH particles in the high-resolution {\sc gadget-2} simulation.  
 Dotted pink line shows the multiple power-law fit to the simulation result, equation (\ref{eq:bgadget}).
 The dark blue dash-dot line shows our theoretical prediction, equation (\ref{eq:bkin}),
 which agrees with the simulation to within a factor 2 for 1~cm\eee$\;<\nh<10^7$~cm\eee\ and becomes higher than the simulation result at higher densities.
 Dashed red line shows the result from Turk et al. (2012) for comparison, where the magnetic field roughly follows $B = B_0 (n / n_0) ^ {0.89}$.  
Green line shows the the relation $B = B_0 (n / n_0) ^ {2/3}$, the evolution expected from pure flux-freezing.   
 }
\label{Bvsn}
\end{figure}

The growth of the field due to both compression and to dynamo action by the end of the {\sc gadget-2} simulation is portrayed in Fig. \ref{Bvsn}, and the structure of the magnetic field is shown in the bottom panels of Fig. \ref{morph_cosmoA}. At the end of the {\sc gadget-2} simulation ($t=-9000$ yr), the results can be fit with 4 power laws:
\beq
B\simeq \left\{\begin{array}{l} 3.6\times 10^{-11}\nh^{2/3}~\,\pm 0.05\mbox{ dex}~~~~(1>\nh>0.01),\\
3.6\times 10^{-11}\nh^{1.43}~\pm 0.10\mbox{ dex}~~~~(250>\nh>1),\vspace{0.1cm}\\
7.3\times 10^{-10}\nh^{0.87}~\pm 0.06\mbox{ dex}~~~~(10^7>\nh>250),\vspace{0.1cm}\\
1.7\times 10^{-8}\nh^{2/3}~~~\pm 0.02\mbox{ dex}~~~~(10^8>\nh>10^7),
\end{array}\right.
\label{eq:bgadget}
\eeq
where $B$ is in G and $\nh$ in cm\eee. 
(Keep in mind that the relative field strengths in this simulation are accurate, but the absolute values are significant only in that they provide continuity with the subsequent {\sc orion2} MHD simulation.)
Dynamo amplification begins at a density of $\nh\simeq 1$~cm\eee, slows at a density of about about 150 cm\eee, and ends at a density of about $10^7$~cm\eee. 
Most of the dynamo amplification occurs in the second of the four stages above; under the assumption that flux-freezing is valid, the dynamo amplifies the field by a factor 53 in this stage, whereas in the third stage the dynamo amplification is only a factor 9. Overall, most of the growth of the field during the {\sc gadget-2} simulation is due to compression: the dynamo amplifies the field by a factor 470 out of the total amplification of $1.0\times 10^8$ for a density increase of a factor $10^8$ (the fact that both numbers are $10^8$ is a coincidence). If flux-freezing is violated, the dynamo is relatively more efficient (see the discussion below eq. \ref{eq:bkin}).

The key assumption in our treatment of the magnetic field in {\sc gadget-2} is that the kinetic energy is significantly greater than the magnetic energy. 
The equipartition field is
\beq
B_{\rm eq}=(4\pi\rho)^{1/2}v_t=5.30\times 10^{-7}\vtf\nh^{1/2}~~~\mbox{G}.
\eeq
The typical velocity dispersion (see Fig. \ref{velprof}) is 1.5~km~s\e. At
the maximum density of $\nh=10^8$~cm\eee\ in the {\sc gadget-2} simulation, the equipartition field is $8\times 10^{-3}$~G, a little more that twice the field in the simulation (eq. \ref{eq:bgadget}).
Since the dynamical effects are proportional to $B^2$, the kinematic assumption is well satisfied over much of the density range and marginally satisfied at the highest density.

\section{Final Stage Collapse ({\sc orion2})}
\label{sec:final}

Before magnetic fields become dynamically significant, we stop the simulation in {\sc gadget-2} and continue it in the {\sc orion2} MHD code to treat the dynamical effects of the fields accurately.
The density and temperature in the {\sc orion2} MHD simulation at the time of sink formation
are shown in Fig. \ref{radprof_nh}. Comparison of the MHD and hydro runs shows that the magnetic field affects the density and temperature only for $r\la 10^{16}$~cm.
The evolution of the density with time is shown in Fig.
\ref{rprof} (panel a).
At the time the sink first forms and
for $10^{15.5}\mbox{ cm}<r<10^{18}\mbox{ cm}$, it is given approximately by
\beq
\nh=1.0\times 10^5 r_{18}^{-2.16}~~~~\mbox{cm\eee},
\label{eq:nh}
\eeq
where $r_{18}=r/(10^{18}\,\mbox{cm})$.
This power law is quite close to the value 2.2 found by \citet{omukai&nishi1998} in their 1D simulation of primordial star formation.  

\begin{figure*}
\includegraphics[width=.47\textwidth]{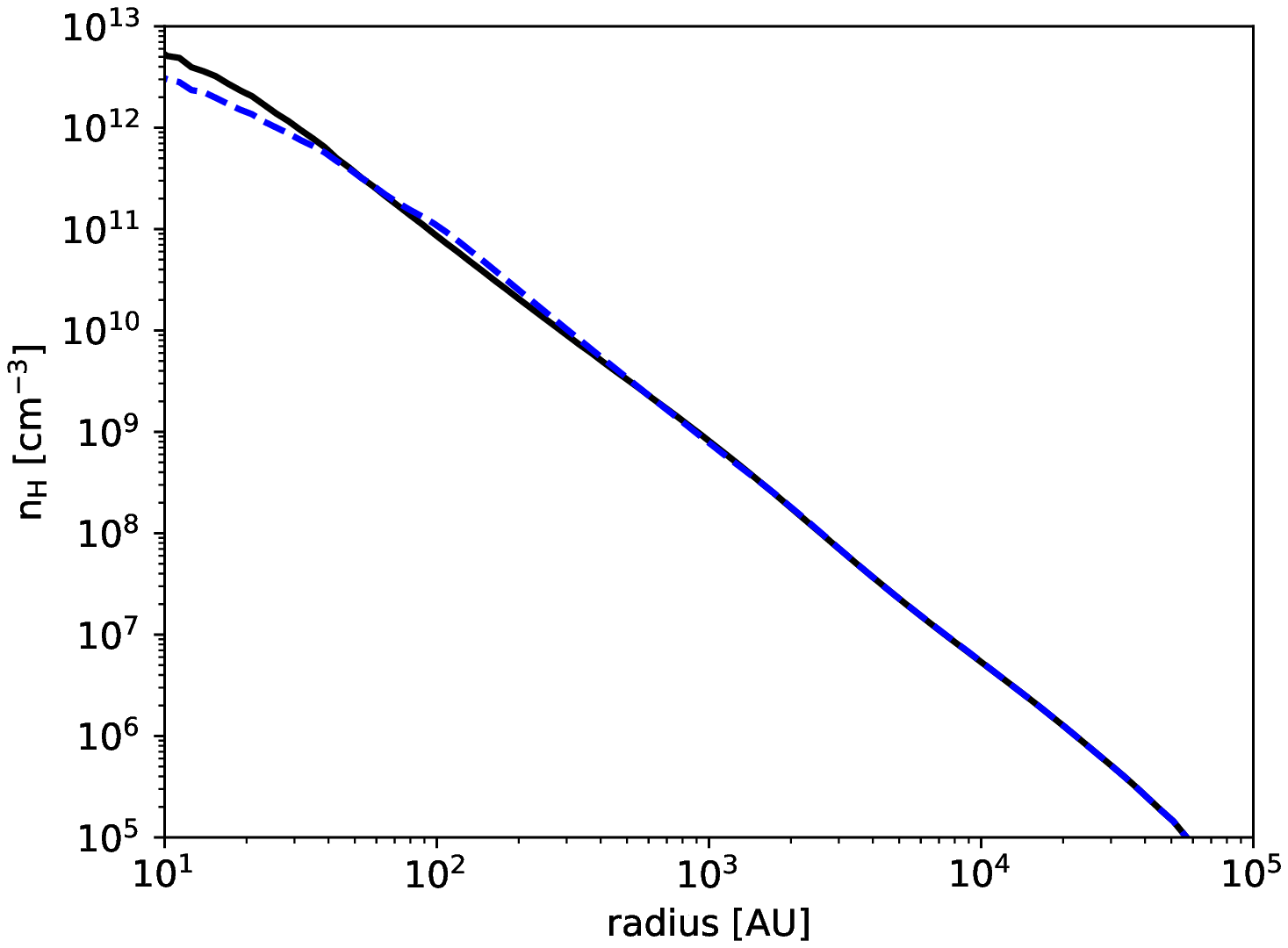}
\includegraphics[width=.47\textwidth]{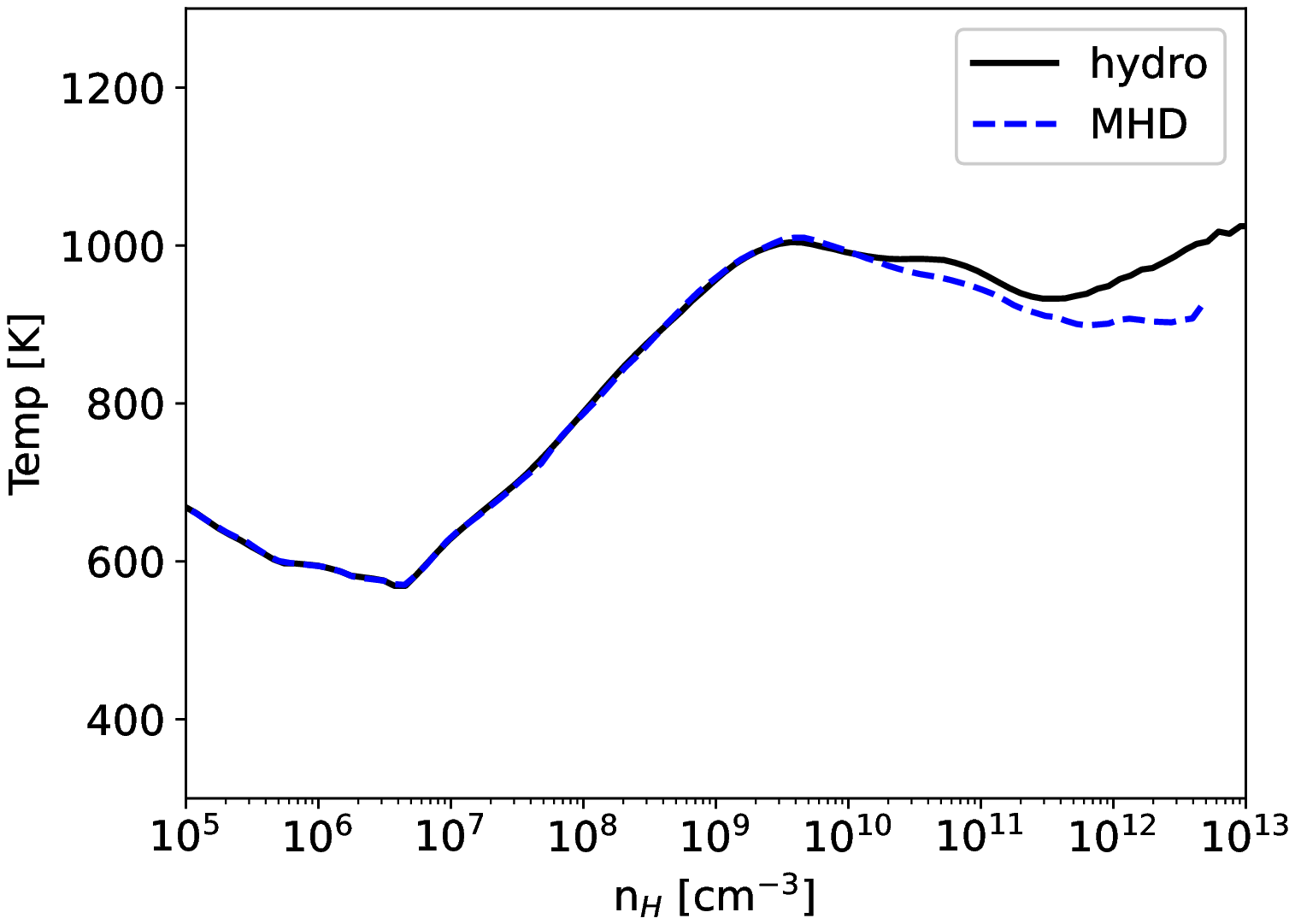}
 \caption
 {
Density versus radius (left) and temperature versus density (right) at the point of initial sink formation 
($t=0$).
Both the hydro and MHD cases
in the {\sc orion2} simulations are shown.
}
\label{radprof_nh}
\end{figure*}

\begin{figure}
\includegraphics[width=.47\textwidth]{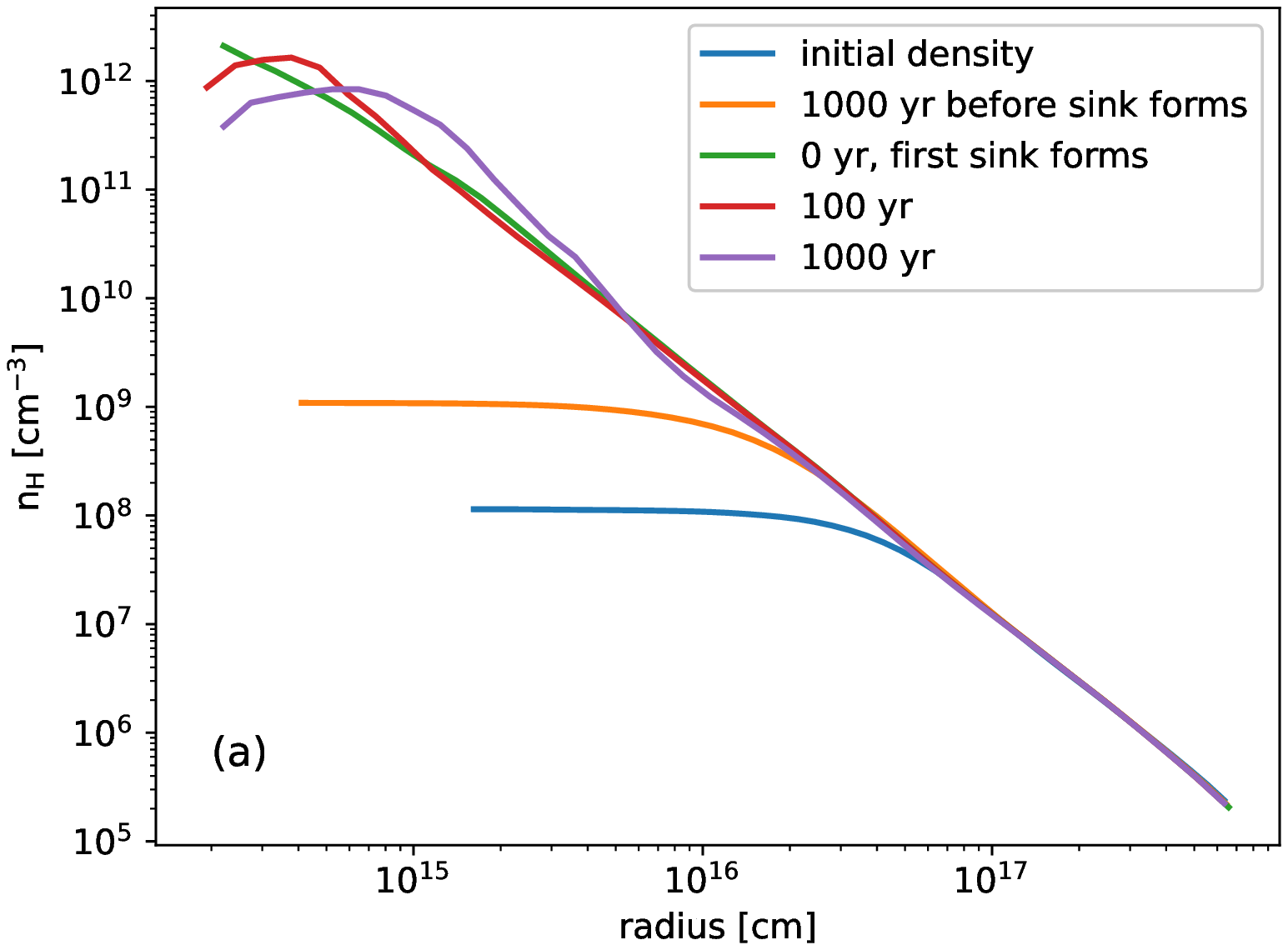}
\includegraphics[width=.47\textwidth]{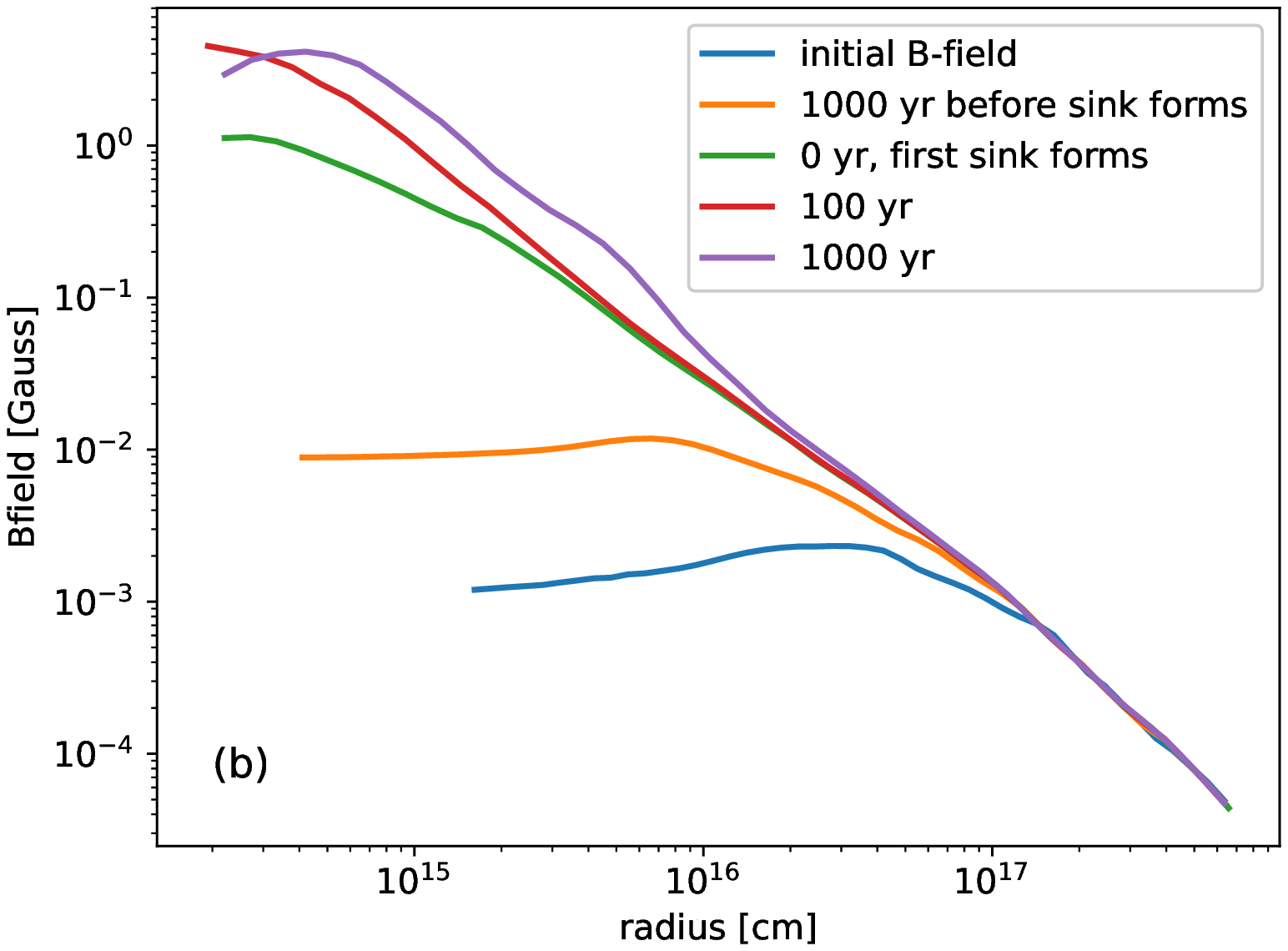}
\includegraphics[width=.47\textwidth]{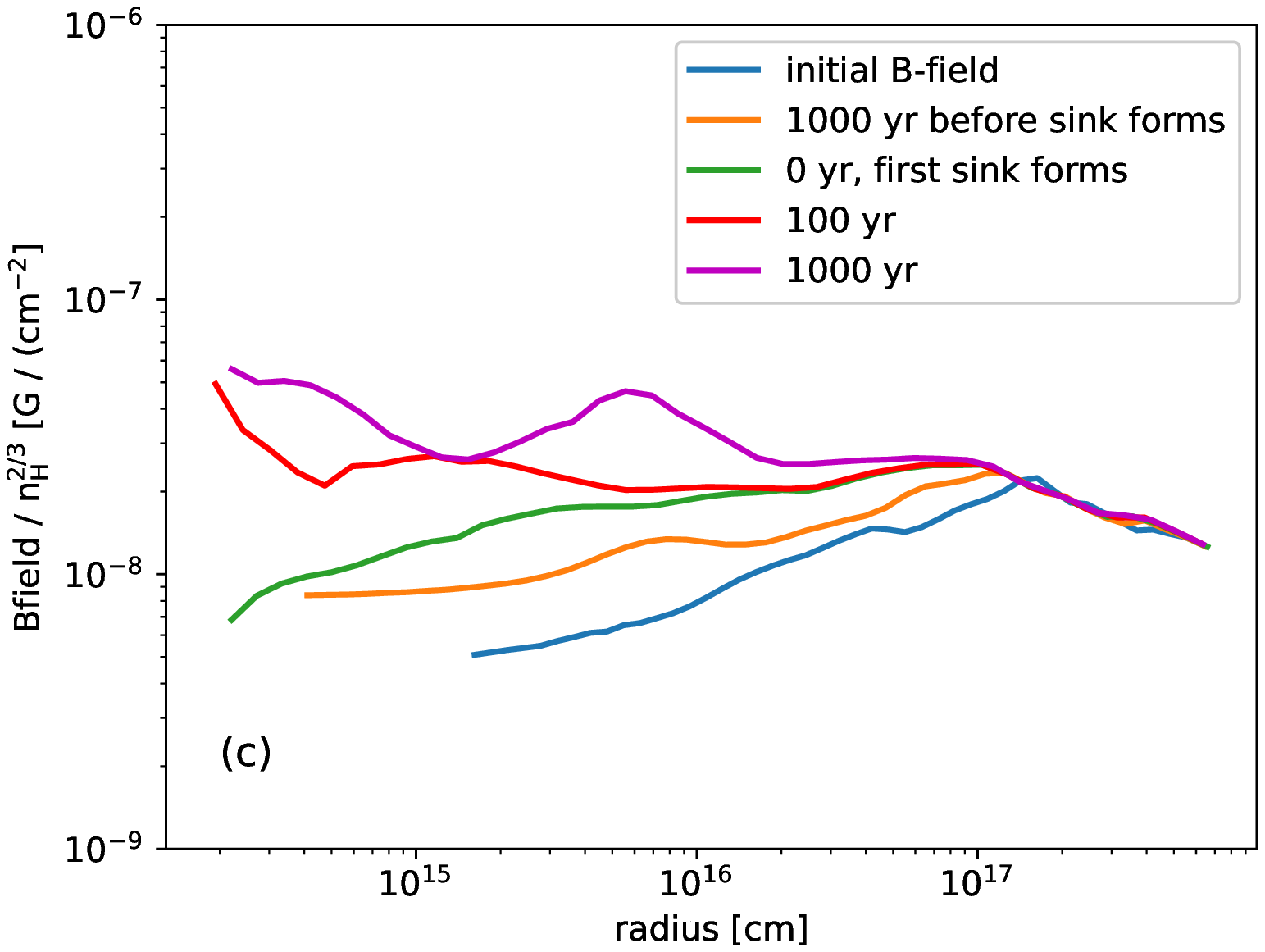}
 \caption
 {
 {\it (a):} Density versus radius at various times throughout the {\sc orion2} simulation.  The initial time is 9000 yr before formation of the first sink.
 {\it (b):}  Magnetic field magnitude versus radius at various times throughout the {\sc Orion2} simulation.
 $B$ is the volume-weighted 
 average of the magnitude of the magnetic field in spherical shells centered on the densest cell, or on the sink cell once one has formed.
 {\it (c):} Normalized magnetic field ($B/\nh^{2/3}$) at various times throughout the {\sc orion2} simulation.
 }
\label{rprof}
\end{figure}

The {\sc orion2} results on the chemical and thermal evolution displayed in Fig. \ref{stuff_vs_n} are consistent with previous work.  This figure shows the state of the central 0.5 pc of gas just before initial sink particle formation.  A fully molecular core of $\sim$ 1000 K gas has formed in the dense region of $\nh > 10^{12}\,\mbox{cm\eee}$ gas.  
The adiabatic index evolved with density from approximately $5/3$ to $7/5$ as the gas transitioned to fully molecular at high densities.

\begin{figure*}
\includegraphics[width=.47\textwidth]{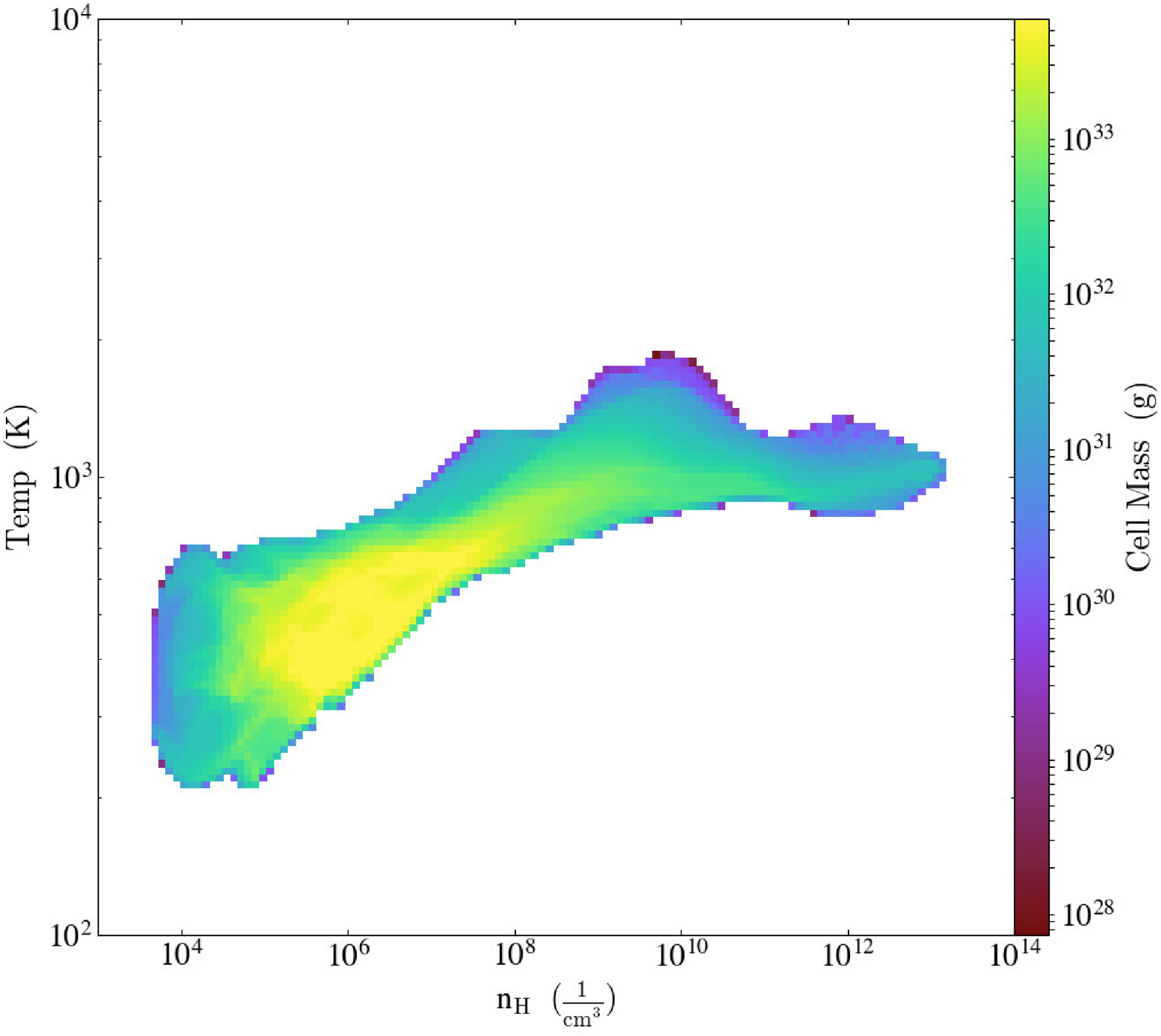}
\includegraphics[width=.47\textwidth]{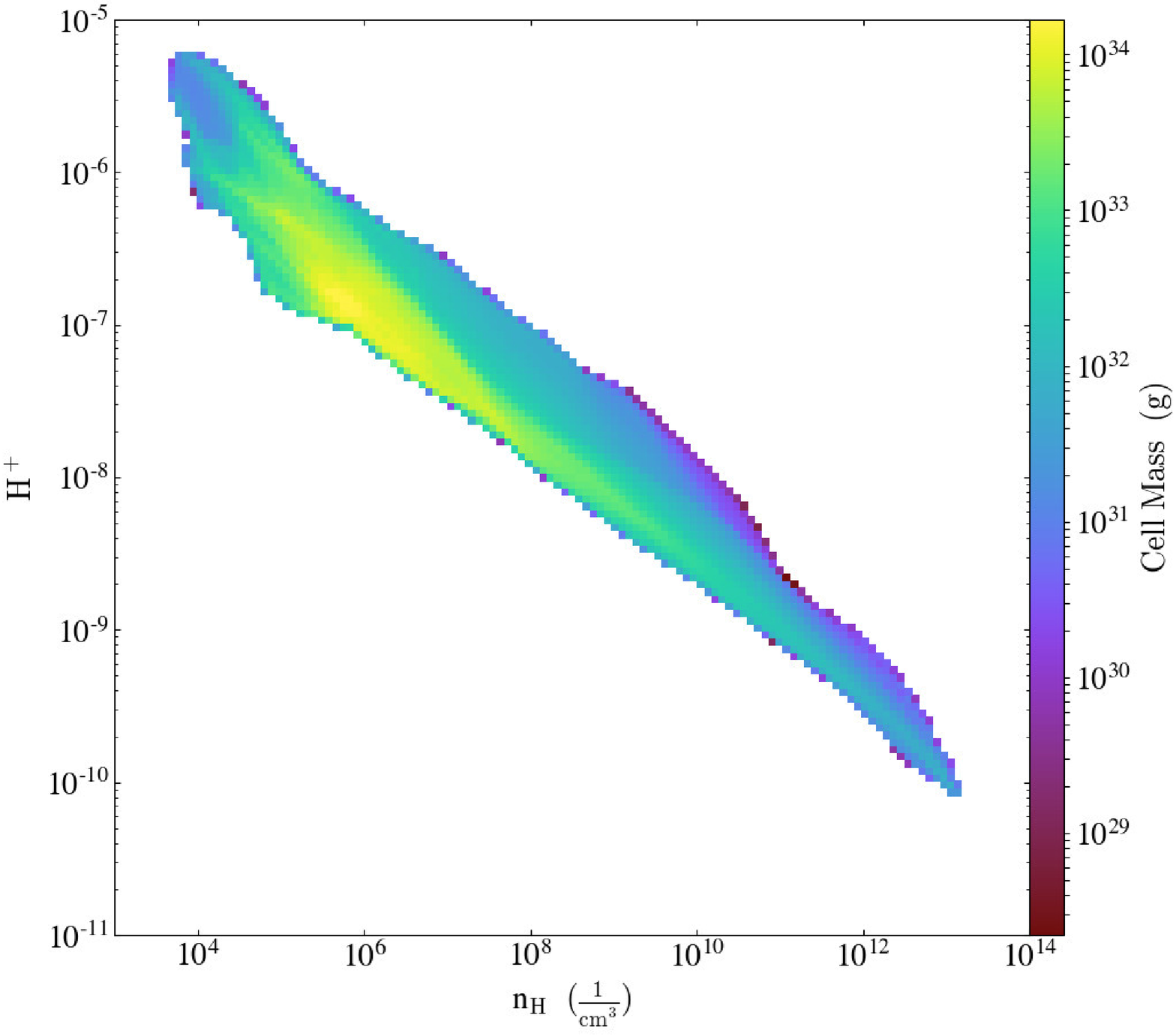}
\includegraphics[width=.47\textwidth]{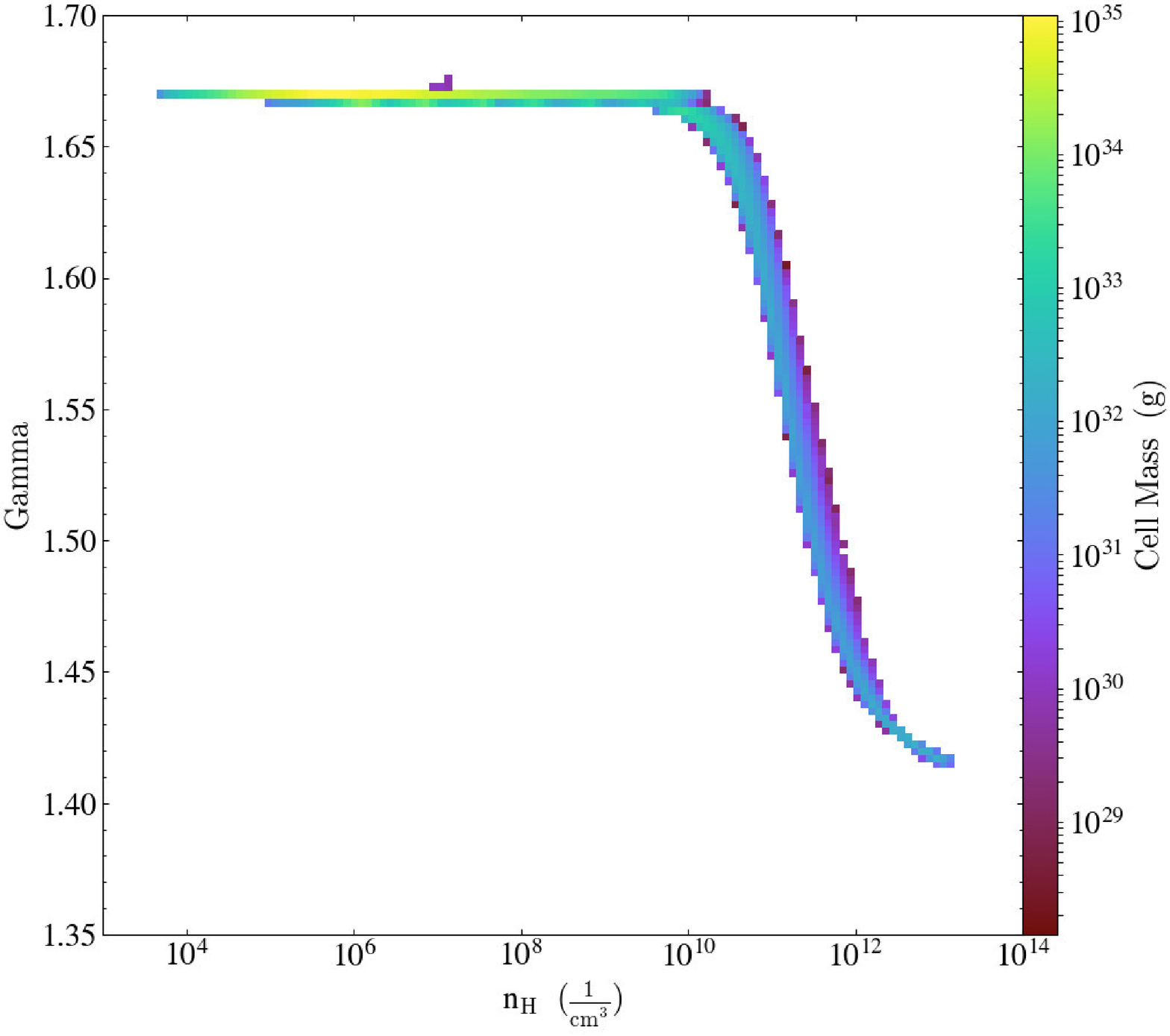}
\includegraphics[width=.47\textwidth]{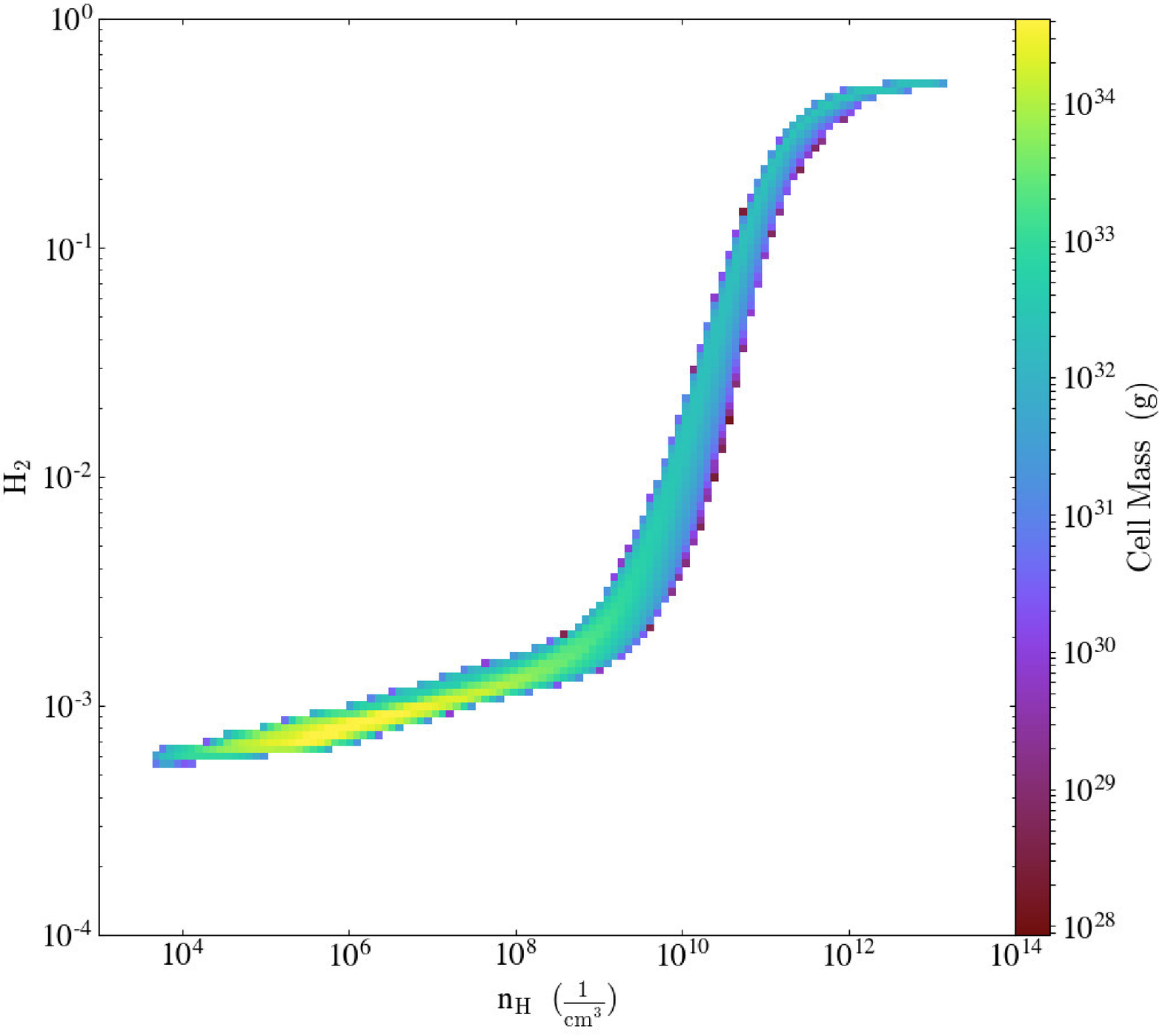}
 \caption
 {
 Variation of different quantities with density in the {\sc orion2} MHD simulation  just prior to initial sink formation ($t=0$).
 {\it Top left:} Temperature versus density.
 {\it Top right:} H$^+$ fraction versus density.
 {\it Bottom left:} Adiabatic index $\gamma$ versus density
 {\it Bottom right:} H$_2$ fraction versus density.
 This primordial gas evolution agrees well with previous work (e.g. Yoshida et al 2006, Greif et al 2012, Stacy and Bromm 2014).
}
\label{stuff_vs_n}
\end{figure*}

The evolution of the field in this stage is shown 
in panels b and c of 
Fig. \ref{rprof} and in Fig. \ref{nprof_bfield}. Most of the growth of the field in this stage is due to compression, as shown by the relative constancy of $B/\nh^{2/3}$ at densities $\nh\ga 10^7$~cm\eee\ in 
Fig. \ref{nprof_bfield}. Up to the point of sink formation ($t=0)$, the maximum value of $B/\nh^{2/3}$ occurs in the density range $\sim  10^{7-8}$~cm\eee. For $3\times 10^5\mbox{ cm\eee}<\nh<10^8\mbox{ cm\eee}$, the {\sc orion2} results for the magnetic field at $t=0$ agree with the {\sc gadget-2} results at $t=-9000$~yr to within 10 percent. This agreement makes sense since the free-fall time is less than 9000 yr only for $\nh>2.5\times 10^7$~cm\eee. At higher densities, a
power-law fit to the field at $t=0$ gives
\beq
B\simeq 
7.7\times 10^{-8}\,\nh^{0.59}~\pm0.05\mbox{ dex}~~~(10^{12}>\nh>4\times 10^7)
\label{eq:59}
\eeq
where again,
$B$ is in G and $\nh$ in cm\eee. It should be noted that the actual slope of the $B-\nh$ relation varies with density: It is closer to $\frac 23$ than to 0.59 for $\nh\la 8\times 10^7$~cm\eee\ and also for a short range around $\nh=10^{10}$~cm\eee. The fact that the slope of the $B-\nh$ relation is less than 2/3 is consistent with the result of \citet{xu20} that flux-freezing is violated in the nonlinear dynamo (see sections \ref{sec:kine} and \ref{sec:nonlinear}).

\begin{figure}
\includegraphics[width=.48\textwidth]{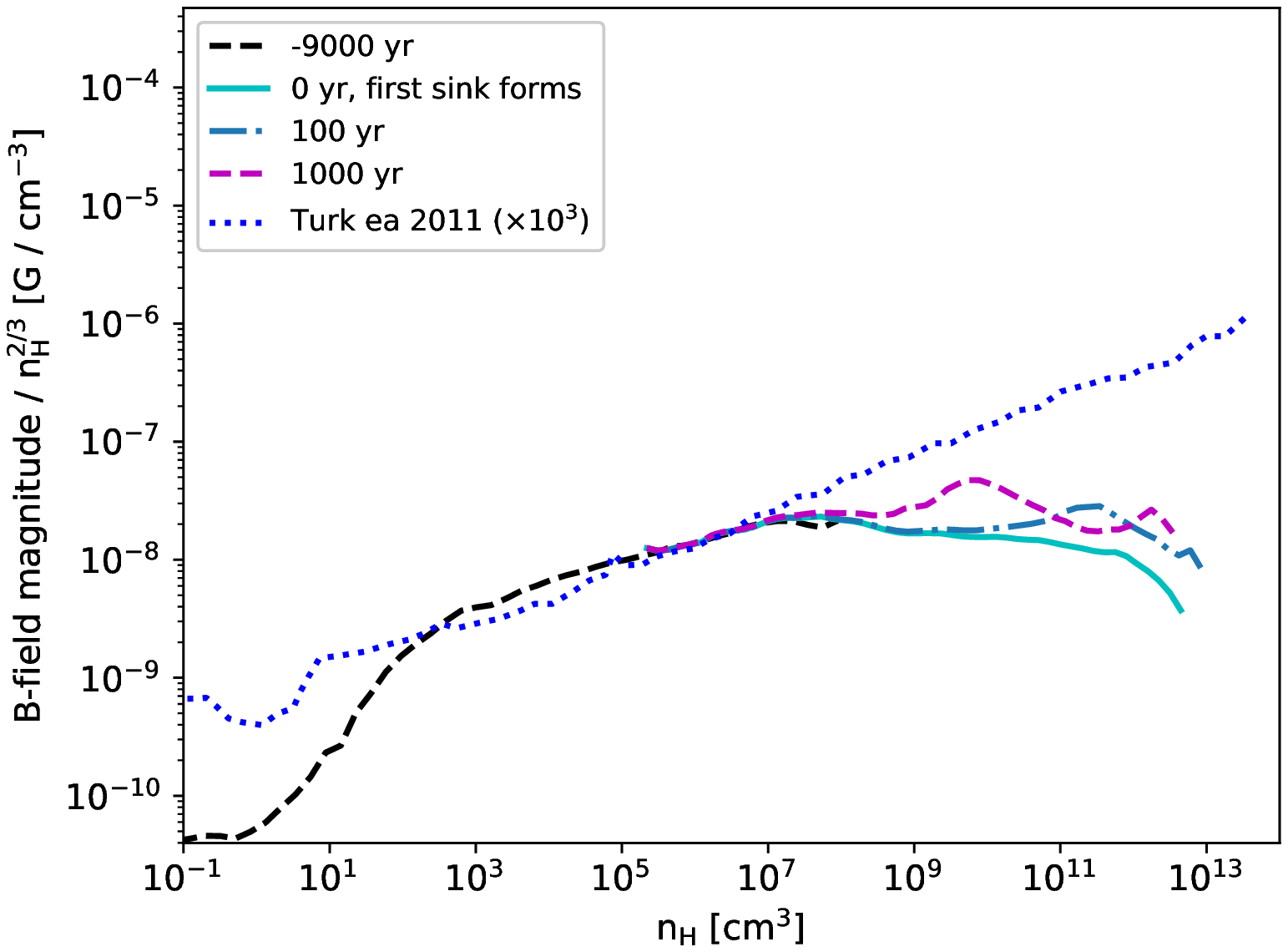}
 \caption
 {
The density-normalized magnetic field, $B/\nh^{2/3}$,
versus density at various times of the gas evolution in {\sc orion2}.  Times are shown relative to when the first sink formed.  The dashed black line is the initial {\sc orion2} magnetic field profile ($t=-9000$ yr).  
The \citet{turketal2012} field continues to increase at high densities because it is about 1000 times weaker than the field in our simulation, so the dynamo remains in the kinematic stage.
 }
\label{nprof_bfield}
\end{figure}

\begin{figure*}
\includegraphics[width=.48\textwidth]{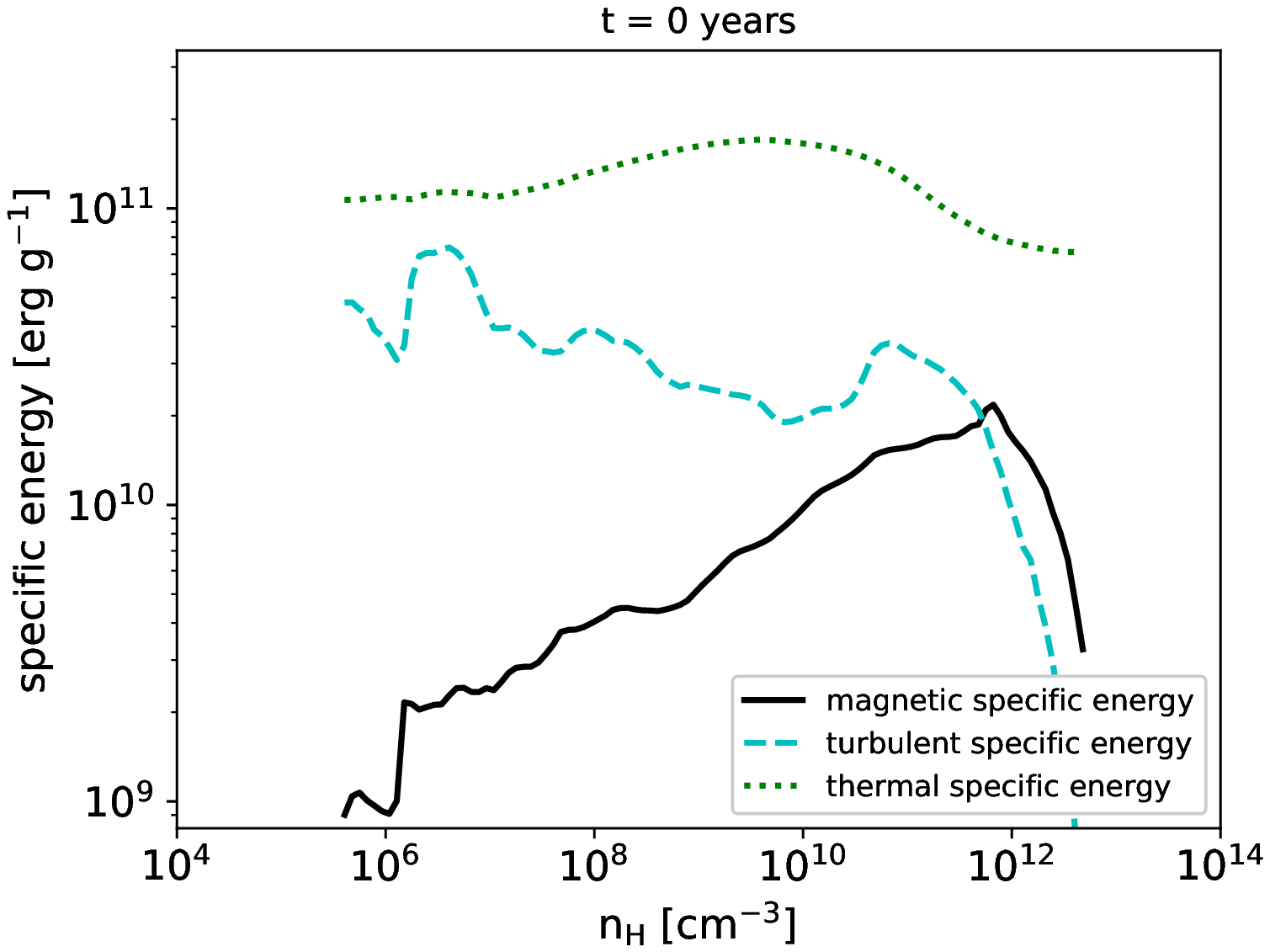}
\includegraphics[width=.48\textwidth]{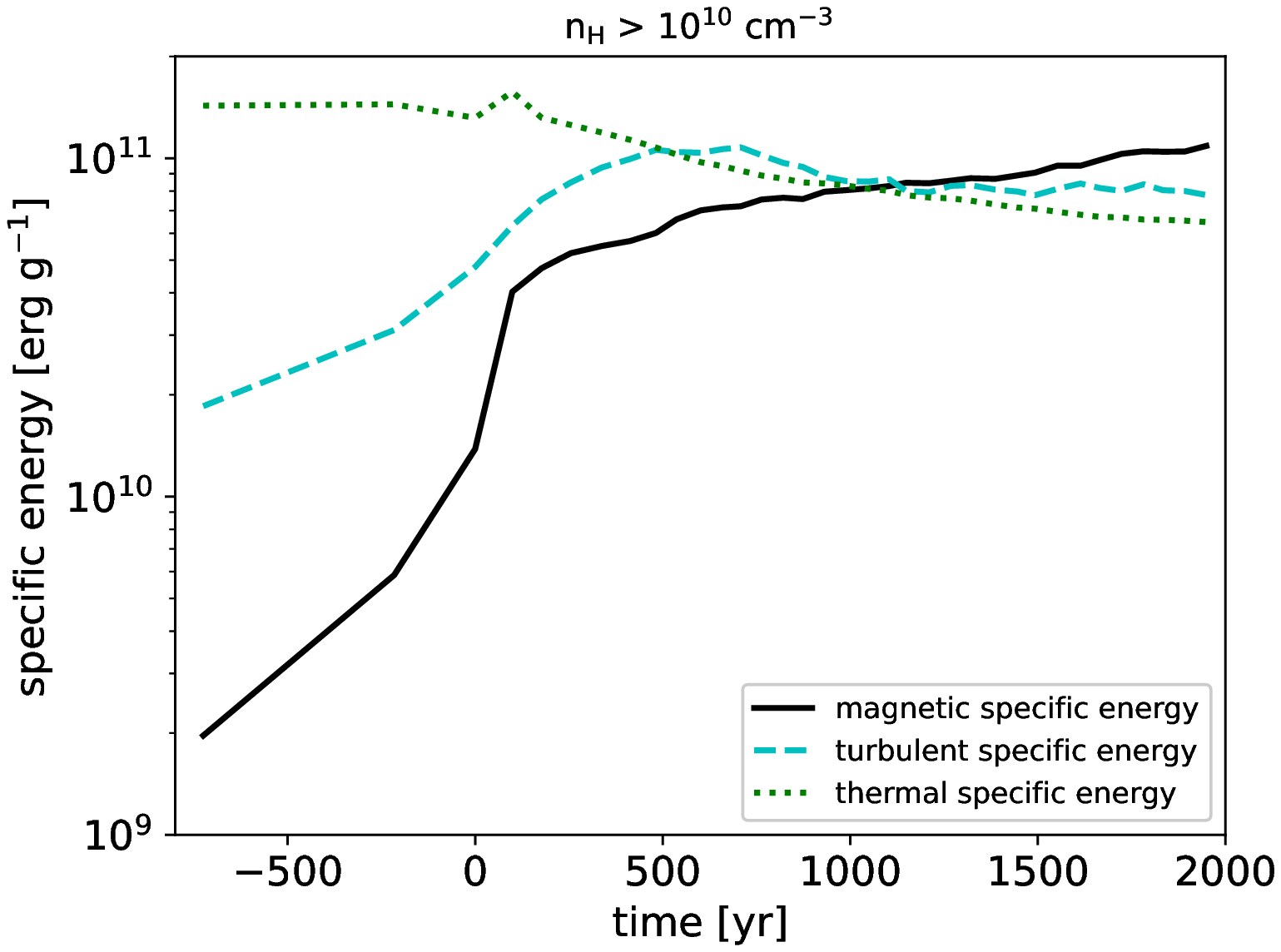}
 \caption
 {
{\it Left:} Comparison of magnetic ($\eb$), turbulent ($\frac 12 v_t^2$), and thermal ($\frac 32 \cs^2$) specific energies in the minihalo immediately before the first sink forms in the {\sc orion2} simulation.  
{\it Right: } Evolution of the magnetic, turbulent, and thermal specific energies in the central gas 
($\nh>10^{10}$~cm\eee) 
over time. 
The first sink forms at $t=0$.  
The magnetic energy quickly becomes comparable in strength to, and then exceeds, the 
thermal and turbulent
energies.
 }
\label{13a_nprof_erat.eps}
\end{figure*}

The slope 0.59 of the $B-\nh$ relation is intermediate between flux-freezing (slope of $\frac 23$) and equipartition (slope of $\frac 12$). As a result, the field gets closer to equipartition as the density increases. The equipartition field is
\beq
B_{\rm eq}=1.06\times 10^{-6}\left(\frac{\vtf}{2}\right)\nh^{1/2}~~~\mbox{G},
\eeq
where we have normalized the turbulent velocity to the typical value for $\nh\ga 10^8$~cm\eee\ in the lower panel of Fig. \ref{velprof}. 
The values of the specific energy of the magnetic field, the turbulence, and thermal motions are portrayed in Fig. \ref{13a_nprof_erat.eps}a.
Just before the formation of the first sink particle, 
the field energy is half the equipartition value at $\nh\simeq 10^{10}$~cm\eee. The field becomes stronger than equipartition for $\nh\ga 10^{12}$~cm\eee\
because numerical viscosity damps the velocities at high densities, which are not well resolved (Fig. \ref{velprof}, lower panel).

\section{Evolution of the Protostar-Disk System ({\sc orion2})}
\label{sec:evoln}

We turn now to the results of the {\sc orion2} simulation at $t>0$, after the first sink has formed. A disk forms around this sink, and further sinks can form when it fragments. These sinks represent low-mass stars. 
To see how the magnetic field affects the formation of these stars, we compare the hydrodynamic {\sc orion2} simulation with the main MHD {\sc orion2} simulation.

\subsection{Disk Fragmentation}
\label{sec:disk}

\subsubsection{Hydrodynamic case}

We begin by discussing the results of the hydrodynamic run with {\sc orion2}. In the absence of rotation,
 fragmentation can occur in regions in which the enclosed mass exceeds the Bonnor-Ebert mass,
\beqa
M_{\rm BE} &\hspace{-0.2cm}=\hspace{-0.2cm}&1.182\,\frac{\cs^3}{(G^3\rho)^{1/2}},\\
&\hspace{-0.2cm}\simeq\hspace{-0.2cm}& 4010  \left(\frac{T}{1000 \rm K} \right)^{3/2} \left(\frac{ n_{\rm H}}{10^4 \,\rm cm^{-3}} \right)^{-1/2}~~\msunm,
\label{eq:mbe}
\eeqa
where the numerical evaluation is for atomic gas.
Regions that grow in mass more quickly are more prone to fragmentation.
The mass inside a radius $r$ grows at a rate
\begin{equation}
\dot{M}(r) = 4 \, \pi r^2 \rho(r) v_r(r) \mbox{.}
\end{equation}
Fig. \ref{radprof_mdot} shows the accretion rate, $\dot{M}(r)$, versus radius at $t=0$.  At this time, the MHD and hydrodynamic cases look similar since magnetic fields do not yet have strong effects in the MHD case.  In both cases $\dot{M}$ ranges between 10$^{-2}$ and 10$^{-1}$ M$_{\odot}$~yr$^{-1}$, consistent with the rate at which the sink mass grows (see Section \ref{sec:sink}).

\begin{figure}
\includegraphics[width=.48\textwidth]{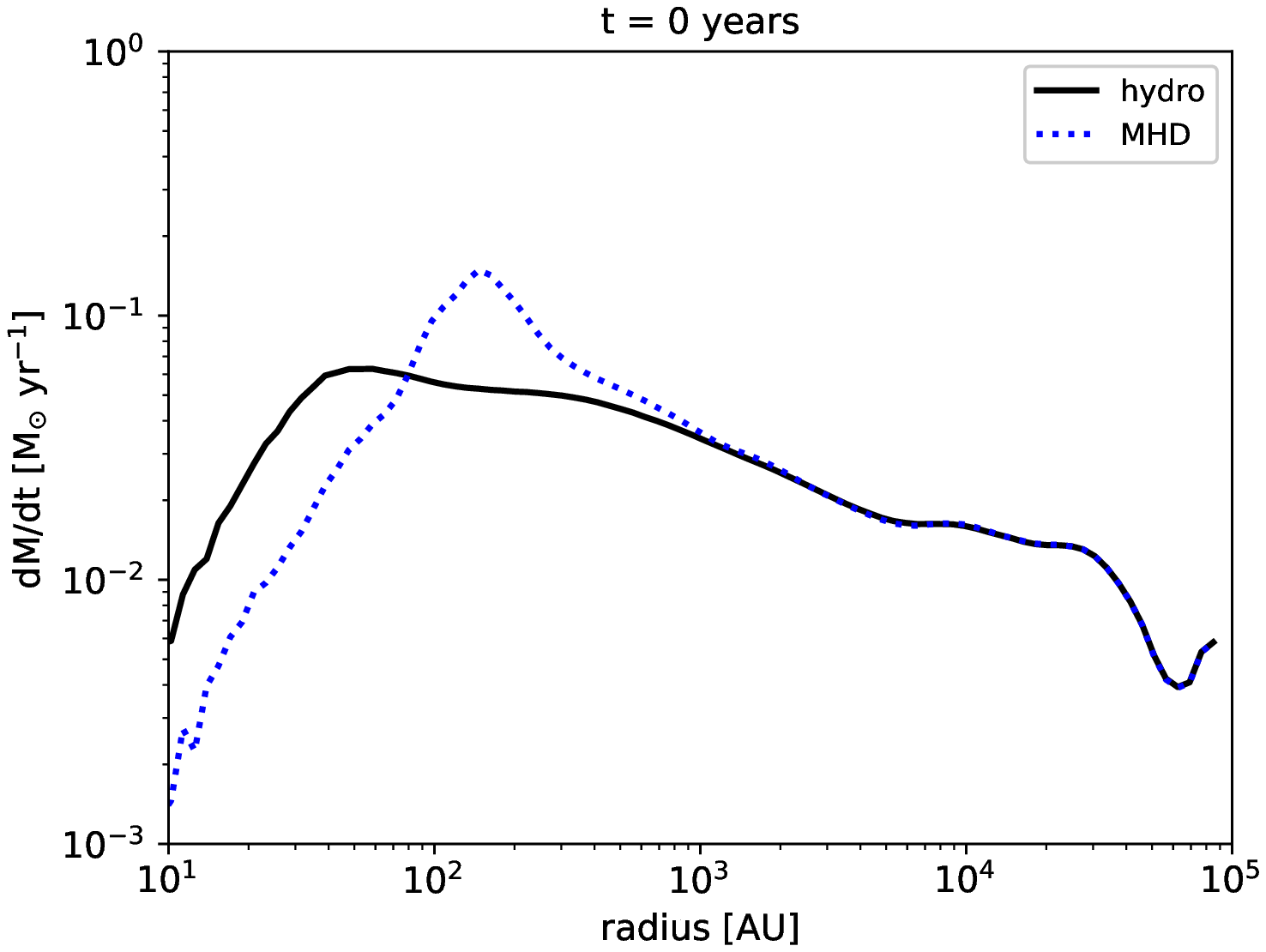}
 \caption
 {
$\dot{M}$ versus radius at $t = 0$ years, the time of initial sink formation.  The hydrodynamic and MHD cases are similar, with the exception of a slight enhancement of  $\dot{M}$ at 100 AU and a deficit at smaller radii in the MHD case.  
  }
\label{radprof_mdot}
\end{figure}

In the hydrodynamic case, secondary sinks form 
throughout the 2000 yr following initial sink formation
in regions of the disk where high accretion rates lead gas cells to become very dense and supercritical.  In the hydrodynamic case, a supercritical grid cell is one in which 
$M>\mbe$,
where $M$ is the mass within the grid cell and $M_{BE}$ is the Bonnor-Ebert mass (eq. \ref{eq:mbe}). Sink cells are created when there are fewer than 4 grid cells per Jeans length at the finest level (level 8; see Appendix \ref{app:refine}), corresponding to one grid cell at level 6. (Recall that refinement to a higher level occurs when there are fewer than 64 cells per Jeans length, quite different from the criterion for sink formation.)
To illustrate when the hydrodynamic case has periods of instability to fragmentation, Fig. \ref{mrat} shows the maximum value of $M/M_{\rm crit}$ 
on the sixth level of refinement (hereafter, the most critical level 6 cell).  $(M/M_{\rm crit})_{\rm max}$ exceeds unity at multiple times after initial sink formation, generally close to times when secondary sinks form. There is not a one-to-one correspondence since sink formation is based on the density and temperature in the cells at the finest level of refinement (level 8), and the maximum density at that level generally exceeds the maximum density at level 6.  Indeed, we found that $(M/M_{\rm crit})_{\rm max}$ at level 8 consistently exceeds unity when there is sink formation.
To gain insight into the fragmentation, we plot $\rho_{\rm max}$, the density in the most critical level 6 cell as a function of time in the left panel of Fig. \ref{beta_rhomax}.  
Times of high densities in these cells often correspond to periods of secondary sink formation.

\begin{figure}
\includegraphics[width=.48\textwidth]{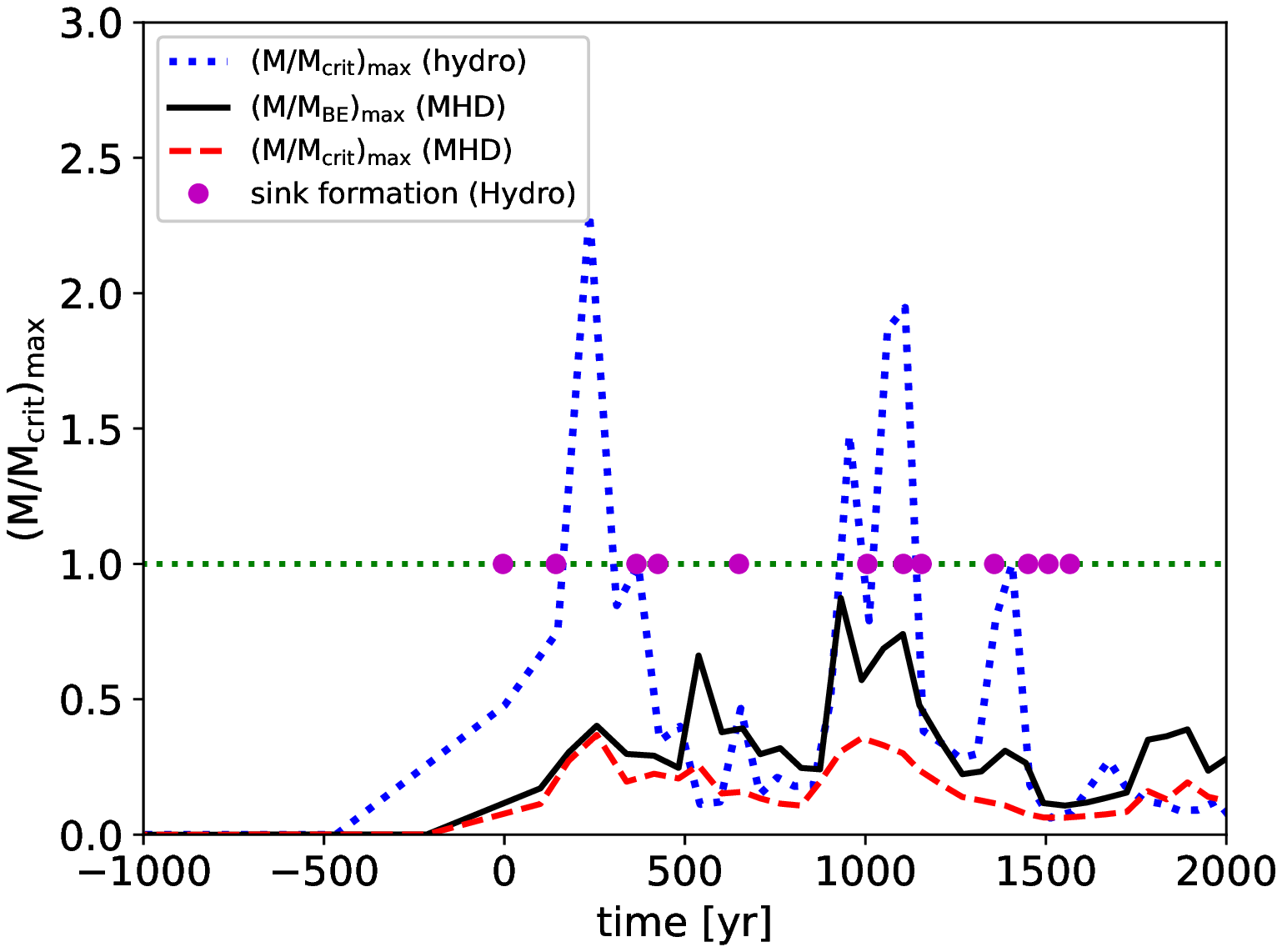}
\caption{
$(M/M_{\rm crit})_{\rm max}$ and $(M/M_{\rm BE})_{\rm max}$ in the most critical
level 6 cell (corresponding to 64 cells at level 8, the finest level), 
where $M/M_{\rm crit}$ is maximum.  In the hydrodynamic case, $(M/M_{\rm crit})_{\rm max}$ exceeds unity at multiple times after initial sink formation, consistent with the formation of secondary sinks.  In contrast, both $(M/M_{\rm crit})_{\rm max}$ and $(M/M_{\rm BE})_{\rm max}$ remain below unity throughout the MHD simulation, consistent with the suppression of secondary sink formation in the MHD case.
}
\label{mrat}
\end{figure}

\begin{figure*}
\includegraphics[width=.47\textwidth]{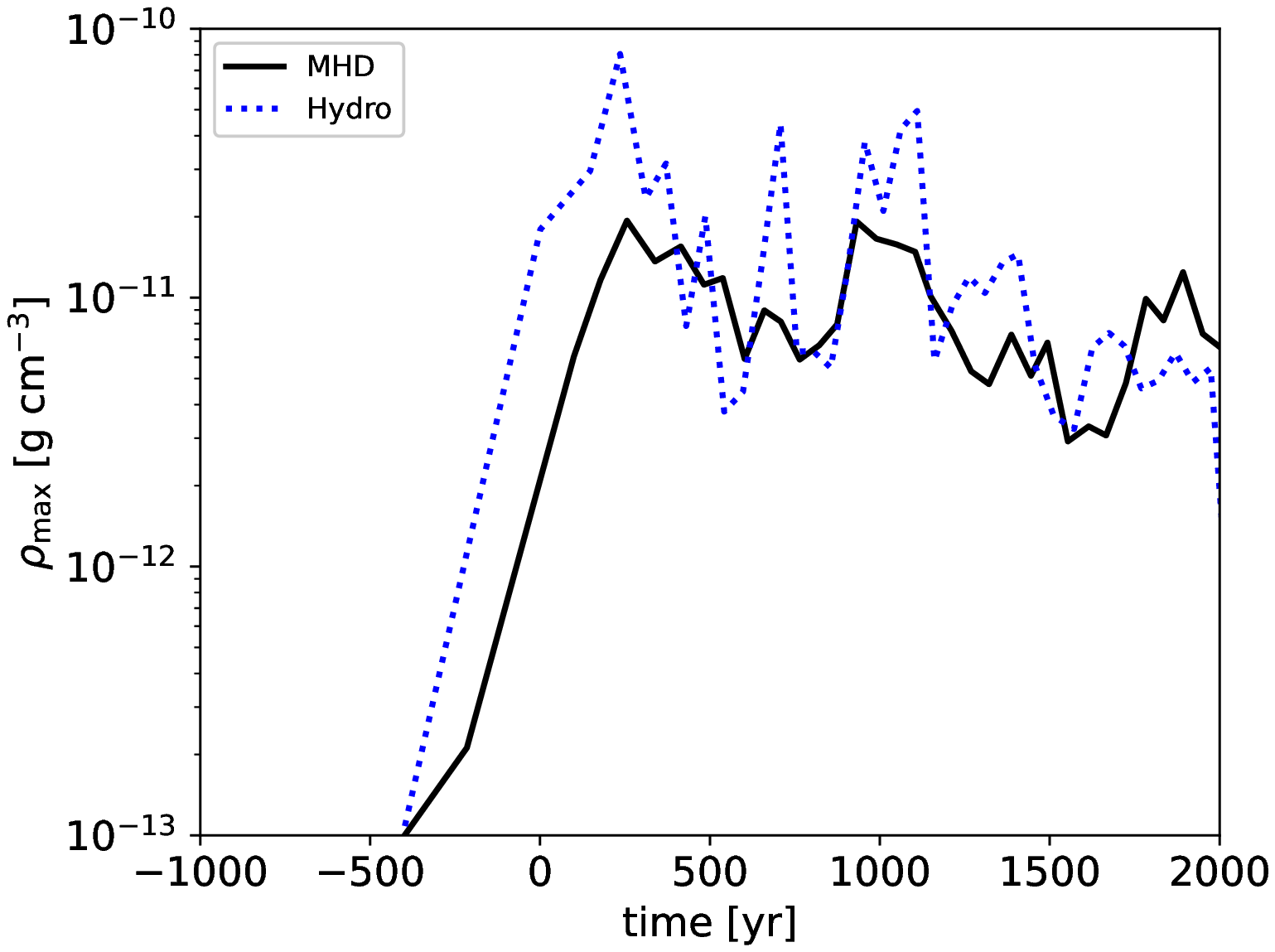}
\includegraphics[width=.47\textwidth]{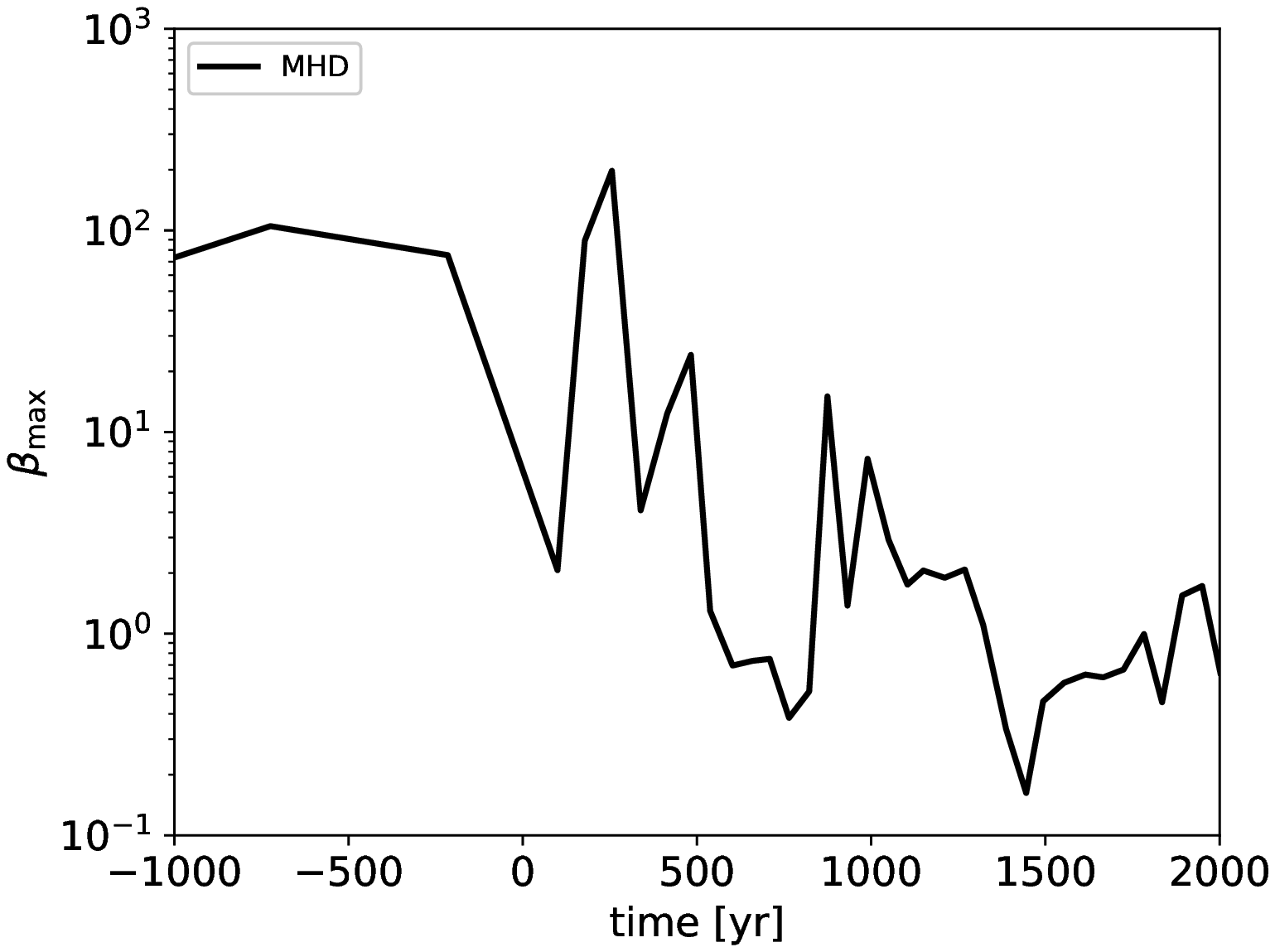}
\caption{
{\it Left:}  
Density $\rho_{\rm max}$ in the most critical level 6 grid cell (corresponding to 64 cells at the finest level), where $M/M_{\rm crit}$ is maximum.
{\it Right:} $\beta_{\rm max}$, the value of $\beta$ in the most critical level 6 grid cell,
where $\beta = 8\pi\rho c_s^2 / B^2$ describes the thermal pressure relative to the magnetic pressure.
Note that $\rho_{\rm max}$ in the hydrodynamic case is generally higher than in the MHD case, indicating that magnetic fields make the gas less subject to fragmentation and the formation of secondary sinks. 
}
\label{beta_rhomax}
\end{figure*}

\begin{figure}
\includegraphics[width=.48\textwidth]{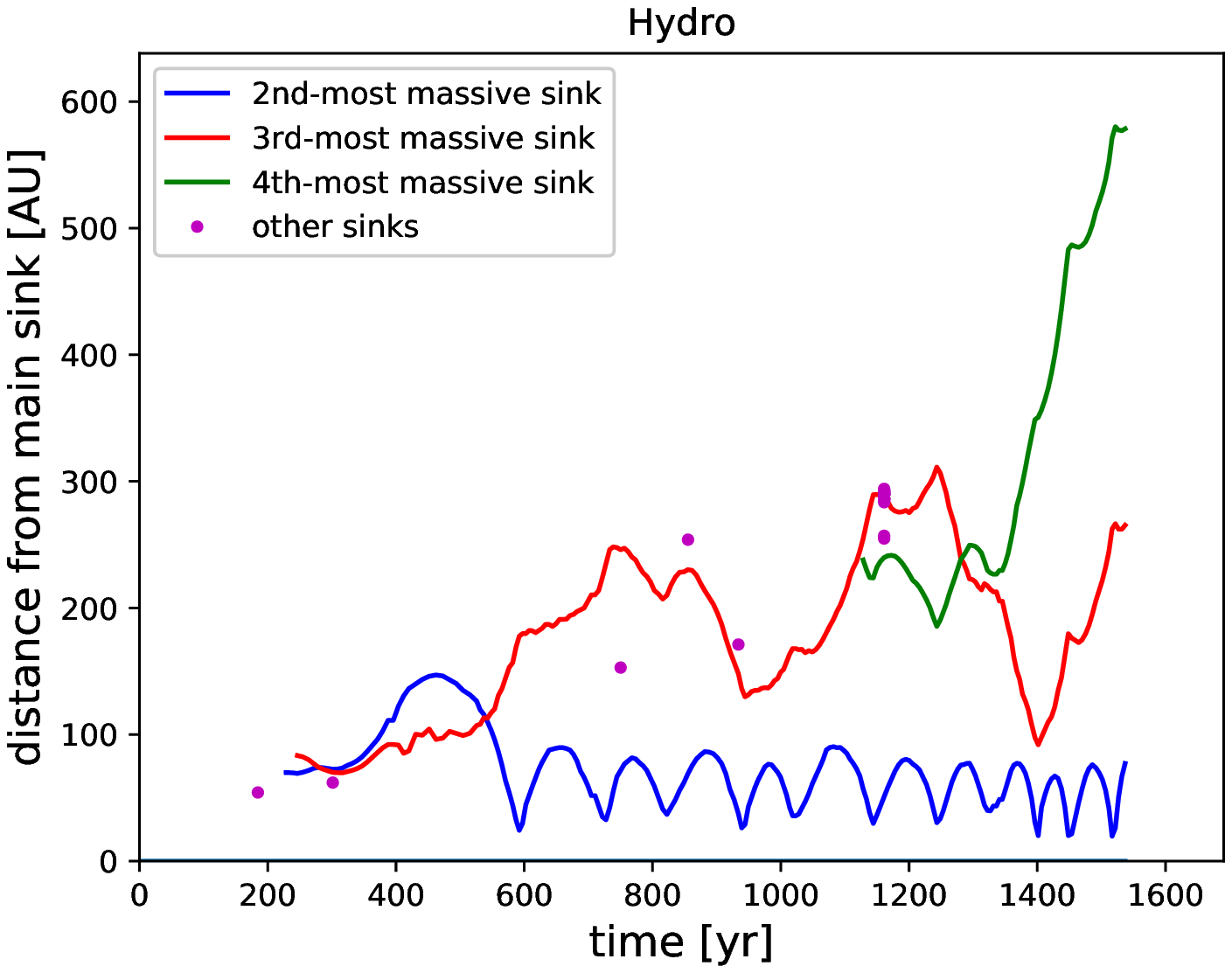}
 \caption
 {
Distance of the three largest secondary sinks from the main sink over time in the hydro simulation.  Magenta-colored dots indicate the time and initial distance at which smaller secondary sinks formed.  Secondary sinks form throughout the hydro run at distances ranging from 50 to 300 au.
  }
\label{sinkdis}
\end{figure}

Secondary sinks form in the disk at locations that range between 50 and 300 au from the initial sink (Fig. \ref{sinkdis}).  
We note a particular period of large $\rho_{\rm max}$ and high $(M/M_{\rm crit})_{\rm max}$ at $t = 1000$ yr, when a burst of sink particle formation occurs (Fig. \ref{sinknum}),
The growth of disk instabilities and evolution of density are shown in Fig. \ref{dens_morph}.
After the secondary sinks form within the densest gas
in the hydrodynamic simulation,
many of them are ejected to larger orbits in lower-density gas before reaching high masses, opening the possibility of long-surviving low-mass Pop III stars.  However, as we discuss in the next section, 
this does not occur in the MHD simulation.

\begin{figure*}
\includegraphics[width=.47\textwidth]{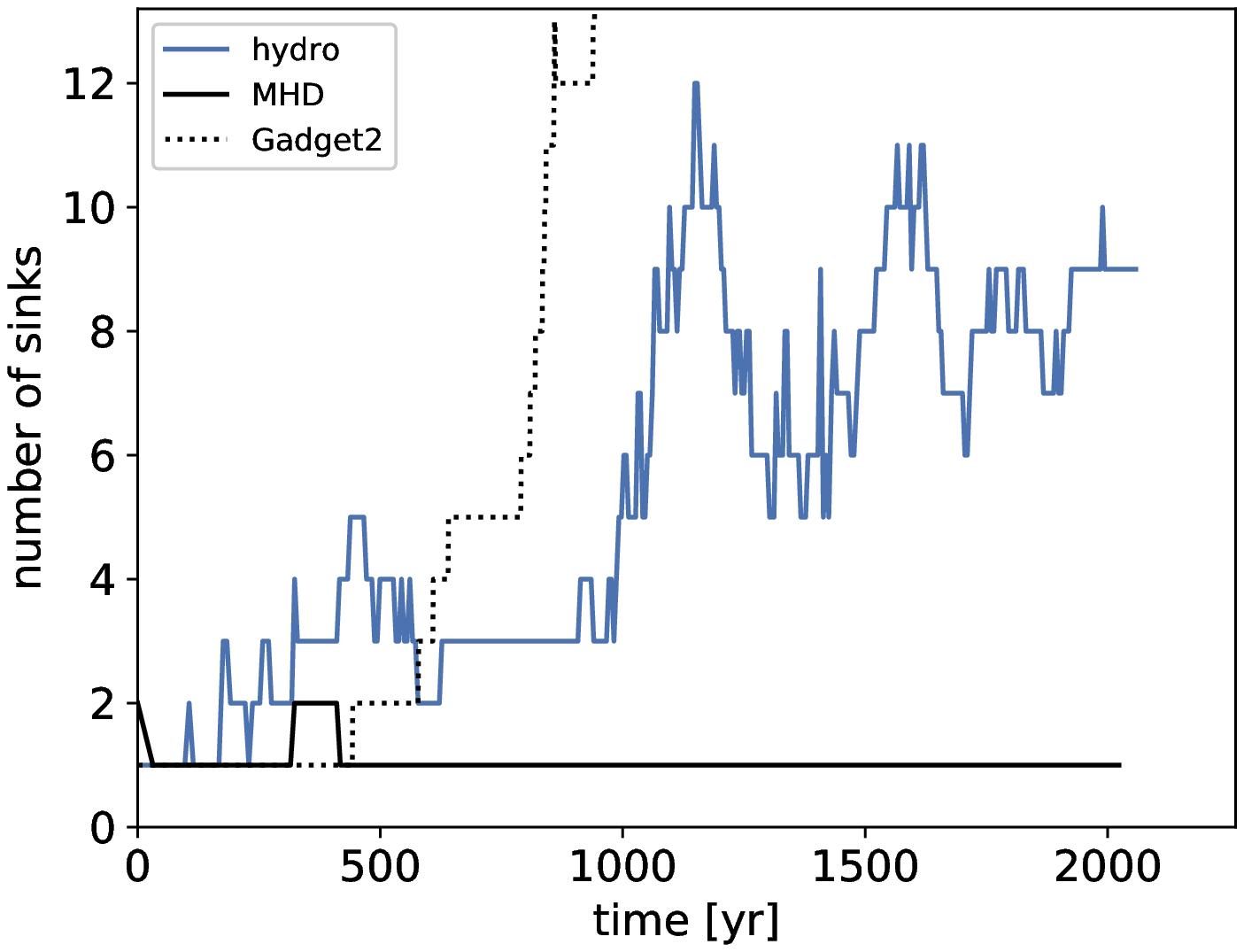}
\includegraphics[width=.47\textwidth]{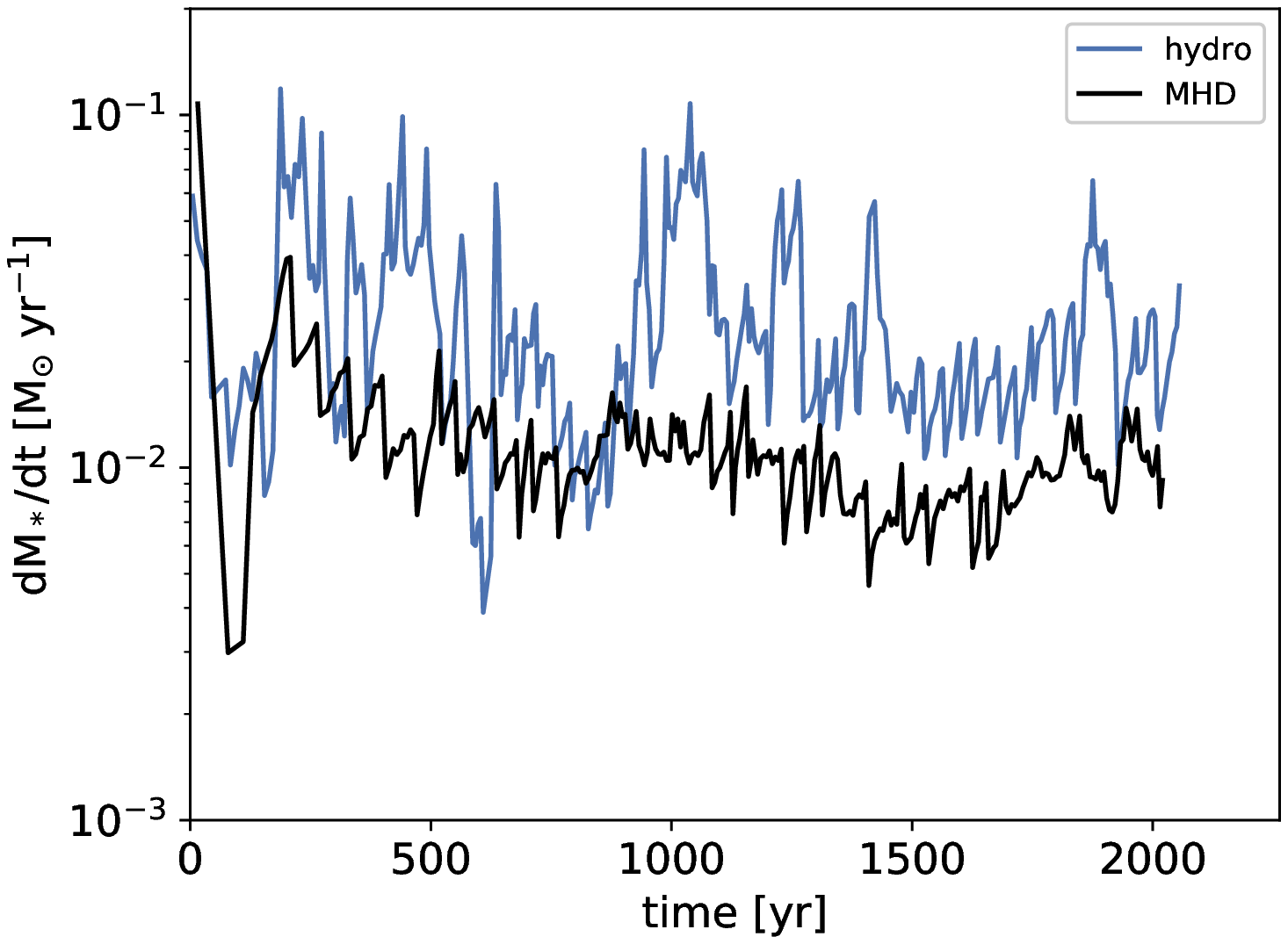}
 \caption
 {
{\it Left:} Number of surviving sinks in the minihalo over time.  Blue line represents the evolution without inclusion of magnetic fields.  Black line depicts the result when magnetic fields are included.  The dotted line shows the number of surviving sinks over time within the same halo in a previous {\sc gadget-2} simulation (Stacy et al. 2016).  A rapid rise in sink number over time after $\sim$ 1000 yr is apparent in both the 2016 {\sc gadget-2} and hydrodynamic {\sc orion2} runs.
{\it Right:} Rate of mass accretion onto sink system over time.  The accretion rate onto the sink system of the hydro case is generally larger than the accretion rate onto the single sink of the MHD case
because there are more accretion sites.
}
\label{sinknum}
\end{figure*}

\subsubsection{Effect of magnetic fields}
\label{sec:effect}

We turn now to the results of the {\sc orion2} MHD simulation.
After initial sink particle formation, the magnetic field increases in strength within the dense gas near the sink (Fig. \ref{rprof}). 
We note here that while sink particles accrete mass, they do not accrete magnetic flux.  For contemporary star formation, the approximation that protostars do not accrete flux is supported by observations that young stellar objects have orders of magnitude less flux than the gas clouds from which they form; theoretically, this is due to the effects of non-ideal MHD (e.g., \citealt{mckee&ostriker2007}). Non-ideal effects are weaker near Pop III protostars since the ionization is higher than near contemporary protostars (e.g., compare the results of \citealp{greifetal2012} for Pop III star formation with those of \citealp{dapp12} for contemporary star formation) and the resistivity therefore lower. However, reconnection diffusion  efficiently allows magnetic flux to diffuse at a rate that is independent of the resistivity \citep{laza14}, so the approximation that most of the magnetic flux is not accreted should be valid for Pop III stars as well.
Because magnetic flux is not allowed to accrete onto the sink particle whereas mass does, the ratio of magnetic energy to mass (the specific magnetic energy) initially increases
(Fig. \ref{13a_nprof_erat.eps}b).
At late times ($t\ga 1000$ yr) the magnetic 
energy slightly exceeds the thermal and turbulent energies
for high-density gas ($\nh \ga 10^{10}$ cm$^{-3}$,
corresponding to $r \la 10^{16}$ cm) after sink formation.

The relative strength of magnetic and gravitational forces is characterized by the magnetic critical mass, at which the magnetic energy of the gas equals its self-gravitational energy,
\beq
M_\Phi=c_\Phi\,\frac{\Phi}{G^{1/2}},
\eeq
where $\Phi$ is the magnetic flux and $c_\Phi\simeq 1/(2\pi)$ \citep{mckee&ostriker2007}. 
Clouds that can
undergo gravitational collapse in the presence of a magnetic field are termed magnetically supercritical ($M>M_\Phi$), whereas those that are magnetically supported so that they cannot collapse are termed magnetically subcritical ($M_\Phi>M)$.
Including the effects of both thermal pressure and magnetic fields, the critical mass for gravitational collapse is \citep{mcke89}
\beq
M_{\rm crit}\simeq M_\Phi+\mbe.
\eeq

The simulation shows that magnetic fields strongly suppress fragmentation:
In contrast to the hydrodynamic run, only two 
secondary sinks form, at approximately 30 and 300 years (Fig. \ref{sinknum}).  Furthermore, these sinks are very short-lived--they form at 30-40 au from the main sink, and they quickly merge with the main sink (see the following section). 
Fig. \ref{mrat} shows that fragmentation is suppressed in the MHD case by both thermal effects and magnetic effects. Thermal effects are measured by $M/\mbe\propto (\rho/T)^{3/2}$, which is generally smaller for the MHD case than the hydrodynamic case (Fig. \ref{mrat}). As shown in Fig. \ref{beta_rhomax}, the lower density in the MHD case plays a significant role in this:  The mass in the most critical level 6 cell never exceeds the Bonnor-Ebert mass, although it comes close at $t\simeq 1000$~yr.  The strength of magnetic fields relative to thermal pressure can be characterized by $\beta=8 \pi \rho c_s^2 / B^2$.
Fig. \ref{beta_rhomax} shows the value of $\beta$ in the most critical level 6 cell, $\beta_{\rm max}$, as a function of time.  This value first drops below unity at $t\simeq 500$~yr. Decreases in $\beta_{\rm max}$ tend to correspond to decreases in $\rho_{\rm max}$ (left panel of Fig. \ref{beta_rhomax}), illustrating that magnetic fields can limit the ability of the gas to collapse to high density.
At late times ($t\ga 1300$~yr), magnetic pressure generally dominates thermal pressure in the most critical level 6 cell.

Magnetic fields also directly suppress fragmentation by significantly increasing the critical mass for $t\ga 500$~yr, when $M_{\Phi}\sim M_{\rm BE}$.  Together, thermal and magnetic effects keep $M/M_\crit<0.4$. Fig. \ref{bfield_morph} shows that the area occupied by strong magnetic fields increases with time.

Fig. \ref{dens_morph} contrasts the MHD run, with no secondary sinks at the times shown, with the hydrodynamic case, which has 7 secondary sinks at the final time. 
While the magnetic fields do not significantly alter the large-scale morphology of the star-forming disk, they are sufficiently strong on small scales to inhibit secondary sink formation.

We note here that radiative feedback, which we did not include in this simulation, has also been found to lower fragmentation rates in Pop III star-forming gas at later times (e.g. \citealt{susa2013, stacyetal2016}).  As the first-forming protostars grow, their Lyman-Werner radiation dissociates H$_2$ and heats the gas, which
reduces fragmentation and lowers the sink accretion rate.
Fragmentation-suppressing effects of radiation have been found in current-day star formation as well, leading to the peak IMF of the Milky Way to lie at about 0.2 $\msunm$ instead of at 0.004 $\msunm$, the opacity limit for fragmentation (e.g. \citealt{krumholzetal2016}).  However, Pop III radiative feedback typically has not been found to prevent fragmentation entirely (e.g. \citealt{stacy&bromm2014}).  Here we have found that magnetic fields can have even stronger fragmentation-suppressing effects at earlier times during sink accretion.

\begin{figure*}
\includegraphics[width=.85\textwidth]{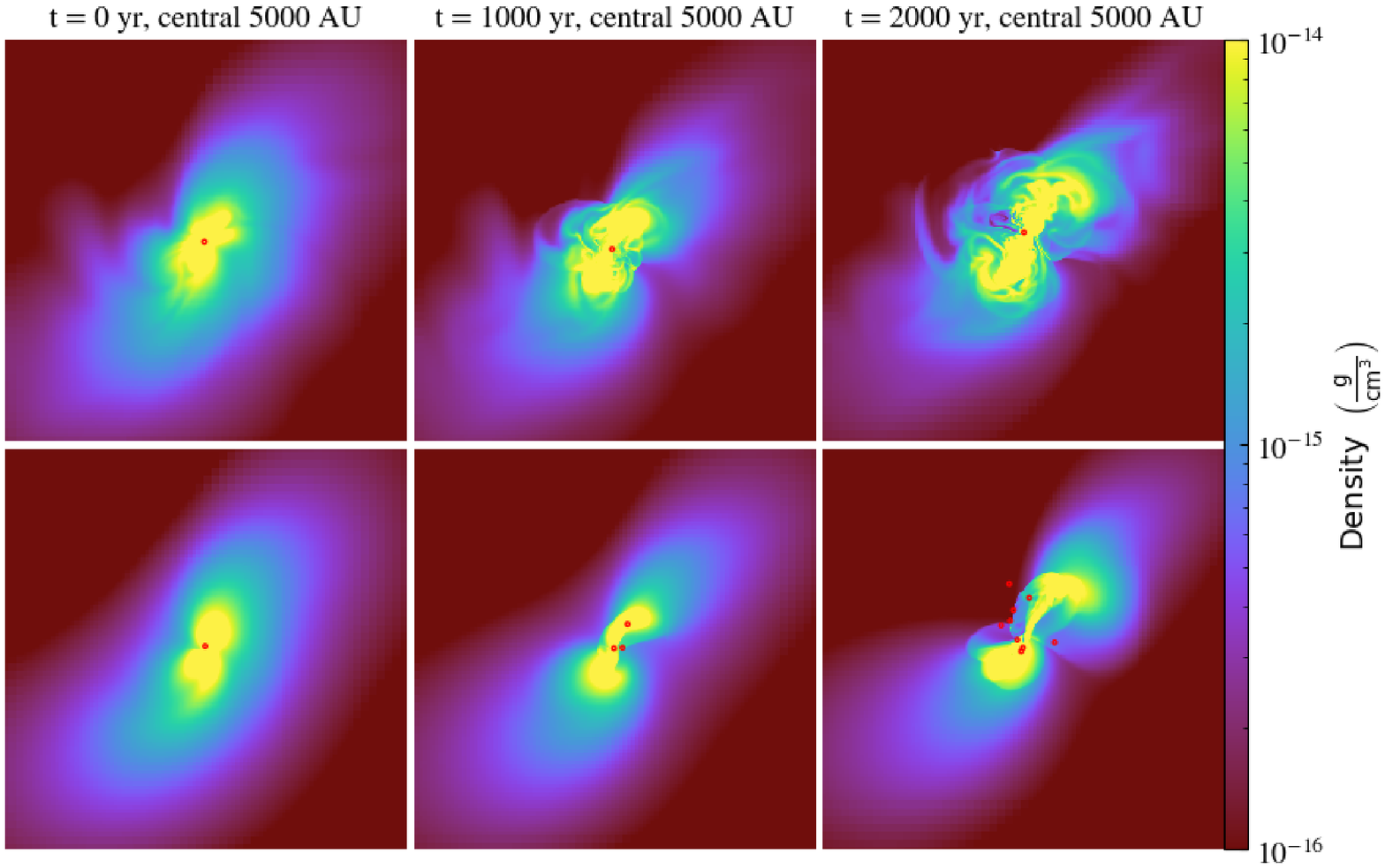}
 \caption
 {
 Slice of density through the most massive sink particle in the minihalo, shown at 0, 1000, and 2000 yr after the initial sink formation.  
 Top row is the MHD simulation.  Bottom row is the hydro simulation.
 Circles denote the location of sink particles.  Box size is 5000 au.  The MHD run has only one surviving sink particles and a less smooth density structure compared the hydro run, in which several sinks form.  
  }
\label{dens_morph}
\end{figure*}


\begin{figure*}
\includegraphics[width=.85\textwidth]{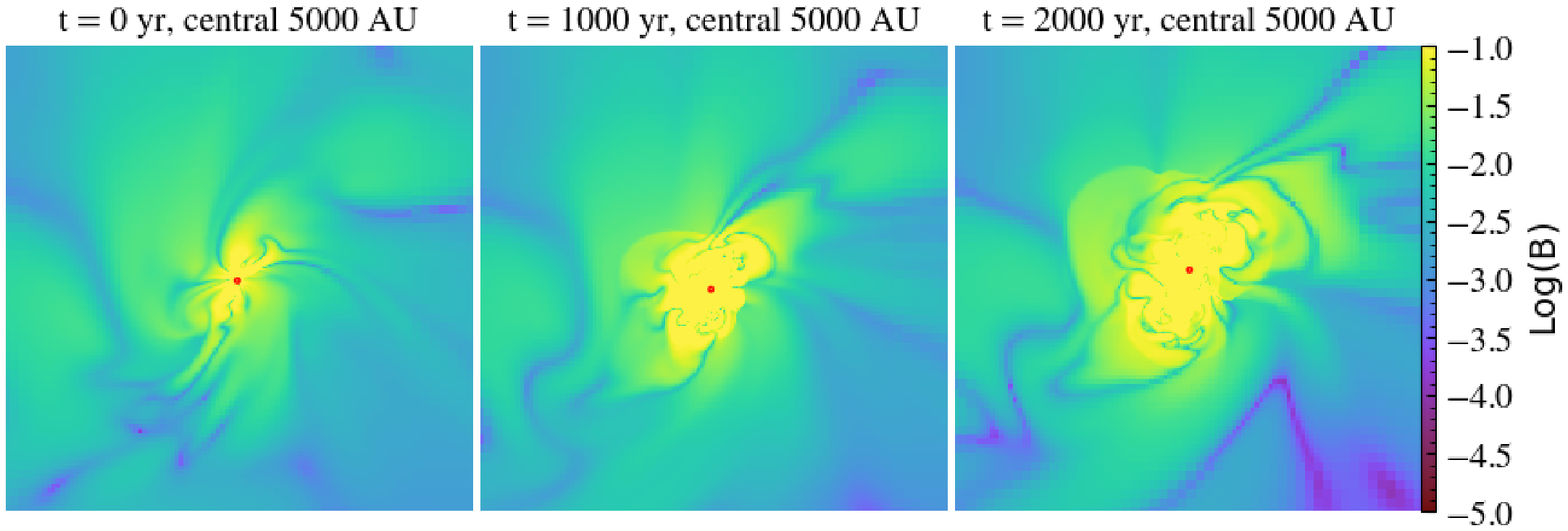}
 \caption
 {
Same as Fig. \ref{dens_morph}, but depicting B-field magnitude for the MHD run.  Regions of high B-fields grow gradually larger over time, extending steadily further from the main sink.
  }
\label{bfield_morph}
\end{figure*}

\subsection{Sink Accretion and Merging: The IMF}
\label{sec:sink}

The growth rate of the total sink mass in the hydro and MHD runs (Fig. \ref{sinkmass}) is similar for the initial 200 yr of accretion. 
Thereafter, the total sink mass increases much more rapidly in the hydro run due to the formation of several new sink particles 
(left panel of Fig. \ref{sinknum}). 
The left panel of Fig. \ref{sinkmass} also compares the hydro run with the equivalent 2016 {\sc gadget-2} run, in which this same cosmological minihalo was simulated  at a resolution length of $\sim$ 1 au.  
The growth of the initial sink in the hydro run is closely consistent with the evolution seen in the 2016 {\sc gadget-2} run
only for the first 200 yr.  
After this, the total sink mass in the hydro run 
temporarily
becomes considerably higher, which is likely due to the 
inclusion of radiative feedback in the 2016 {\sc gadget-2} run. 
In both the hydro and MHD cases, after 200 yr the total sink mass is $M_{\rm tot} \sim$ 7 $\msunm$, while the overall accretion rate is a few times 10$^{-2}$ $\msunm$ yr$^{-1}$ (right panel of Fig. \ref{sinknum}).  
While the total accretion rate remains high in the hydro run, it declines after about 200 yr in the MHD run;
however, if the effect of mergers is eliminated, the accretion rate onto the most massive star in the hydro run is similar to that in the MHD run.

\begin{figure*}
\includegraphics[width=.47\textwidth]{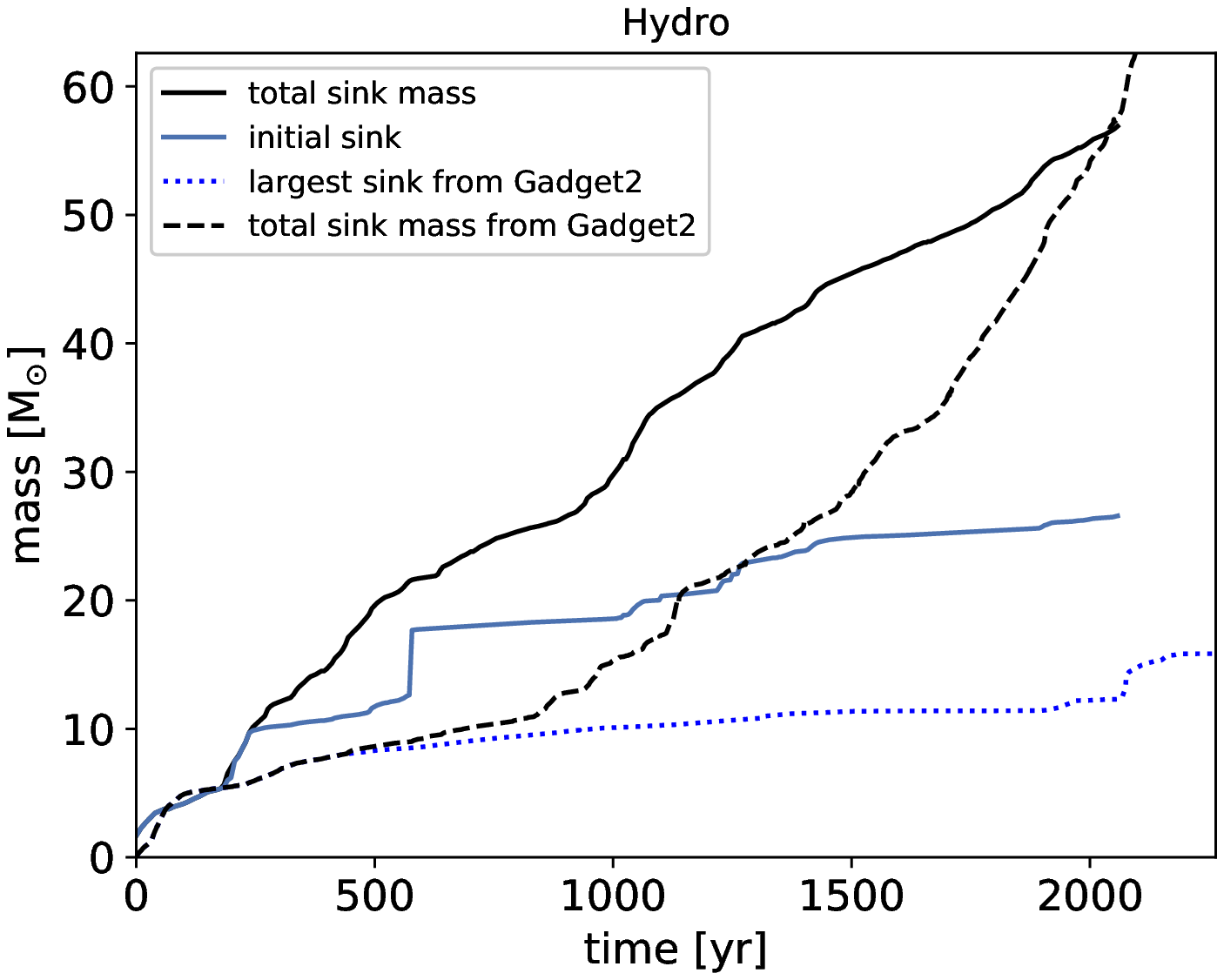}
\includegraphics[width=.47\textwidth]{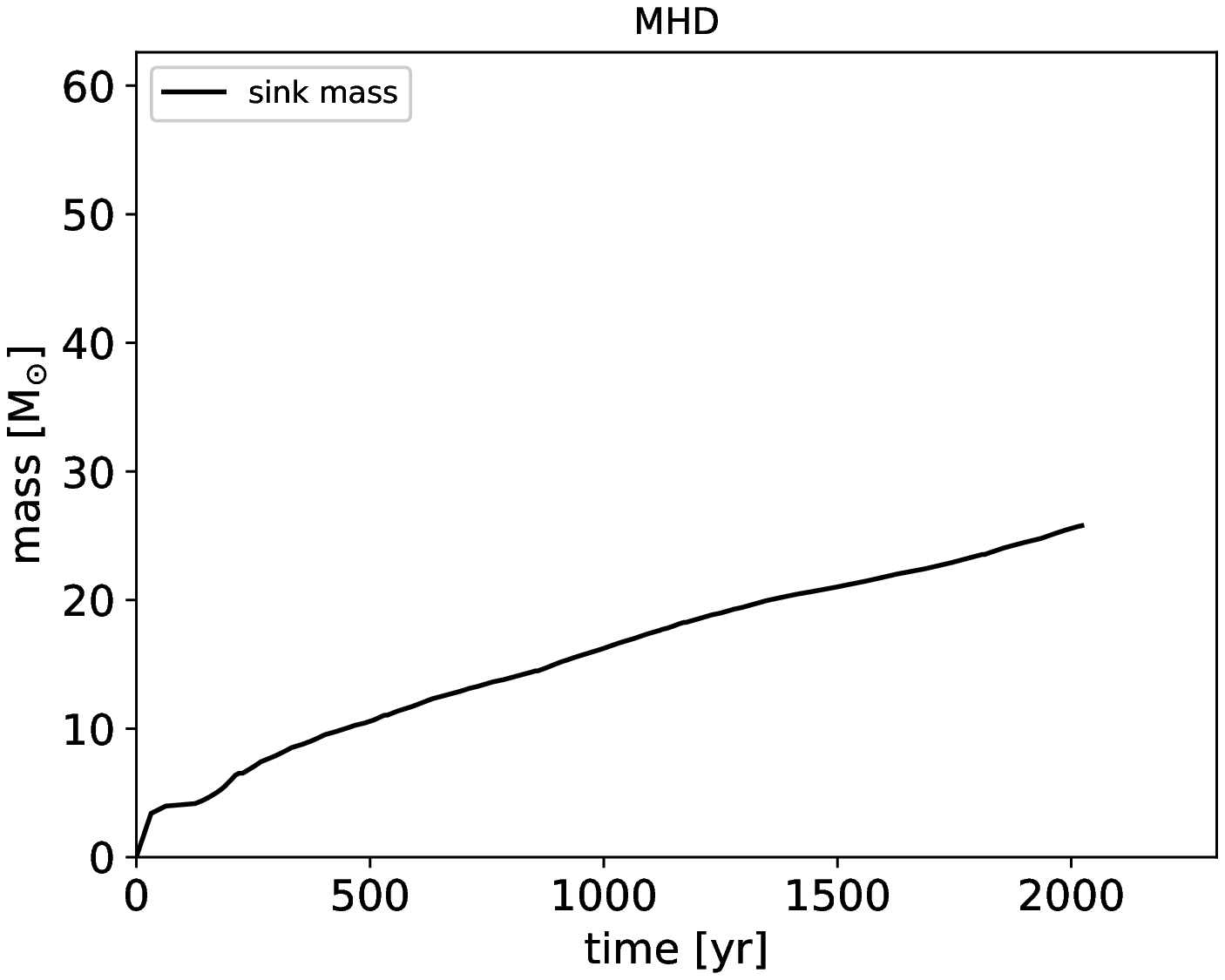}
 \caption
 {
 {\it Left:} Comparison of sink masses in the {\sc orion2} hydrodynamic simulation with the 2016 {\sc gadget-2} simulation, which had higher resolution and radiative feedback.
Solid black line shows the sum of all sinks in the {\sc orion2} simulation; dashed black line is for the same halo in the 2016 {\sc gadget-2} simulation. The solid (dotted) blue line shows 
shows the mass growth of the most massive sink in the {\sc orion2} ({\sc gadget-2}) simulation, respectively.
  {\it Right:} Same as left panel, but for the {\sc orion2} MHD run.  Unlike in the hydro case, the largest sink in the MHD case has no secondary sinks to merge with and thus undergoes smoother growth rate.
  }
\label{sinkmass}
\end{figure*}

These results are in semi-quantitative agreement with the analytic theory of first star formation of \citet{tan04}. They determined the accretion rate onto the star-disk system, with a total mass $m_{*d}=m_*+m_d=(1+f_d)m_*$. They predicted that the disk was significantly larger than the accretion zone around the sink, so we compare our result with their result for the accretion rate onto the star itself,
\beq
\dot m_*=0.070\,\left(\frac{\epsilon_{*d}K^{\prime 3/2}}{1+f_d}\right)t_{\rm yr}^{-3/10}~~~\msunm~\mbox{yr\e}
\eeq
from their equations (9) and (11), where $\epsilon_{*d}$ represents the possible loss of mass from the disk due to outflows and 
\begin{equation}
    K'=\frac{T_\eff}{300\,\mbox{K}}\left(\frac{10^4\mbox{ cm\eee}}{\nh}\right)^{0.1}
\end{equation}
is a normalized entropy parameter for a gas with an effective ratio of specific heats $\gamma=1.1$, which is typical of primordial star-forming gas \citep{omukai&nishi1998}. The effective temperature $T_\eff=(1+\calm^2/3)T$ includes the effects of turbulence. No outflows were observed in our simulations, so $\epsilon_{*d}=1$. 
The low resolution of our simulation prevents an accurate determination of $f_d$, so we shall adopt the \citet{tan04} value, $f_d=\frac 13$. They did not consider fragmentation or magnetic fields, so we compare with the total accretion rate for the hydrodynamic run. The predicted accretion rate, $\dot m_*=0.052K^{\prime 3/2}t_{\rm yr}^{-3/10}\,\msunm$~yr\e, agrees reasonably well with the results in Fig. \ref{sinknum} for $K'$ a little greater than unity. 
Equivalently, for $K'=1.5$ the integral of the accretion rate gives a predicted mass of $27\,\msunm$ at $t=2000$~yr, in agreement with the simulation.

Sink mergers occur rapidly in the hydro run (see Appendix \ref{app:refine} for more detail).  
Comparing the hydro and the 2016 {\sc gadget-2} runs, more sinks form while a smaller number of these sinks merge together in the 2016 {\sc gadget-2} run.  This leads to a significantly higher number of surviving sinks at later times in the 2016 {\sc gadget-2} run (dotted line in left panel of Fig. \ref{sinknum}).  
The difference in sink number may be attributed to the smaller resolution length of 1 au as well as the inclusion of radiative feedback in the 2016 {\sc gadget-2} run.  In both the hydro and 2016 {\sc gadget-2} runs, however, the sink merger rate becomes similar to the formation rate for $t\ga 1000$ yr. These rates cancel each other, yielding a roughly constant sink number at later times, $\sim$ 10 sinks in the hydro run and $\sim$ 40 sinks in the 2016 {\sc gadget-2} run.

We also note a significant merger of two sinks with masses $\sim$ 4 $\msunm$ and $\sim$ 12 $\msunm$ at 600 yr in the hydro run, visible as the jump in the blue line in Fig. \ref{sinkmass}.  In the last simulation snapshot prior to their merger, these two sinks had a relative distance and velocity of 50 au and 20 km s$^{-1}$.  These sinks were thus gravitationally bound.  It is likely that the sink merger represents the formation of an unresolved binary star instead of a true protostellar merger.  In their adiabatic expansion phases, Pop III protostars reach radii exceeding $ 100\, R_{\odot}$; eventually, their radii decrease
as the protostar evolves towards the zero-age main sequence \citep{tan04,hosokawaetal2010}.  However, these protostellar sizes are small compared to the 12 au accretion radius of the sink cell, so we cannot be sure that the stars formed a binary instead of merging.  Because a merger of such massive protostars would have led to significant radiative emission that we do not include, our simulation more closely models the outcome of a tight binary.

Given these rates of sink formation, accretion, and merging, the final mass distribution of sinks
(Fig. \ref{sinkhist}) 
differs considerably between the MHD and hydro runs.  The hydro run has 9 sinks ranging in mass from $\la$ 1 to 27 $\msunm$, and a combined sink mass of 60 $\msunm$.  The MHD run has a single sink of 26 $\msunm$.  Our results suggest that magnetic fields effects thus do not much alter the mass of the primary star, but they do prevent formation of additional lower-mass stars, thereby favoring a top-heavy IMF.

\begin{figure}
\includegraphics[width=.48\textwidth]{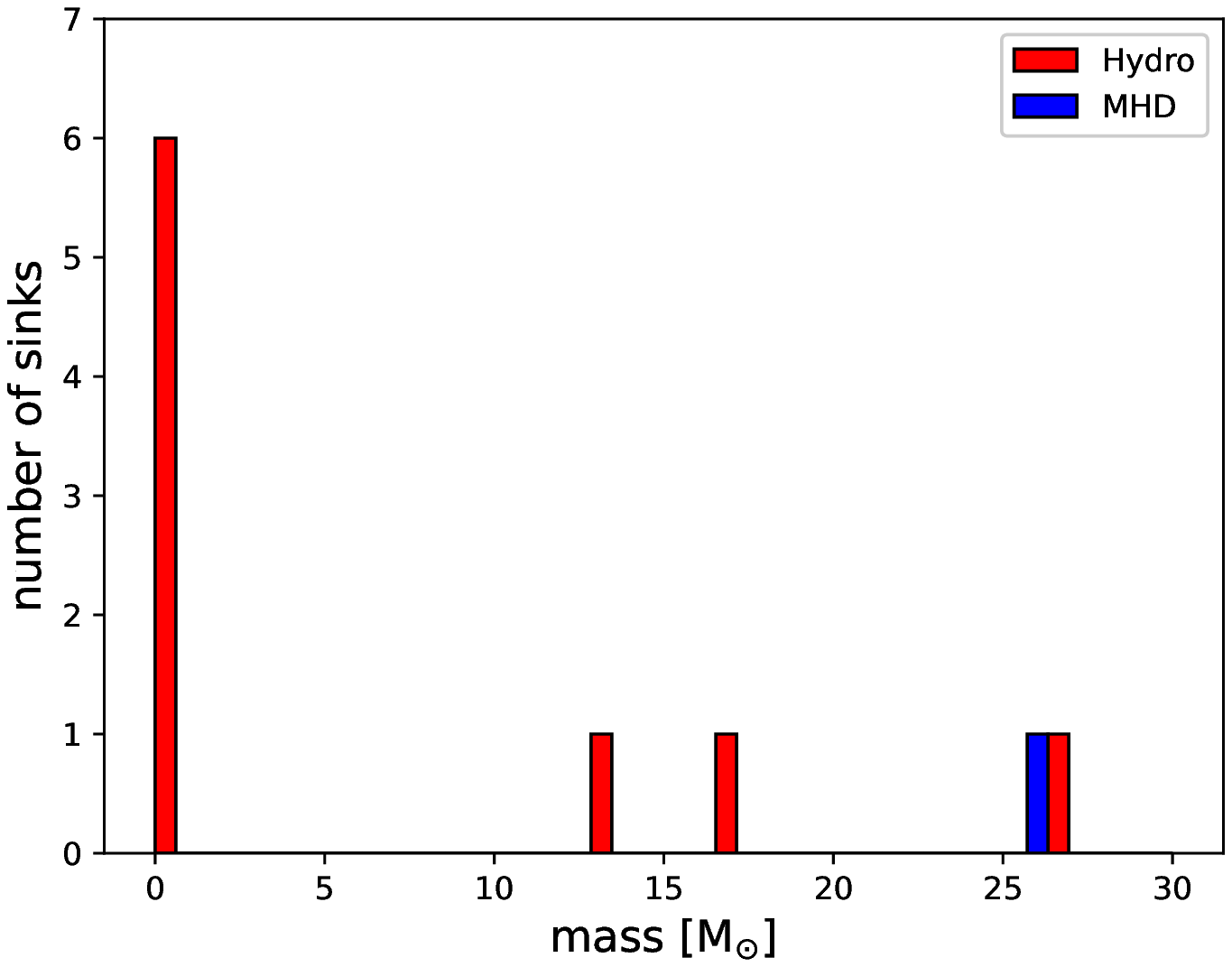}
 \caption
 {
Histogram of sink masses at the time of final simulation output
($t=2000$~yr).
Red depicts the result in the hydro case without magnetic fields, while blue shows the result in the MHD case when they are included.  Only a single 26 $\msunm$ sink forms when magnetic fields are included, while a total of 9 sinks ranging from 1 to 27 $\msunm$ form in the hydro case.  This demonstrates the extent to which magnetic fields suppress secondary sink formation.
  }
\label{sinkhist}
\end{figure}

\begin{table*}
\begin{tabular}[width=.95\textwidth]{crrrrrr}
\hline
Run         &  Sink Number  &   Max. Sink Mass  [$\msunm$]   &    Min. Sink Mass  [$\msunm$]   & Total Sink Mass [$\msunm$]   \\
\hline
Hydro            &   9   &    27    &   1    &    60  \\
MHD             &   1   &    26    &   -     &     26   \\
\hline 
\end{tabular}
\caption{Characteristics of the sink particles at the end of the hydro and MHD runs ($t=2000$~yr).  The hydro run has many sink particles form over a range of masses, while in the MHD run only one large sink particle survives.}
\label{tab1}
\end{table*}

 \section{Theory vs Simulation: Growth of the Magnetic Field}
 \label{sec:predict}
 
 How does the growth of the magnetic field in a gravitationally collapsing cloud compare with that predicted in Paper I? The {\sc gadget-2} simulation, which did not include the Lorentz force, followed the
kinematic stage of the dynamo, whereas the {\sc orion2} simulation primarily followed the nonlinear evolution in which magnetic forces are important. We consider each in turn.

\subsection{Prediction for {\sc gadget-2}: The kinematic dynamo}
\label{sec:kine}

The magnetic field in the gas that formed the first stars quite possibly originated via the Biermann battery and was then amplified by a small-scale dynamo. Field amplification occurred due to the stretching and folding of the field by turbulent motions. For Kolmogorov (i.e., subsonic) turbulence with an outer scale $L$, the properties of a kinematic dynamo (negligible Lorentz forces) are governed by two dimensionless numbers, the Reynolds number, $Re=Lv_L/\nu$, where $v_L$ is the velocity in eddies of scale $L$ and $\nu$ is the kinematic viscosity, and the magnetic Reynolds number, $R_m=Lv_L/\eta$, where $\eta$ is the resistivity. The ratio of these quantities is the magnetic Prandtl number, 
\beq
P_m\equiv \frac{\nu}{\eta}=\frac{R_m}{Re},
\eeq
which is generally large in astrophysical plasmas.
(If the resistivity is due to ambipolar diffusion, it depends on the strength of the field and another dimensionless parameter enters--see Paper I). In Kolmogorov turbulence, the energy dissipation rate, $\epsilon\equiv v_\ell^3/\ell$, is constant in the inertial range. The smallest eddies occur on the viscous scale, $\ell_\nu$, at which the local Reynolds number is unity, so that $\nu=\ell_\nu v_\nu$. It follows that
\beqa
\ell_\nu&=&\left(\frac{\nu^3}{\epsilon}\right)^{1/4}=\frac{L}{Re^{3/4}},\\
v_\nu&=&~(\epsilon\nu)^{1/4}~~=\frac{v_L}{Re^{1/4}}.
\label{eq:vnu}
\eeqa
The eddy turnover rate at the viscous scale is
\beq\Gamma_\nu=\frac{v_\nu}{\ell_\nu}=\left(\frac{v_L}{L}\right) Re^{1/2}.
\eeq

In the kinematic stage of a dynamo, the field is too weak to have any dynamical effects and it grows exponentially, $B\propto \exp(\Gamma t)$.
Astrophysical gases are generally characterized by very large Reynolds numbers.
Furthermore, the magnetic Prandtl number is often large as well, particularly when the field is weak in a dynamo operating after recombination (Appendix A in Paper I). For $Re\gg1$ and $P_m\gg1$, the field is expected to be amplified on the time scale of the fastest eddies, $\sim\Gamma_\nu^{-1}$ \citep{kuls92,sche02b,fede11b}. We therefore express the growth rate of the field as
\beq
\Gamma=C_\Gamma \Gamma_\nu = C_\Gamma\left(\frac{v_L}{L}\right) Re^{1/2},
\label{eq:gamma}
\eeq
where $C_\Gamma$ is a constant to be determined. In the numerical example they worked out, \citet{kuls92} estimated $C_\Gamma=4\pi$. \citet{schoberetal2012a} solved the \citet{kazantsev1968} equation and found $C_\Gamma=37/36\simeq 1$ for Kolmogorov turbulence. The result $\Gamma=\Gamma_\nu$ for $Re,\,P_m\gg 1$ has generally been adopted in subsequent work (e.g., \citealp{xu&lazarian2016}). 
During the operation of a dynamo, the field gets folded on smaller and smaller scales. When the smallest scale reaches the resistive scale (middle panel in Fig. 1 in Paper I), the growth rate of the field is reduced by a factor 3/8 \citep{kuls92, sche02b}, so
$C_\Gamma\simeq 3/8$. \citet{xu&lazarian2016}
suggested that this applies throughout the kinematic stage for $P_m\sim 1$.
On the other hand, the results of numerical simulations \citep{haug04,fede11a,fede11b} are consistent with $C_\Gamma\ll 1$ , presumably because of the greater importance of dissipation at the moderate values of the magnetic Prandtl number in simulations, $P_m\sim 1$ (see Appendix \ref{app:growth}).

The dynamo during star formation occurs in a contracting medium.
In order to treat a dynamo in a time dependent background, the treatment in Paper I followed that of \citet{schleicheretal2010} and \citet{schoberetal2012b} and took the time dependence as separable from the dynamo action. The evolution of the field was expressed as the product of a dynamo amplification factor, $\cala$, and a compression factor based on flux-freezing. When the flux is frozen to the plasma, the field will increase as $B\propto \nh^{2/3}$ for an isotropic, homologous collapse or when the field is randomly oriented on the scale of the collapse. The first condition is approximately satisfied for the collapsing gas in a minihalo, and the second is valid for small scale dynamo in a global collapse.
Flux-freezing breaks down below the resistive scale, but the field on those scales makes only a negligible contribution to the total field. In this paper, we assume flux freezing only for the kinematic stage of the dynamo so that
\beq
B=B_0\xi^{2/3}\akin,
\label{eq:b}
\eeq
where $\xi\equiv\rho/\rho_0$ is the ratio of the density to its initial value and the dynamo  amplification is exponential,
\beq
\akin=\exp \left(C_\Gamma\int\Gamma_\nu dt\right).
\label{eq:cala}
\eeq
Paper I showed that in the case of the formation of the first stars, the dynamo action was rapid so that the compression factor $\xi^{2/3}\simeq 1$. By contrast, the small value of $R_m$ in simulations leads to slower amplification than in reality, and as a result compression is important in increasing the field during the kinematic stage.

\citet{fede11b} showed that the outer scale of the turbulence in a gravitationally collapsing medium is approximately equal to the Jeans length,
\beq
\lj=\left(\frac{\pi \cs^2}{G\rho}\right)^{1/2}=\left(\frac{32}{3}\right)^{1/2}\cs\tff,
\label{eq:lj2}
\eeq
so that
\beq
C_\Gamma \int\Gamma_\nu dt = \left(\frac{3}{32}\right)^{1/2}C_\Gamma \avg{\calm}\int Re^{1/2}\,\frac{dt}{\tff},
\label{eq:intg1}
\eeq
where $\avg{\calm}$ is the time-averaged value of the Mach number. 
The viscosity in SPH codes is $\nu_{\rm SPH}=0.06\cs h_{\rm sm}$ \citep{baue12}, where the smoothing length is given in equation (\ref{eq:hsm}). With the aid of equation (I.80) (equation 80 in Paper I),
this gives $Re=720\vtf/\nh^{1/6}$, where $\nh$ is measured in cm\eee. In Paper I, we defined the integral
\beq
I_q(\xi_1,\xi)=\frac{1}{\phiff\tffo}\int_{t(\xi_1)}^{t(\xi)} \xi^q dt,
\label{eq:iq}
\eeq
where $\phiff$ is defined in equation (\ref{eq:phiff}). Now, when the outer scale of the turbulence is the Jeans length, which varies as $\xi^{-1/2}$, it follows that $Re\propto \xi^{-1/6}$ from equation (I.C21); 
then $q=5/12$ since $\tff\propto \xi^{-1/2}$. Figures \ref{phi_ff} and \ref{velprof} show that $\phiff$, which enters the integral implicitly through its effect on the rate of collapse, and $v_t$ both vary with density and therefore time. However, there is no systematic variation of these quantities, and furthermore the evaluation of $I_q$ is based on the assumption that $\phiff=\,\mbox{const}$.
We therefore take average values for these quantities and the temperature, which gives
\beqa
C_\Gamma\int\Gamma_\nu dt&=&3.17\left(\frac{C_\Gamma \vtf^{3/2}\phiff}{T_3^{1/2}\nho^{1/12}}\right)I_{5/12}(1,\xi),\\
&\simeq&58\,C_\Gamma I_{5/12}(1,\xi),
\label{eq:58}
\eeqa
where we took $\vtf=v_t/(10^5\mbox{ km s\e})=1.5$, $\phiff=7$ (section \ref{sec:initc}), $T_3=T/(10^3\mbox{ K})=0.5$ and $\nho=1$~cm\eee, so that $\xi=\nh/$(1 cm\eee). Equations (\ref{eq:b}) and (\ref{eq:cala}) then imply
\beq
B=3.6\times 10^{-11}\,\nh^{2/3}\exp\left[58 C_\Gamma I_{5/12}(1,\xi)\right]~~~\mbox{G},
\label{eq:bkin}
\eeq
where we took the value of $B_0$ from equation (\ref{eq:bgadget}).
Evaluating $I_{5/12}(1,\xi)$ numerically, we find that equation (\ref{eq:bkin})   reproduces the simulation results over the entire range of dynamo amplification, $1\mbox{ cm\eee}<\nh<10^7$~cm\eee, to within a factor 2 for $C_\Gamma=0.066$. Note that this agreement covers the growth of the field by a factor in excess of $10^7$ and includes the density ranges that were fit by the quite different power laws in equation (\ref{eq:bgadget}), $B\propto \nh^{1.43}$ and $B\propto\nh ^{0.87}$. The fit is very sensitive to the value of $C_\Gamma$: If it is increased by 10 percent, the agreement with the simulation is degraded from a factor 2 to a factor 3.5. 

While $C_\Gamma$ is accurately determined for the parameters that we have chosen, it is actually the product of $C_\Gamma$ and the factors that enter the number 58 in equation (\ref{eq:58}) that is well determined; $C_\Gamma$ itself is as uncertain as those factors. Furthermore, we have followed \citet{schleicheretal2010} and \citet{schoberetal2012b} in assuming that compression obeys flux-freezing, so that $B\propto \xi^{2/3}$ in equation (\ref{eq:b}). In section \ref{sec:nonlinear} below, we shall see that our simulation is consistent with $B\propto\xi^{0.54-0.58}$ for the nonlinear dynamo, which violates flux-freezing \citep{xu20}. It is possible to get good fits to the data for the kinematic stage with exponents that differ from 2/3; for example, setting $B\propto \xi^{0.5}$ in equation (\ref{eq:b}) gives agreement with the data to within a factor 1.7 for $C_\Gamma=0.088$. The higher value of $C_\Gamma$ would result in a greater efficiency for the dynamo. The results of \citet{fede11b} are consistent with $B\propto\xi^{2/3}$ due to compression, but they did not investigate other possibilities.

In Paper I, we assumed that $C_\Gamma=3/8$ for simulations based on an extrapolation of theoretical results for high values of $P_m$ to $P_m\simeq 1$ \citep{xu&lazarian2016}, but here we find that $C_\Gamma$ is smaller by a factor $4-6$. 
As discussed in Appendix \ref{app:growth}, the results of \citet{fede11a}'s simulation of a static turbulent box correspond to $C_\Gamma\simeq 0.026$, whereas those of \citet{fede11b} for a cloud  undergoing a more rapid collapse than
the cloud in our simulation correspond to $C_\Gamma\simeq 0.1$. The value of $C_\Gamma$ we have found is intermediate between these values, but closer to that of the collapsing cloud. We attribute the small growth rate of the simulated kinematic dynamo (i.e., the small value of $C_\Gamma$) to the fact that the simulations have $P_m\sim 1$, whereas theoretical expectations are based on the assumption that $P_m\gg 1$.
Since the growth rate varies as $C_\Gamma Re^{1/2}$, simulated small-scale dynamos are slower than natural ones because both the coefficient $C_\Gamma$ and the Reynolds number, $Re$, are smaller in simulations. We emphasize that the reduction in the dynamo growth rate that we find applies only to simulations dominated by numerical resistivity and viscosity, with $P_m\sim 1$. As shown in Paper I, the kinematic dynamo in the formation of the first stars operates at low densities and high $P_m$, and we have no evidence that $C_\Gamma$ deviates from the theoretically expected value of 1 there.

As shown in section \ref{sec:kinesim}, the magnetic energy is less than about 20 percent of the kinetic energy during the {\sc gadget-2} simulation. However, nonlinear effects first set in on small scales, when the energy density of the field matches the kinetic energy in viscous scale eddies \citep{xu&lazarian2016}. This occurs at a magnetic field $B_\nu$ given by equation I.95,
\beq
B_\nu=(4\pi\rho)^{1/2}v_\nu=1.00\times 10^{-7}(h_f m_\sph'^{1/3})^{1/4}\vtf^{3/4}\nh^{13/24}~~~\mbox{G},
\label{eq:bnu1}
\eeq
where $h_f m_\sph'=1.12$ (eq. \ref{eq:hsm}). For the average velocity of 1.5 km s\e, this equals the simulated field summarized in equation (\ref{eq:bgadget}) at a density $\nhn\simeq 10^7$ cm\eee. Our kinematic assumption begins to fail on small scales shortly before we terminate the {\sc gadget-2} simulation.
It should be noted that $B_\nu\propto Re^{-1/4}$ (eq. \ref{eq:vnu}), and since it is not possible to simulate the large Reynolds numbers found in Nature, the transition to a nonlinear dynamo occurs later in a simulation than in reality.

\citet{haug04} showed that the dynamo ceases to operate below a critical value of the magnetic Reynolds number, $R_{m,\,\crit}=220$ (see Appendix \ref{app:growth}). 
For a fixed value of $P_m$, this corresponds to a critical value of $Re$. Since
$Re$ in an SPH simulation varies as $\nh^{-1/6}$ as shown above, this critical value of $R_m$ corresponds to a maximum density for operation of the dynamo, $\nhm$. Our simulation shows that dynamo amplification ceased at $\nh\simeq 10^7$~cm\eee\ (eq. \ref{eq:bgadget}). For our SPH simulation, we expect
 (eq. I.82),
\beq
\nhm=1.14\times 10^3\,\vtf^6 P_m^9~~~\mbox{cm\eee}.
\eeq
Even though $\vtf$ and $P_m$ are uncertain by factors of only about 1.5 or so, the high powers to which they enter into this equation make the maximum density for the dynamo quite uncertain. In Paper I, we adopted $P_m=1.4$; for the average turbulent velocity of 1.5 km s\e\ estimated above, this corresponds to $\nhm=2.7\times 10^5$~cm\eee. On the other hand, if $P_m=2$, as adopted by \citet{fede11a}, this becomes
$\nhm=6.6\times 10^6$~cm\eee, close to our simulation result. As a result, the best that can be said is that the simulation is consistent with the theoretical expectation for the value of $\nhm$.

The predicted evolution of an SPH simulation of a small-scale dynamo was shown in Figure 4a in Paper I. Since $\nhm\simeq\nhn\simeq 10^7$~cm\eee, the trajectory of the simulated dynamo is intermediate between the curves for $\phiff=1$ and 2 in the figure: it intersects the line for $\nhm$ at about the same point that the line labeled $B_\nu$ does. This is to be expected, since the product
$C_\Gamma\phiff=0.46$ in our simulation is between the values (3/8, 3/4) for the $\phiff=(1,\,2)$ curves in the figure.

\subsection{Prediction for {\sc orion2}: The nonlinear dynamo}
\label{sec:nonlinear}

We switched from {\sc gadget-2} to {\sc orion2} when the peak density reached $\nh=10^8$ cm\eee. As just noted, nonlinear effects become important on small scales when $B=B_\nu$. Since the numerical viscosity for grid-based codes differs from that for SPH codes, so does the value of $B_\nu$:
\beq
B_\nu=1.11\times 10^{-7}\left(\frac{J_{\max}}{1/64}\right)^{1/3}\vtf\nh^{1/2}~~~\mbox{cm\eee},
\eeq
where $J_{\max}={\max}(\Delta x/\lj)$ is the Jeans number (Appendix \ref{app:refine}), which we set equal to 1/64 as recommended by \citet{fede11b}. For a typical value of the turbulent velocity in the {\sc orion2} run of 2~km~s\e, the simulated field reaches $B_\nu$ at a density $\nh=7\times 10^6$~cm\eee. Since we focus on densities $\ga 10^8$~cm\eee\ in the {\sc orion2} simulation, the dynamo is always in the nonlinear stage in this simulation. We note that \citet{turketal2012} assumed a much weaker initial field than we did,
$B_0=10^{-14}$G, and as a result their dynamo simulation was entirely in the kinematic stage.

The treatment of the nonlinear dynamo in Paper I followed the treatment of \citet{schleicheretal2010} and \citet{schoberetal2012b} in assuming that the dynamo effect was independent of a change in the mean density so that the two effects added (eq. I.43). \citet{xu20} have given a self-consistent derivation of the equation for a nonlinear dynamo in a time-dependent medium in which they drop this assumption. We generalize their treatment and follow the evolution of the dynamo during the late stages of a gravitational collapse, beginning when the compression ratio is $\rho_1/\rho_0=\xi_1$. The integrated form of the \citet{kazantsev1968} equation is
\beq
\eb=\cale_{B,\refer}\left(\frac{k_p}{k_\refer}\right)^{5/2}\exp \left(\frac 34\int_{t_1}^t \Gamma dt\right),
\label{eq:kaz}
\eeq
where the subscript ``$\refer$" denotes a reference value and $k_p$ is the wavenumber at which the turbulent eddies in the inertial range are in equipartition with the field \citep{xu&lazarian2016}. In the absence of dynamo action, $\eb=B^2/(8\pi\rho)$ would increase as $\rho^{1/3}$ in an isotropic compression, so that $\cale_{B,\refer}=\ebone(\xi/\xi_1)^{1/3}$
\citep{xu20}. They couple the dynamo to the compression by assuming that the reference wavenumber scales inversely with the Jeans length, $k_\refer\propto\lj^{-1}\propto \xi^{1/2}$ for isothermal contraction. We refer to this as the Jeans-regulated case. Since the outer scale of the turbulence in a gravitationally collapsing cloud is the Jeans length \citep{fede11b}, it follows that the velocity, and therefore the rate of change of the field, responds to the Jeans length. However, it is also plausible that the reference length responds in a homologous fashion to the contraction so that
$k_\refer r=$~const.
To allow for a range of possibilities, we write
\beq
k_\refer=(\xi/\xi_1)^{\alpha_\refer} k_{\refer,1},
\eeq
where  $\alpha_\refer=(\frac 12, \,\frac 13,\, 0)$ for $k_\refer\lj=$~const (Jeans-regulated, \citealp{xu20}), $k_\refer r=$~const (homologous), and
 $k_\refer=$~const (flux-freezing, Paper I), respectively.

We now follow the derivation of \citet{xu20}. The growth rate of field is $\Gamma=v_p/\ell_p$, where $v_p$ is the eddy velocity on the scale $\ell_p$. Since
the field is in equipartition with eddies on that scale, $\eb=\frac 12 v_p^2$. It follows that
\beq
\Gamma\eb=\frac 12\left(\frac{v_p^3}{\ell_p}\right)=\frac 12\epsilon.
\label{eq:ge}
\eeq
They assume that the turbulent velocity is independent of time, which is a good approximation for the {\sc orion2} stage of the collapse. Since the dissipation rate is independent of scale in the inertial range, it follows that $\epsilon=v_p^3/\ell_p=v_L^3/L\propto\lj^{-1}$ so that
$\epsilon=\epsilon_1(\xi/\xi_1)^{1/2}$. Evaluating $d\eb/d\ln t$ from both equations
(\ref{eq:kaz}) and (\ref{eq:ge}) and solving for $\eb$ gives
\beq
\eb=\left(\frac{\xi}{\xi_1}\right)^a\ebone+\frac{\chi\epsilon_1\xi^a}{\xi_1^{\frac 12}}
\int_{t_1}^t\xi(t')^{\frac 12 -a} dt',
\label{eq:eb}
\eeq
where
\beq
a=\frac 13-\frac{10}{19}\,\alpha_\refer.
\eeq
Here we have replaced their factor $3/38\simeq 0.079$ in equation (\ref{eq:eb}) by a parameter, $\chi$. We generally set $\chi=1/16\simeq 0.067$, which is intermediate between 3/38 and 0.05, the result of the numerical simulation by \citet{beresnyak2012}; it is close to 0.07, the result of the simulation of \citet{choetal2009}. \citet{xu20} chose the normalization $\xi_1=1$. They set $\chi= 3/38$, which corresponds to $\alpha_\refer=\frac 12$ and $a=4/57$; equation (\ref{eq:eb}) then agrees with their result.
For the flux-freezing case, $\alpha_\refer=0$ (so that $a=\frac 13$), $\xi_1=\xi_\nu$, and $\ebone=\ebn$ (the onset of the nonlinear stage), equation (\ref{eq:eb}) agrees with equation I.44 in Paper I.

The first term in equation (\ref{eq:eb}) represents compression of the initial field, whereas the second represents the effect of the dynamo. 
As emphasized by \citet{xu20}, their result that $a=4/57$ implies that compression has a weaker effect on the field than in the case of flux-freezing, $B\propto\xi^{0.5+2/57}=\xi^{0.54}$ vs. $B\propto\xi^{2/3}$. The homologous case is intermediate,
$B\propto \xi^{0.5+3/38}=\xi^{0.58}$.

Our goal here is to predict the field in the {\sc orion2} simulation just before the first sink forms. As discussed above, the flux becomes disconnected from the gas in the sink, so it is not possible to follow the field evolution after sink formation. We focus on the gas at densities above the initial maximum of $10^8$~cm\eee\ since gas at lower densities does not have time to evolve much before the sink forms. Since the initial density was $n_0\simeq 1$~cm\eee, it follows that $\xi>10^8$.
The integral that enters equation (\ref{eq:eb}) is proportional to the quantity $I_{1/2-a}$
defined in equation (\ref{eq:iq}); for large $\xi^{1/3}$ this quantity is given by equation I.B17 ,
\beq
\int_{t_1}^t\xi^{\frac 12-a}dt\simeq
\frac{2\phiff\tffo}{3\pi a}\left(\frac{1}{\xi_1^a}-\frac{1}{\xi^a}\right).
\eeq
Evaluating the dissipation rate $\epsilon_1$ with the aid of equation  (\ref{eq:lj2}), we obtain
\beq
\eb =\left(\frac{\xi}{\xi_1}\right)^a\ebone+
\frac{\chi\phiff v_t^3}{2\surd 6\pi a\cs}\left[\left(\frac{\xi}{\xi_1}\right)^a-1\right].
\label{eq:eb2}
\eeq
We can re-express this in terms of
the nonlinear dynamo amplification factor beginning at $\xi_1$, $\cala_1$, by factoring out the effect
of compression of the initial field,
\beq
\eb=\left(\frac{\xi}{\xi_1}\right)^a\ebone\cala_1^2,
\label{eq:eb3}
\eeq
so that
\beq
\cala_1^2=1+0.0081
\,\frac{\phiff\calm}{a}\left(\frac{\chi}{1/16}\right)\left[1-\left(\frac{\xi_1}{\xi}\right)^a\right]\frac{\frac 12 v_t^2}{\ebone}.
\label{eq:cala1}
\eeq
Dynamo amplification ($\cala_1^2\gg 1$) can be important only if the field is initially well below equipartition 
($\ebone\ll \frac 12 v_t^2$).

The one remaining complication is that $\xi_1$ is a function of position and therefore of $\xi$ for a medium with a spatially variable density as we are considering (see Fig. \ref{rprof}). By contrast,
\citet{xu20} considered a uniform medium and took $\xi_1=1$. The late stages of collapse are described by equation (\ref{eq:vr}), which implies
\beq
\left(\frac{r_1}{r_0}\right)^{3/2}-\left(\frac{r}{r_0}\right)^{3/2}=\frac{3\pi}{4\phiff}\left(\frac{\Delta t}{\tffo}\right),
\label{eq:r1}
\eeq
where $\Delta t=t-t_1$.
In terms of the compression ratio, $\xi=(r_0/r)^3$, this becomes
\beq
\xi_1^{-1/2}-\xi^{-1/2}=\xi_{1u}^{-1/2},
\label{eq:xi1}
\eeq
where $\xi_{1u}$ is the upper limit on $\xi_1$, corresponding to the initial compression of the gas that reaches the origin at a time $t$. The {\sc orion2} simulation begins at $t_1=-9000$~yr. We evaluate the field at $t=0$, when gas with an initial density of
$10^8$~cm\eee, corresponding to $\xi=10^8$, reaches the origin. It follows that
$\xi_{1u}=10^8$, and one can show that the right-hand sides of equations (\ref{eq:r1}) and (\ref{eq:xi1}) agree for $\phiff=4.7$ (see below eq. \ref{eq:deltat}) and $\Delta t=9000$~yr.

The field in a nonlinear dynamo implied by equation (\ref{eq:eb3}) is
\beq
B=B_1\cala_1\left(\frac{\xi}{\xi_1}\right)^{(1+a)/2},
\eeq
where $B_1(\xi_1)$ is
the value of the field at time $t_1$; in our case, $t_1=-9000$~yr. According to equation (\ref{eq:xi1}),
$\xi_1$ decreases from $\xi_{1u}=10^8$ for $\xi\rightarrow\infty$ to $0.25\xi_{1u}$ at $\xi=\xi_{1u}$.
Equation (\ref{eq:bgadget}) gives this field in our simulation as 
\beq
B_1=1.7\times 10^{-8}\xi_1^{2/3}~~~\mbox{G}
\label{eq:bone}
\eeq
for $\xi_1$ in the range $10^{7-8}$, corresponding to
$\ebone=5.15\times 10^6\xi_1^{1/3}$~cm$^2$~s\ee.
In the {\sc orion2} simulation we have $\vtf\simeq 2$, $T_3\simeq 1$ (corresponding to $\calm\simeq 0.8$), and $\phiff=4.7$. 
With the aid of equations (\ref{eq:cala1}) and (\ref{eq:xi1}), we evaluate this field and plot its ratio to the result of our simulation in Fig. \ref{b_bsim}.

\begin{figure}
\includegraphics[width=.48\textwidth]{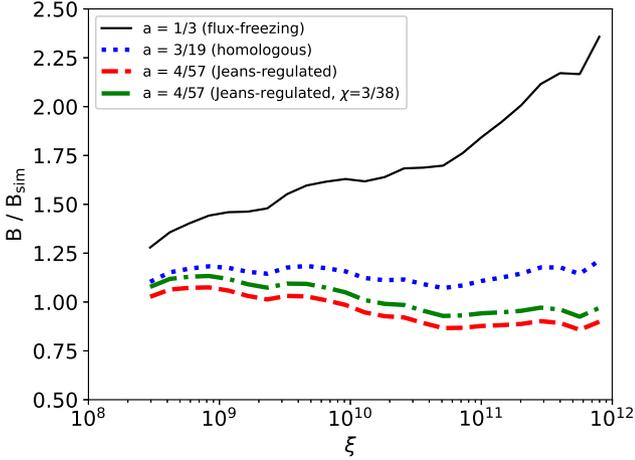}
 \caption
 {
 Ratio of the theoretically predicted magnetic field to the simulated one for the flux-freezing case ($a=1/3$, solid black line), the homologous case ($a=3/19$, dotted blue line), and the two Jeans-regulated cases ($a=4/57$; $\chi=1/16$, dashed red and $\chi=3/38$, dash-dot green).  The homologous case and both Jeans-regulated cases show remarkably good fits.
  }
\label{b_bsim}
\end{figure}
Four cases are considered. The first three have $\chi=1/16$: Flux freezing ($a=1/3$); homologous response ($a=3/19$); and  Jeans-regulated ($a=4/57$). The fourth case is the Jeans-regulated case with $\chi=3/38$, as found by \citet{xu20}. A summary of the results over the range $3\times 10^8<\xi<6\times 10^{11}$ is
\beqa
a&=&\frac 13~:~~~1.28<B/B_{\rm sim}<2.17,\\
a&=&\frac{3}{19}:~~1.07<B/B_{\rm sim}<1.19,\\
a&=&\frac{4}{57}\left\{\begin{array}{l}0.86<B/B_{\rm sim}<1.08,\\
0.93<B/B_{\rm sim}<1.14~~~\left(\chi=\dis\frac{3}{38}\right),\end{array}\right. 
\eeqa
where $\chi=1/16$ for the first three cases.
The first case (flux-freezing) is clearly the worst and can be rejected. The homologous case
($a=3/19$) has the smallest dispersion, whereas the Jeans-regulated cases ($a=4/57$) are closest to unity, and they also maintain their fits over a somewhat larger range. All three are remarkably good fits when one considers that (1) the value of $\ebone\propto B_1^2$ that entered had an uncertainty of 0.04 dex $\simeq 10$ percent from equation (\ref{eq:bgadget}); (2) the theory assumes perfect spherical symmetry, which is not satisfied by the simulation; 
and (3) the density is approaching the maximum above which $R_m$ is too small
for the dynamo to operate, $\nhm=1.2\times 10^{12}$~cm\eee\ (Paper I). (The value of $\nhm$ here is much larger than for the {\sc gadget-2} simulation, $10^{5-7}$cm\eee, because we have chosen a higher resolution for the {\sc orion2} simulation.)

Finally, (4) the field is approaching equipartition, but this was not taken into account. Equipartition is reached when $\eb=\frac 12 v_t^2$, which occurs at
\beq
\xi_\eq=\xi_1\left(\frac{1+\dis\frac{\chi\phiff\calm}{\surd 6\pi a}}{\dis\frac{2\ebone}{v_t^2}+\dis\frac{\chi\phiff\calm}{\surd 6\pi a}}\right)^{1/a}.
\label{eq:eq}
\eeq
The simulation was close to equipartition at $\xi\sim 10^{12}$, but the resolution was not high enough for an accurate treatment. We therefore consider the predicted value of $\xi$ for the case in which the field energy is half the turbulent energy, which is given by equation (\ref{eq:eq}) with the 1 in the numerator replaced by 0.5. The Jeans-regulated case reaches this point at $\xi\simeq 2\times 10^{11}$ and the homologous case does so at $\xi\simeq 1.7\times 10^{10}$; the value given by the simulation is $\xi=3.6\times 10^{10}$, intermediate between these two predicted values.

As found in Paper I, the nonlinear dynamo is very inefficient in simulations that are possible now. Over the density range $10^8-10^{12}$~cm\eee, the simulation shows that the field grows by a factor 210; for $a=4/57$, compression accounts for a factor 140 of this and the dynamo only a factor 1.5. Equations (\ref{eq:cala1}) and (\ref{eq:bone}) imply that
\beq
\cala_1^2\rightarrow 1+\frac{0.25}{aT_3^{1/2}}\left(\frac{\phiff}{4.7}\right) 
\left(\frac{\vtf}{2}\right)^3\left(\frac{\chi}{1/16}\right)\left(\frac{10^8}{\xi_{1,u}}\right)^{1/3}
\eeq
for $(\xi/ 10^8)^a\gg 1$.
For small $a$, $\cala_1$ does not reach its asymptotic value by $\xi=10^{12}$, so for $a=(1/3,\,3/19,\,4/57)$ and $\chi=1/16$, the nonlinear amplification is $\cala_1=(1.3,\,1.5,\,1.6)$ at this value of $\xi$. 
In Paper I we found that the nonlinear dynamo could provide an amplification of about a factor 10 in the actual formation of the first stars: Since $\cala_1^2\propto 1/\ebone\leq 1/\ebn\propto Re^{1/2}$ for $\cala_1\gg 1$ (see below eq. \ref{eq:bnu1}), the large Reynolds number of natural dynamos enables them to be much more efficient.

We conclude that, as predicted in Paper I, the nonlinear dynamo is ineffective at amplifying the field in current numerical simulations of the formation of the first stars.  We find here that flux-freezing is not a good approximation. Instead, we find good agreement with the theory developed by \citet{xu20} or a variant thereof, in which compression increases the field by a factor in the range $\xi^{0.5+2/57}-
$  $\xi^{0.5+3/38}$ instead of $\xi^{2/3}$.

\section{Discussion}
\label{sec:discussion}

\subsection{Implications for Pop III stars and their detection}

The simulations we present here indicate that 
magnetic fields generated by a small-scale dynamo 
suppress fragmentation and the formation of lower-mass secondary stars.  This would reduce the overall star formation efficiency within minihalos and skew the Pop III IMF to be more top-heavy.

The true Pop III IMF will become better constrained with future observations.  Detection of low-mass Pop III stars within our own Milky Way or its satellites would provide direct confirmation that the Pop III IMF extends to $\la$ 1 $\msunm$,
and this in turn would suggest that magnetic fields within minihalos are not as strong as theory
and our simulation predict. However, detecting Pop III stars is very challenging: under the reasonable assumption that Pop III stars would have a spatial distribution similar to that of extremely metal-poor (EMP) and ultra metal-poor (UMP) stars, \citet{magg19} conclude that there are fewer than 1650 Pop III stars in the entire halo of the Milky Way. 

Discovering a Pop III star is complicated by the possibility that it may accrete metal-enriched gas over its lifetime before becoming incorporated into the Galaxy (e.g. \citealt{frebeletal2009, johnson&khochfar2011, komiyaetal2015,  shenetal2016}).  The mass of metals gained depends on the path and accretion rate of the Pop III star as well as the strength of its stellar winds, and sufficient metal accretion may mask Pop III stars as Pop II.  
\cite{tanakaetal2017} find that the combination of the stellar wind and magnetosphere of the Pop III star would prevent accretion, weakening metal-accretion as an explanation for the lack of detected Pop III stars today.

Since we terminated our simulation 2000 years after the formation of the first sink, we did not address the high-mass end of the Pop III IMF.
However, observations should shed light on this. First, pair instability SNE (PISNe) as well as core collapse SSNe (CCSNe) may be detectable by {\it JWST} out to $z \sim 10-15$, or even up to $z \sim 20$ for Type IIN SNe (e.g., see the review by 
\citealt{tomaetal2016}).   
Second, Pop III GRBs along with the metal-abundance ratios of their afterglow spectra may further constrain the high-mass end of the IMF.  
Abundances in EMP and UMP stars within the Galaxy and its satellites such as Segue 1 may provide additional clues about the mass of the Pop III stars, as Pop III supernovae may have enriched these later generations of stars (e.g. \citealt{heger&woosley2002, frebeletal2014, jietal2015, fraseretal2017}).

How do the effects of magnetic fields in the formation of the first stars compare with those in contemporary star-forming regions? We addressed this problem theoretically in Paper I, where we found that the nonlinear dynamo could amplify the field to equipartition by a compression $\xi\sim 10^{3-4}$, corresponding to a density 
$\sim 10^{3-4}$~cm\eee. As a result, the field strength would be comparable to that in contemporary star formation throughout much of the gravitational collapse; the principal difference would be that the field in contemporary star formation is much more ordered than that produced by a small-scale dynamo. What we have found in this paper is that numerical dissipation makes it impossible to accurately simulate the growth of the field in a star-forming region over cosmological time scales: Even though we began with an artificially large field of $4.5\times 10^{-11}$~G at $z\simeq 50$, the field energy reached half the equipartition value only at a compression  $\xi=4\times 10^{10}$.

\subsection{Caveats}
\label{sec:caveat}

The principal caveat of our work is that the cosmological magnetic field remains unknown.  The upper limit on the comoving field on a comoving scale of 1 Mpc set by Planck observations is 4.4 nG \citep{plan16}. As earlier discussed, we assume that the magnetic field at $z\simeq 50$ is uniform on the scale of our cosmological box (1.4 Mpc comoving), and that it has a physical magnitude of $4.5\times 10^{-12}$ G, 
far smaller than the Planck upper limit of 11.4 $\mu$G at that redshift.  In Paper I, we showed that the Biermann battery could create a field $B\sim 10^{-16}$~G in a cosmic minihalo at a redshift of 25, and that this could be amplified by a small-scale dynamo to approximate equipartition by the time that the gas density had increased to $\nh\sim 10^{3-4}$~cm\eee, as just noted. Observational confirmation of dynamically significant magnetic fields at high redshifts would be valuable. 

A second important caveat is that the numerical viscosity and resistivity in our simulation--indeed, in any simulations possible with current computers--are orders of magnitude larger than the physical values. In Paper I, we showed how this altered the evolution of the field in a small-scale dynamo. In order to have the field become dynamically significant, we had to begin with a field strength much greater than that expected from the Biermann battery.
Furthermore, our simulation assumed ideal MHD, whereas in reality ambipolar diffusion determines the resistivity (Paper I); as a result, the resistivity varies as $B^2$, although it remains less than the numerical resistivity.

Other physics must also be included to predict the Pop III mass distribution in the minihalos we simulate.  In particular, we do not include radiative feedback due to photodissociation and photoionization or possible protostellar outflows.  
For this reason we do not follow the evolution beyond when the most massive star surpasses 
$\sim$ 20 $\msunm$,
since we expect radiative feedback to become important by this stage (\citealt{stacyetal2016}, see also \citealt{mckee&tan2008, hosokawaetal2011, hiranoetal2014, susaetal2014}).

We also do not include the effects of the relative streaming motion between DM and baryons (\citealt{tse&hirata2010}), which may delay Pop III star formation by $\Delta z \sim 4$ and increase the turbulent velocity dispersion of the collapsing gas \citep{stacyetal2011}.  Higher turbulent velocities increase the rate at which the dynamo enhances minihalo magnetic fields.  Relative streaming may additionally alter overall minihalo statistics and supermassive black hole formation (e.g., \citealt{maioetal2011,naozetal2012,tanaka&li2014,scha19}).

Next, we note that the chemothermal rates used in our model are not perfectly constrained, and variation in these rates will also lead to differences in the collapse and fragmentation of primordial gas.  In particular, \cite{turk11b} find that using different published rates of three-body H$_2$ formation (e.g. \citealt{pallaetal1983, flower&harris2007, glover&abel2008}) leads to significant differences in long-term disk stability and fragmentation.  
As mentioned in Section 2.2 and discussed in Appendix \ref{app:chem}, we use the rates of \cite{forrey2013}, which fall between the higher rates published by \cite{flower&harris2007} and lower rates published by \cite{abeletal2002}, allowing us to avoid the high and low extremes of proposed H$_2$ formation rates.  

Finally we note that we have carried out only one MHD simulation, so it is not possible to draw firm conclusions on effect of magnetic fields on the IMF. \citet{shar21} found that multiple stars form in 2/3 of their MHD simulations, whereas that was not the case in our MHD simulation. It should be noted that our results are not directly comparable to theirs since they began with an isolated turbulent cloud, whereas our initial conditions came from a cosmological simulation.

\section{Summary and Conclusions}
\label{sec:summ}

The purpose of this paper is two-fold: First to
examine the impact of magnetic fields on the formation and growth of the first stars through a set of simulations initialized on cosmological scales, and second to compare the growth of the field from an initial low value with the theory developed in Paper I.  This represents one of the first simulations in which magnetic fields self-consistently evolved from cosmological scales were included in the subsequent formation and evolution of a Pop III system. We first employed the SPH code {\sc gadget-2} for the cosmological portion of the calculation, following the formation of a minihalo and dense central clump. This simulation extended from $z=100$ to $z=27.5$. The simulation was hydrodynamic, and we developed a method based on the deformation tensor to follow the kinematic evolution of the magnetic field in SPH (Appendix \ref{app:mag}). We then extracted a 0.5 pc$^3$ box from this simulation and mapped it onto an {\sc orion2} AMR grid with full MHD (Appendix \ref{app:mapping}).   
We chose the cosmological seed field to have the value $B_0 = 4.5\times 10^{-12}$ G
at $z=54$
in order that the dynamical effects of the magnetic field were $\la 10$ percent at the end of the {\sc gadget-2} run. 
With both hydro and MHD runs in {\sc orion2}, we continued the evolution to densities of $\sim$ 10$^{14}$ cm$^{-3}$ (although the accuracy declined above about $10^{12}$~cm\eee), replaced the resulting gravitationally unstable regions with sink particles (Appendix \ref{app:refine}), and followed the sink accretion stage for the next 2000 years.

{\it Evolution of the magnetic field.} We found that the magnetic field within the collapsing minihalo gas grew through compression and dynamo action to 
$\sim 10^{-5}$ G at a density of $10^4$~cm\eee, within an order of magnitude of field strengths observed in regions of contemporary star formation (e.g., \citealt{heiles&troland2005}). By the end of the {\sc gadget-2} simulation, the field reached
$\sim 3.5$~mG at a density $\nh=10^8$~cm\eee, a little less than half the equipartition value.  In the subsequent {\sc orion2} simulation, the field continued to grow to 1.0 G at a density $\nh\simeq 10^{12}$~cm\eee and was very close to equipartition.

Paper I presented a prediction of the results of the simulation based on determinations of the numerical viscosity in both SPH and AMR. Here we have found that two changes are needed to the theory in Paper I: First, we found that the growth rate of the field in the kinematic phase of a small-scale dynamo is 4-6 times less than that expected theoretically for the case of very high magnetic Prandtl numbers (section \ref{sec:kine} and Appendix \ref{app:growth}). Previous simulations of kinematic dynamos also showed showed lower growth rates than predicted by theory; our results are intermediate between those of \citet{fede11a} for a turbulent box and \citet{fede11b} for a collapsing cloud. We attribute this reduction to the fact that the magnetic Prandtl number in numerical simulations is of order unity, so that resistivity plays an important role. With this adjustment in the growth rate, we found that theory and simulation agreed to within a factor 2 over 7 orders of magnitude increase in density. The kinematic phase of evolution of the actual dynamo that occurs in the formation of the first stars operates at high magnetic Prandtl numbers, and we have no evidence that the growth rate of the field there differs from the theoretically expected value. Second, we found that the nonlinear dynamo violates flux-freezing, as proposed by \citet{xu20}. Our results are consistent with both their Jeans-regulated model ($B\propto\xi^{0.5+2/57}$) and a homologous model ($B\propto\xi^{0.5+3/38}$). The simulated nonlinear dynamo is very inefficient: Over the range $10^8<\nh<10^{12}$~cm\eee, it amplified the field by less than a factor 1.6. Real nonlinear dynamos are more efficient: We estimated that the nonlinear dynamo amplified the field in a cosmological minihalo by an order of magnitude in Paper I.

{\it Effect of magnetic fields on star formation.}
Our principal result is that magnetic fields amplified by a small-scale dynamo in collapsing minihalos can suppress fragmentation in the formation of the first stars, thereby leading to a top-heavy IMF.
In our magnetic simulation, two secondary sinks form but quickly merge with the primary sink, while only a single massive sink survives.
When magnetic fields are not included, a similarly massive sink still forms, but about 10 
secondary sinks survive as well, so that the total stellar mass is about twice as large.  The total accretion rate in the hydrodynamic case is close to the prediction of \citet{tan04}.

We conclude that a small-scale dynamo generates a magnetic field that can significantly reduce the number of low-mass Pop III stars that form and thereby create a top-heavy IMF, in agreement with \citet{shar21}.  As a result, magnetic fields may contribute to the rarity of low-mass Pop III stars, none of which has been observed to date.

\section*{Acknowledgments}
We thank Christoph Federrath, Alex Lazarian, Daniel Price, and Siyao Xu for
valuable discussions, and we thank Andrew Cunningham for sharing data analysis routines with us.
We thank Chalence Safranek-Shrader for significant contributions to the {\sc orion2} chemothermal network, and also for valuable discussions when developing this work. 
We thank the referee for feedback which helped to significantly improve the content and clarity of this paper.
CFM acknowledges the hospitality of the Center for Computational Astrophysics of the Flatiron Institute in New York, where he was a visiting scholar at the end of this work.
This research was supported in part by the NSF though grant AST-1211729  (A.S., C.F.M. and R.I.K.), by NASA through ATP grants NNX13AB84G, NNX17AK39G, and 80NSSC20K0530 (C.F.M and R.I.K), and by the US Department of Energy at the Lawrence Livermore National Laboratory under contract DE-AC52-07NA 27344 (R.I.K).
This research was also supported by grants of high performance computing resources from the National Center of Supercomputing Application through grant TGMCA00N020, under the Extreme Science and Engineering Discovery Environment (XSEDE), which is supported by National Science Foundation grant number OCI- 1053575, the computing resources provided by the NASA High-End Computing (HEC) Program through the NASA Advanced Supercomputing (NAS) Division at Ames Research Center (LLNL-JRNL-737724).
Some figures in this work were generated using the yt toolkit \citep{turk11a}.

\section*{Data Availability Statement}
The data underlying this article will be shared on reasonable request to the corresponding author.

\bibliographystyle{mnras}
\bibliography{PapII}{}


\appendix

\def\ead		{\eta_{\rm AD}}
\def\gad		{\gamma_{\rm AD}}
\def\he		{{\rm He}}
\def\hhp		{{\rm H\,H^+}}
\def\nuni		{\nu_{ni}}
\def\odon		{\omega_{d0,\nu}}
\def\rad		{R_{\rm AD}}
\def\sat		{{\rm sat}}
\def\sigv		{\avg{\sigma v}}
\def\visc		{{\rm visc}}
\def\vtf			{v_{t,5}}
\def\xif			{x_{i,-4}}

\section{Magnetic Field Calculation in SPH}
\label{app:mag}

\subsection{Kinematic SPH Formulation}
\label{appsub:kine}

Here we describe a method of determining the evolution of a kinematic magnetic field using the deformation tensor.
This enables us to follow the growth of a weak magnetic field in a hydrodynamic cosmological {\sc gadget-2} simulation.  
For each time step of the {\sc gadget-2} simulation, we follow the relative motion of the SPH particles from $z=100$ until the time when the particle properties are mapped onto the {\sc orion2} uniform grid, which occurs when the maximum density is $\nh\simeq 10^8$~cm\eee.  
For particles that begin at $\bf{x_0}$ and move to $\bf{x}$ at a later time, the deformation tensor $\bf{D}$ is defined as
\begin{equation}
D_{ij} = \frac{\partial x_i}  {\partial x_{0j}}. 
\end{equation}
During this time, the density at a particle transitions from $\rho_0$ to $\rho$ as follows: 
\begin{equation}
\rho(\vecx ) = \frac{\rho_0( \vecx_0 )} { | \bf{D} |}  \mbox{.}
\end{equation}
Conservation of magnetic flux determines that the particle's magnetic field correspondingly evolves from $\vecB_0$ to $\vecB$ as
\begin{equation}
\frac{ {B_i} (\vecx) } { \rho(\vecx) } = D_{ij} \frac{ B_{0j}( \vecx_0 ) }  { \rho_0( \vecx_0)  }  
\label{Btensor}
\end{equation}
(see  \citealt{walen1947, newcomb1962, zweibel&mckee1995}).  

To accurately calculate the components of $\bf{D}$, we calculate the gradient as suggested in \cite{price2012a} by solving 
\begin{equation}
 \chi^{ij} \frac{\partial  A_a} {\partial {\bf r}^i} =\sum_{b} m_{b} (A_b - A_a) \nabla^j W_{ab}
\end{equation}
where $\chi$ is the matrix quantity
\begin{equation}
 \chi^{ij} \equiv \sum_{b} m_{b} ({\bf r}_b - {\bf r}_a)^i \nabla^j W_{ab} \mbox{,}
\end{equation}
and $W_{ab}$ is the smoothing kernel as defined in equation (\ref{eq:kernel}).
In equation A5, $a$ represents the particle in question, its set of kernel neighbors are denoted by $b$, ${\bf r}_a$ and ${\bf r}_b$ are the positions of particles $a$ and $b$, 
and $i$ and $j$ each represent a component along one of the three Cartesian axes.  Each element of the three-by-three matrix $\chi$ is a kernel-weighted average of the neighbors of particle $a$ (see equation 72 of \citealt{price2012a}), and we determine the elements of $\chi$ for each particle at
each
time step.
From Equation A4 we find
\begin{equation}
\frac{\partial  A_a} {\partial {\bf r}^i} =  \left(\chi^{-1}\right)^{ij} \sum_{b} m_{b} (A_b - A_a) \nabla^j W_{ab}
\end{equation}
where $\chi^{-1}$ is the inverse of the matrix $\chi$.  
We next replace $A$ with $\bf{r}_{1}$, the particle position at the subsequent time step. The deformation tensor is then
\begin{equation}
D_{ij}=\frac{\partial {\bf r}_{1}^i} {\partial {\bf r}^j} =  \left(\chi^{-1}\right)^{jk} \sum_{b} m_{b} ({\bf r}_{1b} - {\bf r}_{1a})^{i} \nabla^k W_{ab}.
\end{equation}

Note that this method follows only the advection of the magnetic field, and thus is appropriate only for dynamically insignificant fields.  After being mapped onto the {\sc orion2} grid, the subsequent evolution of the the field to dynamically significant strengths is followed with the {\sc orion2} MHD.  

As the magnetic field is evolved in {\sc gadget-2}, we additionally apply the hyperbolic/parabolic divergence cleaning scheme of \cite{dedneretal2002}.  We apply this to the SPH formulation as described in \cite{tricco&price2012} (see also \citealt{price&monaghan2005}).  
This scheme adds a divergence cleaning term to the time-evolution of the magnetic field:
\begin{equation}
\left( \frac{{d }{\vecB} }{ {d} t} \right)_{\psi} = -  \grad \psi \mbox{,}
\end{equation}
where
\begin{equation}
\frac{{d} \psi }{ {d} t}  = -c_h^2 \grad \cdot {\vecB} - \frac{\psi}{\tau} \mbox{.}
\end{equation}
In the above equation, the first term is the hyperbolic term where the wave speed $c_h$ is typically set to the maximum signal propagation speed.  The second term is the parabolic damping term where we set 
\begin{equation}
\frac{1}{\tau} = \frac{\sigma c_h} {h_{\rm sm}} \mbox{.}
\end{equation}
Here $\sigma$ is a dimensionless quantity specifying the damping strength (\citealt{tricco&price2012}), and it can range from zero to one.  We use $\sigma=1$ since that best minimized the remaining divergence error.

The first term of the equation for ${\rm d} \psi / {\rm d} t$ is discretized for a particle `$a$' and neighbor particles `$b$' in SPH as described in
\cite{tricco&price2012} eq. (30):
\begin{equation}
\frac{{d} \psi_a }{ {d} t} = c_h^2 \frac{1}{\Omega_a\rho_a} \sum_{b} m_b \left( {\rm {\vecB_a - \vecB_b}} \right) \cdot \grad_a W_{ab}(h_a) \mbox{,}
\end{equation}
where $\Omega_{a}$ is a term arising due to gradients in the smoothing length (see \citealt{tricco&price2012} for details).
We set $\Omega_a=1$ in order to simplify and increase the speed of the calculation.
A direct calculation of $\Omega_{a}$ for a subset of the SPH particles at the final snapshot shows that $\Omega_{a}$ varies between 0.7 and 1.0 over the entire density range, while over 99\% of the particles had  $0.78 < \Omega_{a} < 0.85$.  Setting $\Omega_{a} = 1$ thus had minimal affect other than effectively increasing the parameter $c_h^2$ by $\sim 25\%$, thereby slightly increasing the rate of divergence cleaning.
With that approximation,
the term for $ { d }{\vecB}  /{d} t$ is discretized as in \cite{tricco&price2012} eq. (34):
\begin{equation}
\left( \frac{d {\vecB_a}} {{d} t} \right)_{\psi} = -\rho_a \sum_{b} m_b \left[ \frac{\psi_a}{\rho_a^2}\grad_a W_{ab}\left(h_a \right)  + \frac{\psi_b}{\rho_b^2}\grad_a W_{ab}\left(h_b \right) \right].
\end{equation}

We also apply a switch for artificial resistivity as described in \cite{tricco&price2013}, particularly their equations (6) and (8):
\begin{equation}
\left( \frac{d {\vecB_a}} {{d} t} \right)_{\rm diss} = \rho_a \sum_{b} m_b \frac{\alpha_{B,ab} v_{\rm sig}^B}{\rho^2_{ab}} \left( {\rm {\vecB_a - \vecB_b}} \right) \hat{r}_{ab} \cdot\grad_a W_{ab}
\end{equation}
with
\begin{equation}
\alpha_{B,a} = \frac{h_{\rm sm, a} |\grad {\vecB_a}|}{|{\vecB_a}|} \mbox{.}
\end{equation}

\subsection{The Cosine Whirl Test}
\label{appsub:whirl}

\begin{figure*}
\includegraphics[width=.32\textwidth]{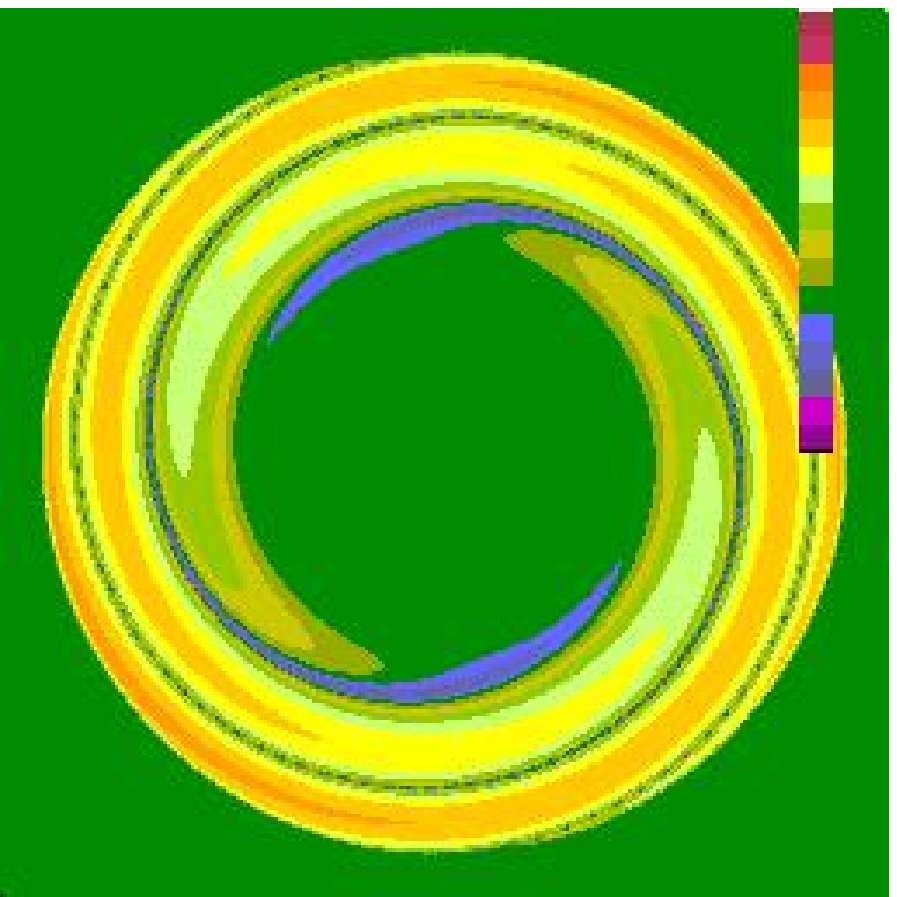}
\includegraphics[width=.32\textwidth]{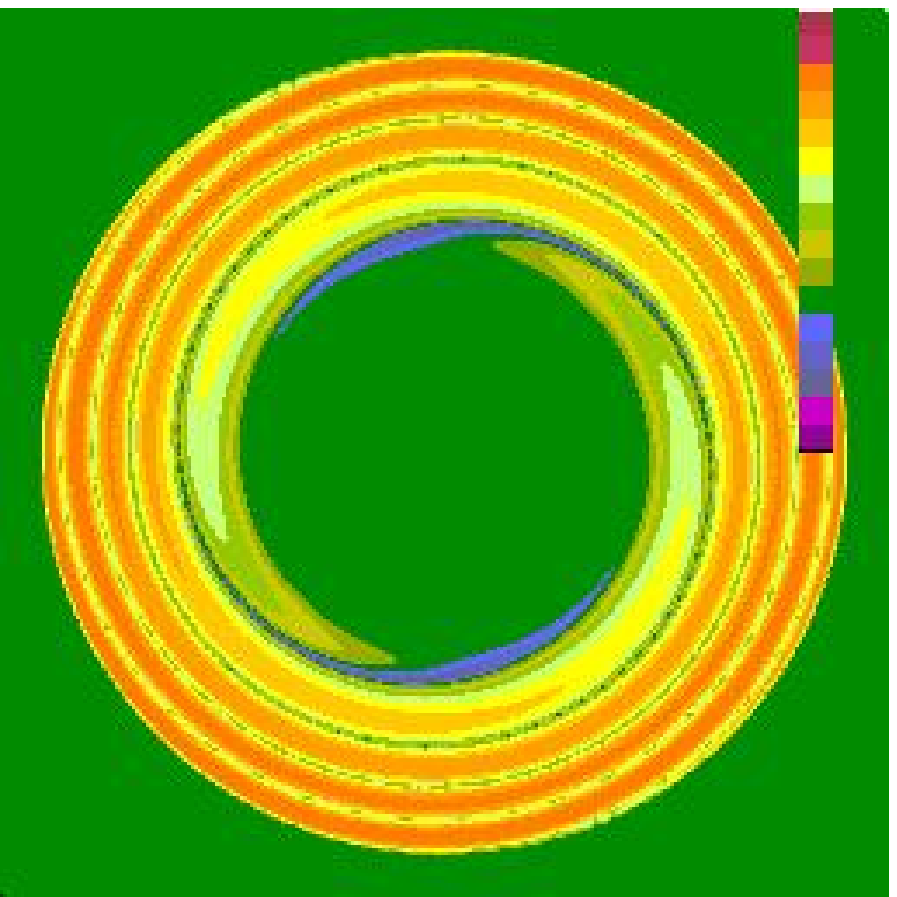}
\includegraphics[width=.32\textwidth]{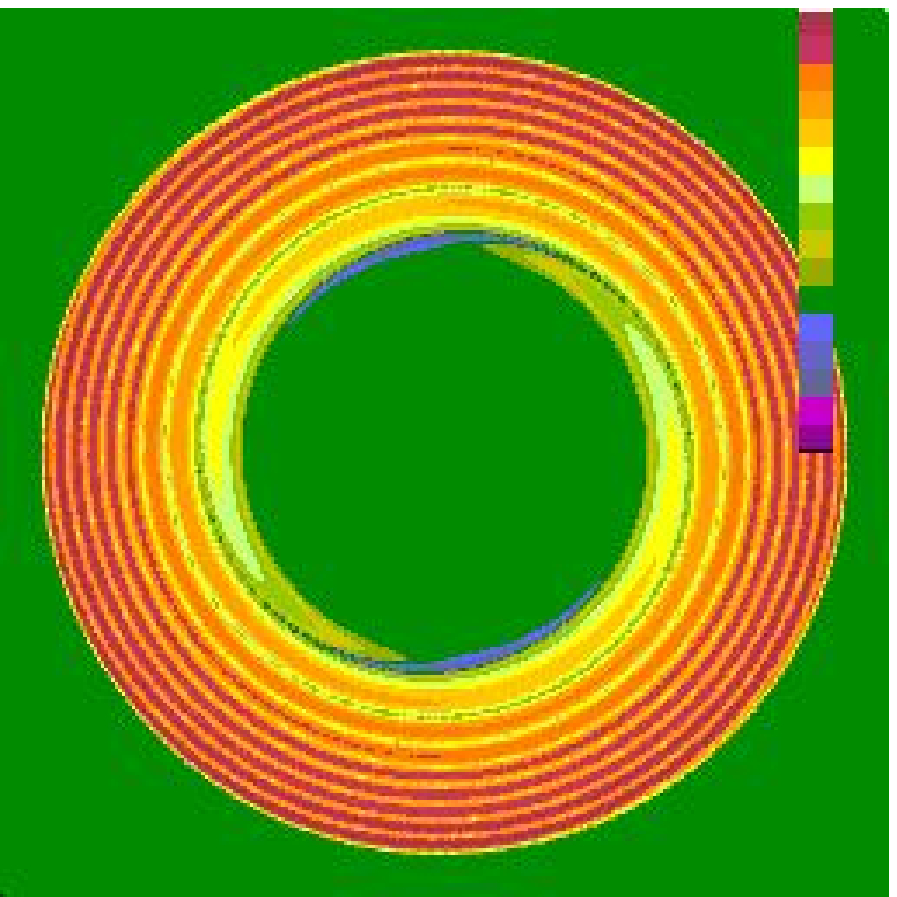}
 \caption
 {
Magnitude of magnetic field strength $B$ after one, two, and four rotations of the cosine whirl, as determined from the analytic solution.  
Color bar ranges over a factor 1000.
 }
\label{whirl}
\end{figure*}

To test our method of following magnetic field growth using the deformation tensor, we obtain an analytic solution for the field amplified by a steady azimuthal flow in the $x_1 - x_2$ plane.  Let $\phi$ be the azimuthal angle, and assume $v_\phi = v_\phi(r)$.  The initial location of a point is denoted by $(x_{0,1}, x_{0,2})$.  Since it remains at constant radius $r = (x_{0,1}^2 + x_{0,2}^2) ^ {1/2}$, the initial location in cylindrical coordinates is $(r, \phi_0)$.  Subsequently, we have $\phi = \phi_0 + \dot{\phi}(r)t$, so that in cartesian coordinates
\beqa
x_1 &=& r \, {\rm cos}\phi = r \, {\rm cos} \left[  {\rm cos}^{-1} \left(  \frac{x_{0,1}} {r} \right) + \dot{\phi}(r)t \right],
\label{x_def}\\
x_2 &=& r \, {\rm sin}\phi = r \, {\rm sin} \left[  {\rm sin}^{-1} \left(  \frac{x_{0,2}} {r}  \right) + \dot{\phi}(r)t \right].
\label{y_def}
\eeqa
For the components of Equation \ref{Btensor} we find
\beqa
\frac{\partial x_1} {\partial x_{0,1} }& =& \frac{1} {r^2} \left[ x_{0,1} x_1 + x_{0,2} x_2 - a x_{0,1} x_2 \right],\\
\frac{\partial x_1} {\partial x_{0,2} } &=& \frac{1} {r^2} \left[ x_{0,2} x_1 - x_{0,1} x_2 - a x_{0,2} x_2 \right],\\
\frac{\partial x_2} {\partial x_{0,1} } &=& \frac{1} {r^2} \left[ x_{0,1} x_2 - x_{0,2} x_1 + a x_{0,1} x_1 \right],\\
\frac{\partial x_2} {\partial x_{0,2} } &=& \frac{1} {r^2} \left[ x_{0,2} x_2 + x_{0,1} x_1 + a x_{0,2} x_1 \right],
\eeqa
where
\begin{equation}
a \equiv r \left[ \frac{\partial \dot{\phi} (r) } {\partial r} \right] \, t
\label{a_def}
\end{equation}
Note that $a=0$ for uniform rotation.  One can show that det$(D) = 1$, so the density remains constant.

\begin{figure*}
\includegraphics[width=.47\textwidth]{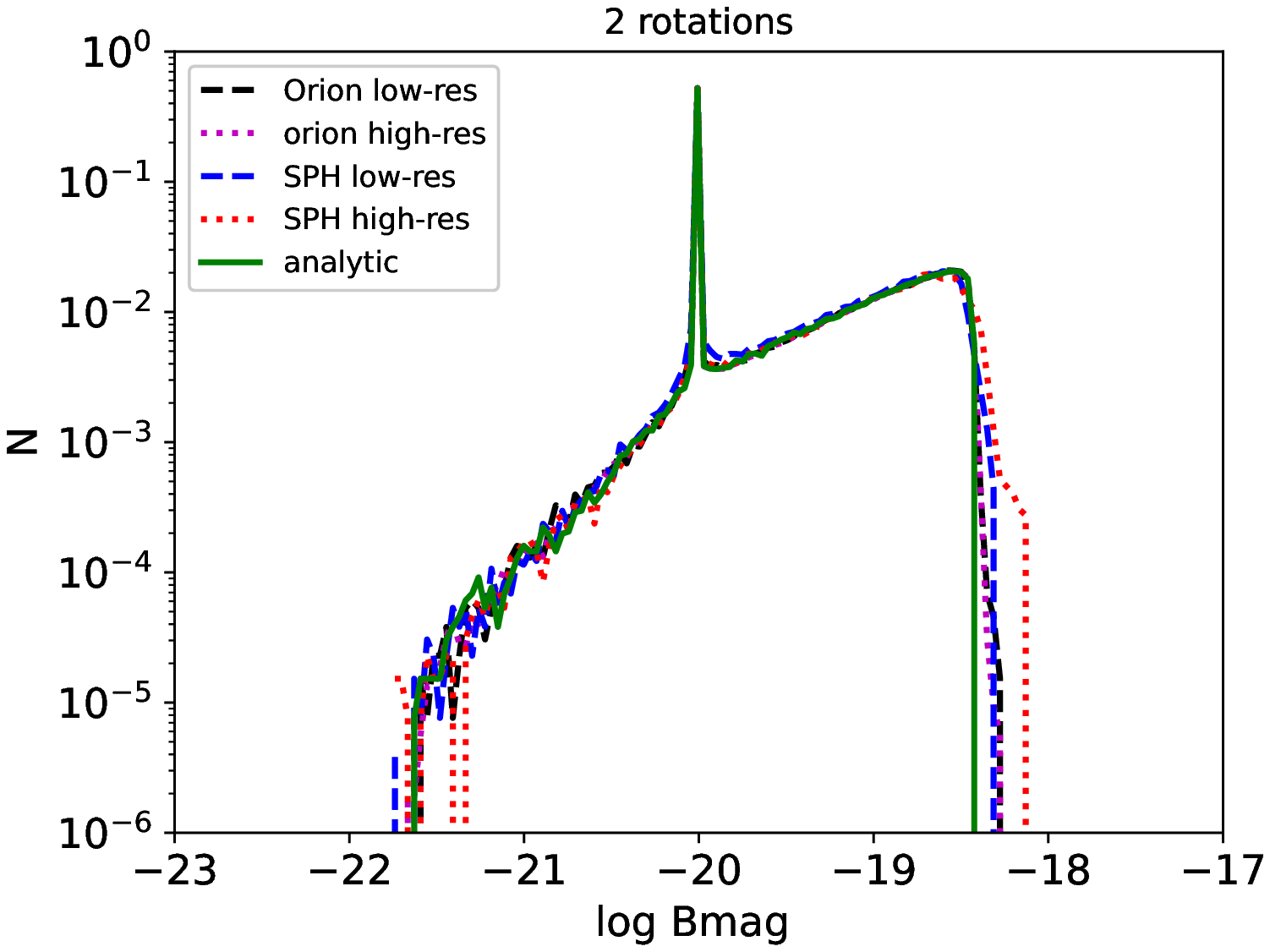}
\includegraphics[width=.47\textwidth]{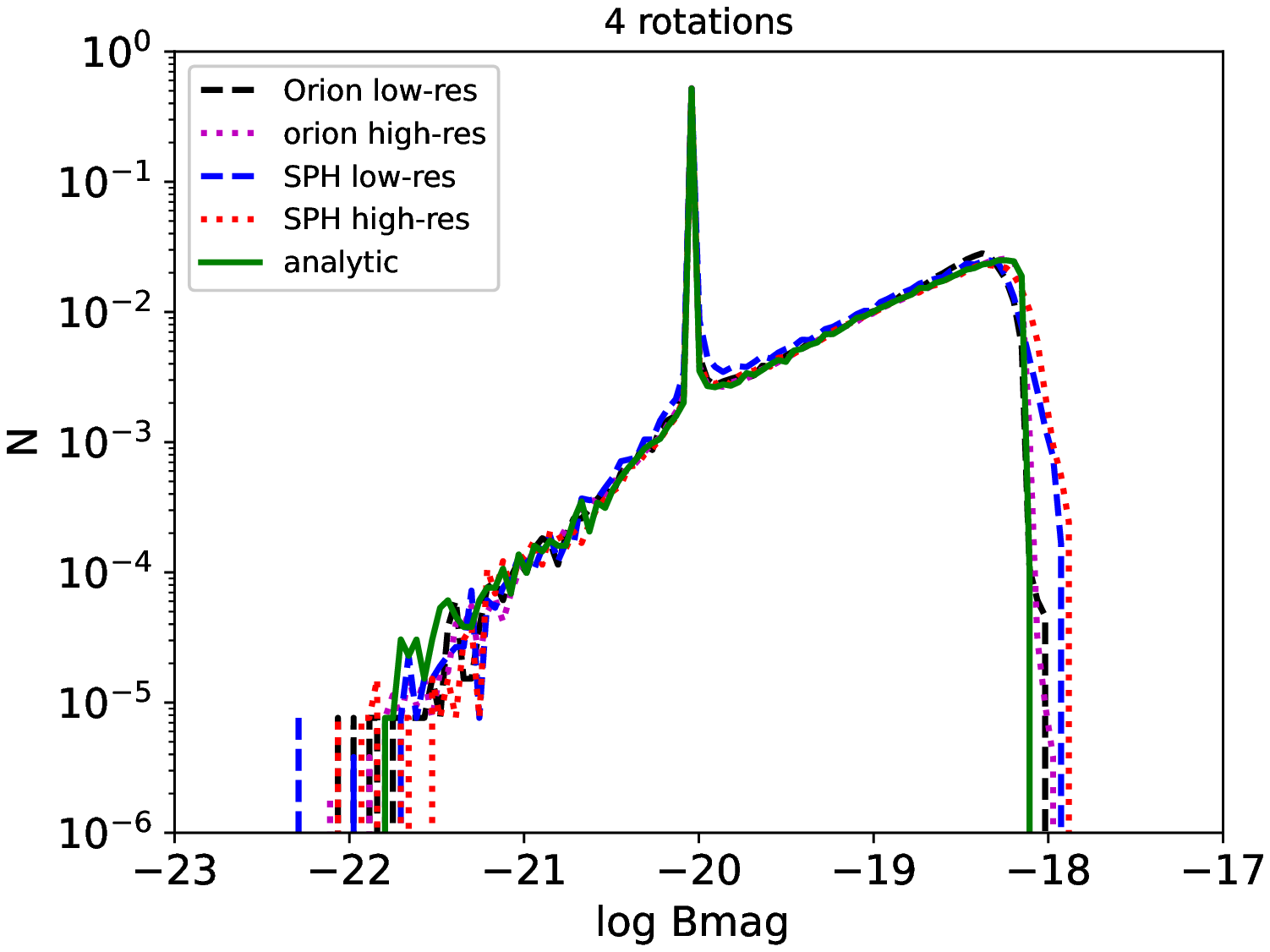}
\includegraphics[width=.47\textwidth]{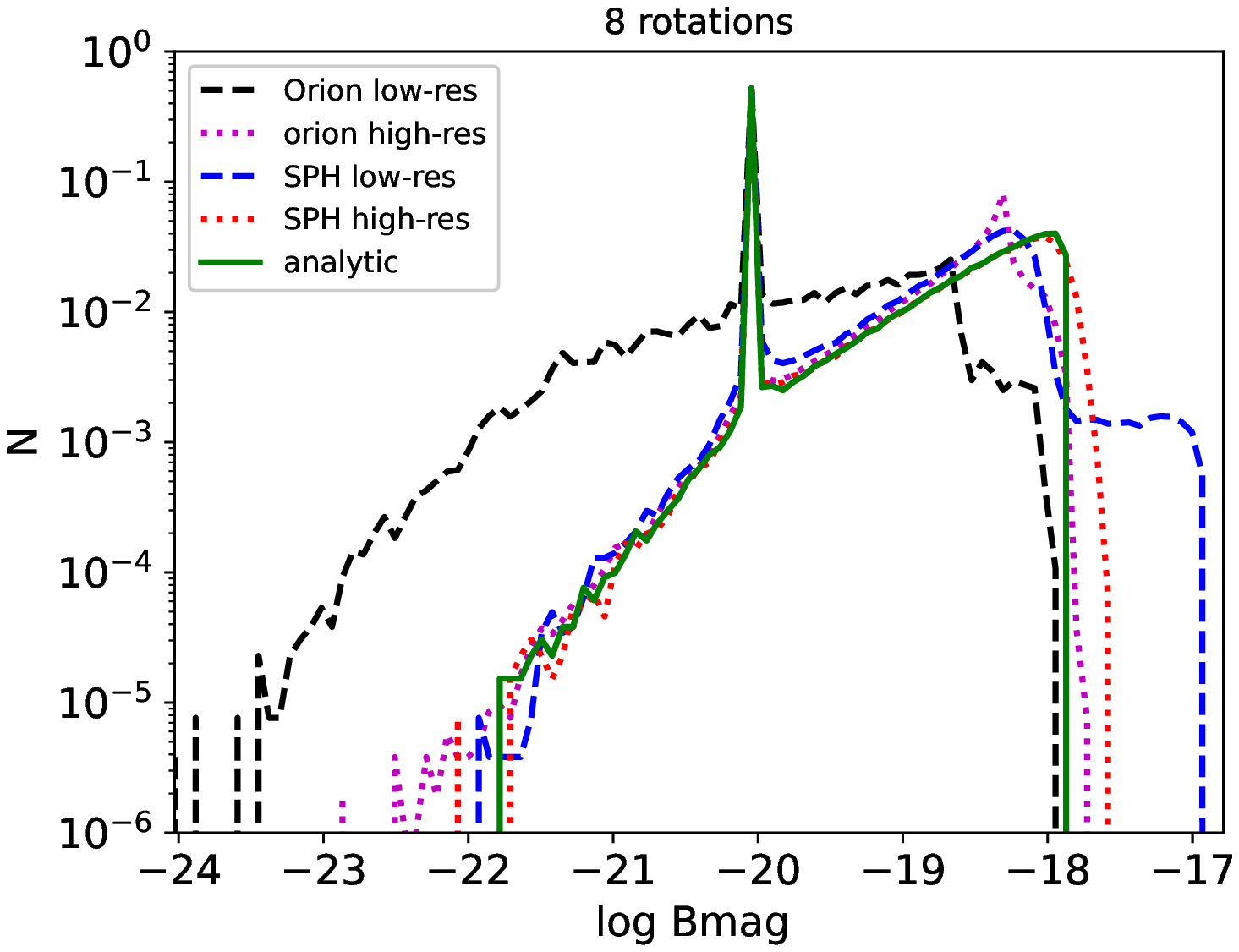}
\includegraphics[width=.47\textwidth]{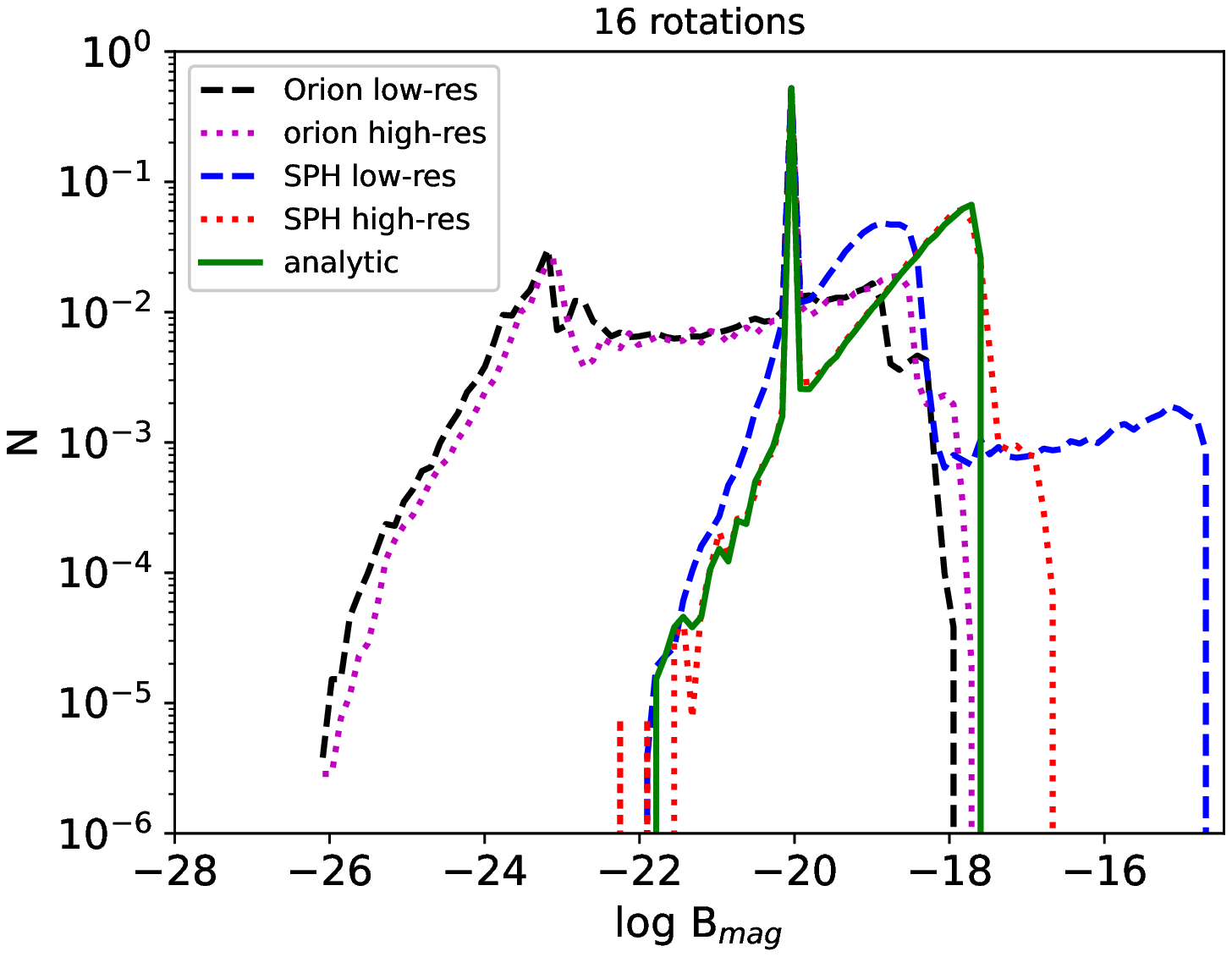}
 \caption
 {
 Probability distribution function of magnitude of $B$ after two, four, eight, and 16 rotations of the cosine whirl.  Peak at $B=10^{-20}$ G results from the portion of the fluid whose initial magnetic field remains unchanged.
 Solid green line shows the analytic solution.  Dashed black and dotted magenta lines depict gridLR and gridHR.  Dashed blue and dotted red lines represent sphLR and sphHR.
 The resolution requirements implied by these results for both SPH and grid simulations are given in equation (\ref{eq:resolution}).
 }
\label{pdf_rot}
\end{figure*}

\begin{figure}
\includegraphics[width=.48\textwidth]{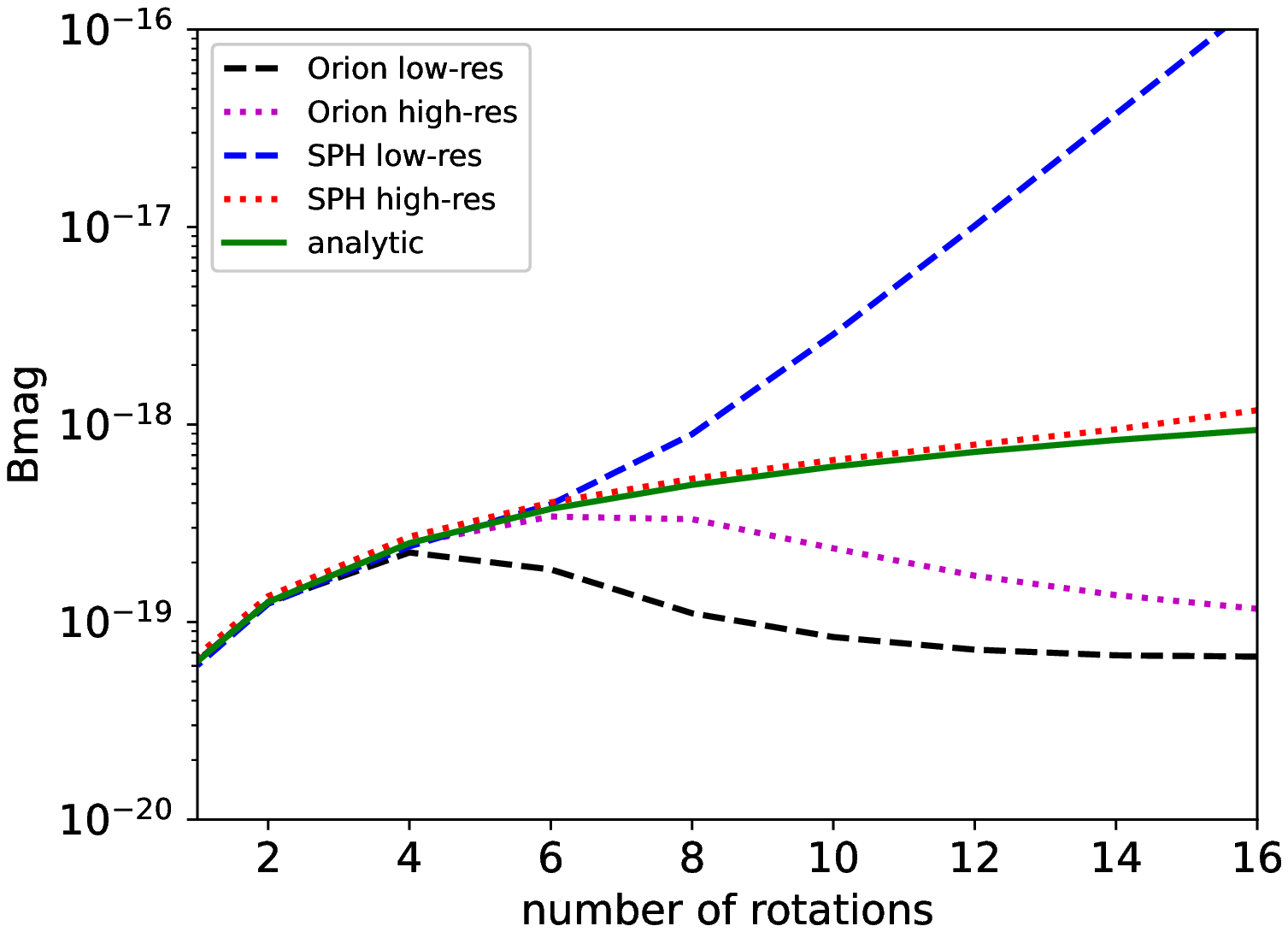}
 \caption
 {
Evolution of rms 
field
with increasing number of rotations in the cosine whirl test.  Line colors and style have same meaning as in previous figures.  
 }
\label{b_evol}
\end{figure}

Now let us consider an initial field ${\vecB_0} = B_0  \hat{\boldsymbol{x}}_1$.  
In this case we find
\begin{equation}
\begin{split}
{\vecB} &= B_0 \left( \frac{\partial x_1}{\partial x_{0,1}}  \hat{\boldsymbol{x}}_1   + \frac{\partial x_2}  {\partial x_{0,1}}  \hat{\boldsymbol{x}}_2  \right) \\
&= \frac{B_0}{r^2}  
\left( x_{0,1}x_1 + x_{0,2}x_2 - ax_{0,1}x_2 \right)\hat{\boldsymbol{x}}_1 \\
&+  \frac{B_0}{r^2} 
\left( x_{0,1}x_2 - x_{0,2}x_1 + ax_{0,1}x_1 \right)\hat{\boldsymbol{x}}_2 
\end{split}
\end{equation}
The magnitude of the field satisfies
\begin{equation}
B^2 = B_0^2 \left[ 1 - 2a\left(\frac{x_{0,1}x_{0,2}} {r^2} \right) + a^2\left( \frac{x_{0,1}^2} {r^2} \right)\right]
\end{equation}
and is portrayed in Fig. \ref{whirl}.

To see the pattern of field reversals, consider the field along the $x_2$ axis.  From equations (\ref{x_def}) and (\ref{y_def}) 
with $x_1=0$ and $x_2=r$
we find
\beqa
x_{0,1} &=& r \, {\rm cos} \left( \frac{\pi}{2} - \dot{\phi} t  \right) = r\sin(\dot\phi t)\mbox{,}\\
x_{0,2} &=& r \, {\rm sin} \left( \frac{\pi}{2} - \dot{\phi} t  \right) =r\cos(\dot\phi t)\mbox{,}
\eeqa
so that
\begin{equation}
B_1 = B_0 \left[ {\rm cos}\left(\dot{\phi} t \right) - a {\rm sin} \left( \dot{\phi} t \right) \right] \mbox{.}
\end{equation}
Writing $a = {\rm tan} \, \delta$ yields
\begin{equation}
B_1 = \frac{B_0}{{\rm cos} \delta} {\rm cos} \left( \dot{\phi} t + \delta \right).
\end{equation}
One can furthermore show that ${\rm cos \, \delta} = (1 + a^2)^{-1/2}$, so that this becomes
\begin{equation}
B_1 = B_0 \, \left( 1 + a^2 \right)^{1/2} {\rm cos} \left( \dot{\phi} t + {\rm tan}^{-1} \, a \right).
\end{equation}
One can similarly show that
\begin{equation}
B_2 = B_0 \left( \frac{x_{0,1}}{r} \right) = B_0 {\rm sin} \left( \dot{\phi} t \right) \mbox{.}
\end{equation}
Note that at late times the magnitude of $B_1$ increases linearly with time, whereas $B_2$ varies sinusoidally at all times.

To test the code we need a velocity pattern that is rigidly rotating near the origin and is motionless at the boundaries.  A pattern that satisfies this and has a continuous first derivative is:
\begin{align}
\dot{\phi} &= \dot{\phi}_1    & {\rm for} \; 0  \leq r < r_1 \\
  &= \frac{\dot\phi_1}{2} \left[ 1 + {\rm cos} \pi \left( \frac{r - r_1}{r_2 - r_1} \right)\right]  & {\rm for} \; r_1 \leq r \leq r_2 
 \end{align}
It follows that the quantity $a$ defined in Equation \ref{a_def} is
\begin{align}
a &= 0   & {\rm for} \; 0  \leq r < r_1 ,\\
& = - \frac {\dot{\phi} t} {2} \left( \frac{\pi r} {r_2 - r_1} \right) {\rm sin} \pi \left( \frac{r-r_1}{r_2 - r_1} \right)  & {\rm for} \; r_1 \leq r \leq r_2 .
\end{align}
We may furthermore write $\dot{\phi} t$ in terms of the number of rotations:
\begin{equation}
N = \frac{\dot{\phi} t } {2 \pi } \mbox{.}
\end{equation}
Then $B_1$ and $B_2$ depend on $2 \pi N$, and at late times there are $2N + \frac{1}{2}$ field reversals in $B_1$ and $B_2$ between $r = r_1$ and $r = r_2$.

In our test we use a use a uniform distribution of 600$^2$ and 1500$^2$ SPH particles and evolve them according to the above cosine-whirl velocity structure within a simulation box of unit length, and the initial magnetic field is set to be uniform along the $x$-axis, $B = B_1 = 10^{-20}$ G.  
We set $r_1 = 0.25$ and $r_2 = 0.45$.
We refer to these two tests as `sphLR' and `sphHR,' and the particles in these tests have smoothing lengths of $h_{\rm sm} = 5\times10^{-3}$ and $1.5\times10^{-3}$, respectively.  
We additionally evolve uniform 512$^2$ and 1024$^2$ grids in {\sc Orion2} in the same way (`gridLR' and 'gridHR').  

All test simulations remain consistent with the analytic solution after the central uniformly-rotating region has undergone four complete rotations.  This is evident in the probability distribution functions and rms of the the magnitude of $B$, as shown in Figs. \ref{pdf_rot} and \ref{b_evol}.  We find that after approximately 6 rotations, sphLR (blue line in Fig. \ref{pdf_rot}) begins to deviate from the analytic solution, while sphHR (red line) slightly deviates after 16 rotations.  

As described above, there are approximately $2N$ field reversals between $r_1$ and $r_2$.  This
allows us to estimate the resolution needed to follow the whirl between $r_1$ and $r_2$:
\begin{equation}
c_{\rm sph} h_{\rm sm} = \frac{r_2 - r_1} {2N} \mbox{,}
\end{equation}
where $c_{\rm sph}$ is the required number of smoothing lengths per field reversal.  Approximating $N \sim 6$ and $N \sim 16$ for sphLR and sphHR, then in both cases $c_{\rm sph} \sim 4$.

Similarly, for the AMR simulations, gridLR and gridHR no longer resolve the analytic solution after 
$N \sim$ 6 and 12 rotations, respectively.
This leads to the analogous relation,
\begin{equation}
c_{\rm grid} l_{\rm grid} = \frac {r_2 - r_1} {2N} \mbox{,}
\end{equation}
where $l_{\rm grid}$ is the length of one grid cell and 
$c_{\rm grid} \sim  8$
is the number of grid cells needed to resolve a field reversal.  

In more general terms, to resolve a field
$B_x = {\rm sin} \left(2 \pi y/\lambda \right)$
(i.e., one reversal),
we would require either a grid or smoothing length of
\begin{equation}
l \le \frac{\lambda}{2c_{\rm res}} \mbox{,}
\label{eq:resolution}
\end{equation}
where for our grid test 
$c_{\rm res}\sim 8$
and for our SPH test $c_{\rm res}~\sim4$.

\subsection{Divergence Cleaning Test}
\label{appsub:div}

We test our SPH divergence cleaning using a test similar to that presented in \cite{price&monaghan2005} and \cite{tricco&price2012}.  We introduce a magnetic field with artificial divergence within a cube of dimension [1,1,1] with periodic boundary conditions.  The box is uniformly filled with 100$^3$ particles with density $\rho=1$, pressure $P=1$, and adiabatic index $\gamma=5/3$.  The velocity field is [1,1,1].  In addition, each particle has $B_z = 10^{-19} / \sqrt{4\,\pi}$.  A perturbation to the $x$-component of the magnetic field is introduced as follows:
\begin{equation}
B_x = \frac{1} {\sqrt{4\,\pi}} \left[ \left( \frac{r}{r_0}\right)^8 - 2 \left(\frac{r} {r_0} \right)^4 + 1 \right]    \quad {\rm for} \quad r < r_0 \mbox{,}
\end{equation}
where $r$ is the distance from the point [0.2, 0.2, 0.2] and $r_0 = 0.1$.

Fig. \ref{monopole} shows the result when the divergence is advected, when undamped cleaning is applied (purely hyperbolic), and when damped cleaning is applied (both hyperbolic and parabolic).  For damped cleaning we apply $\sigma=1.0$, and the divergence is quickly removed.

\begin{figure*}
\includegraphics[width=.9\textwidth]{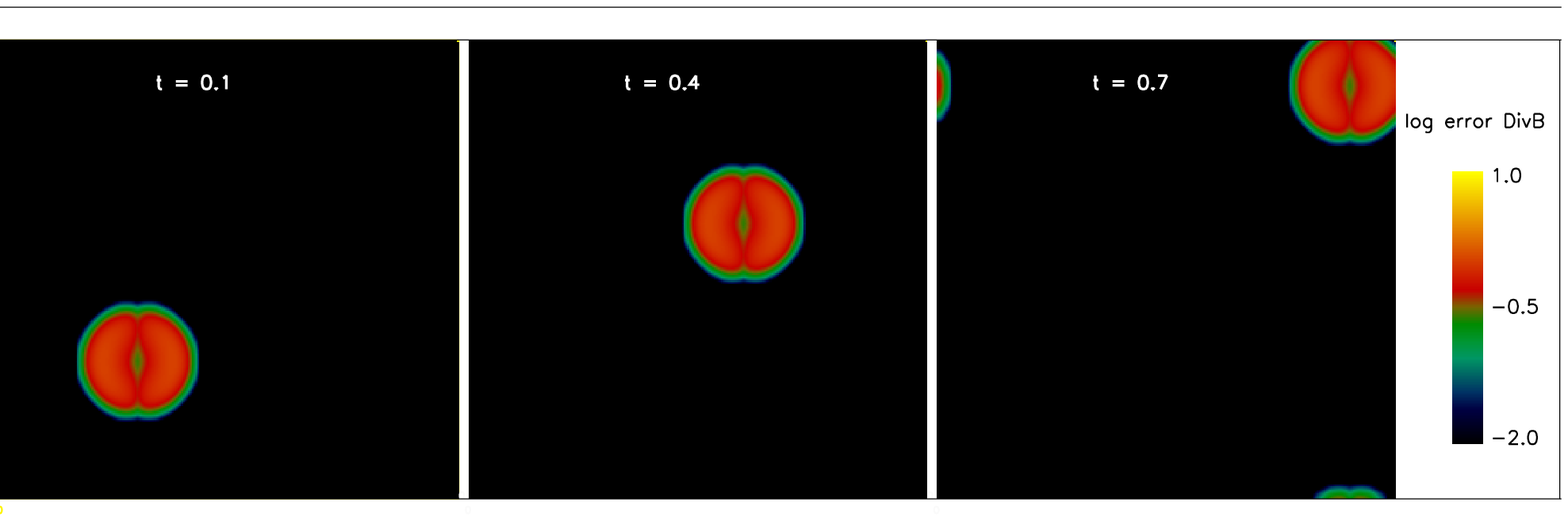}
\includegraphics[width=.9\textwidth]{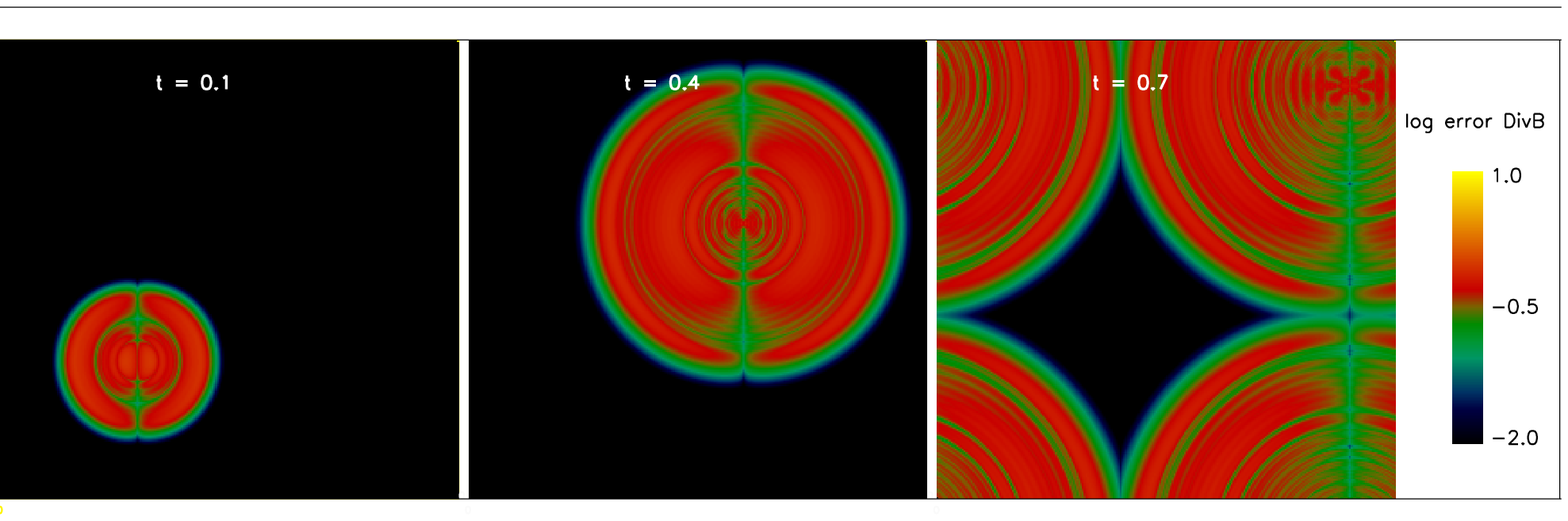}
\includegraphics[width=.9\textwidth]{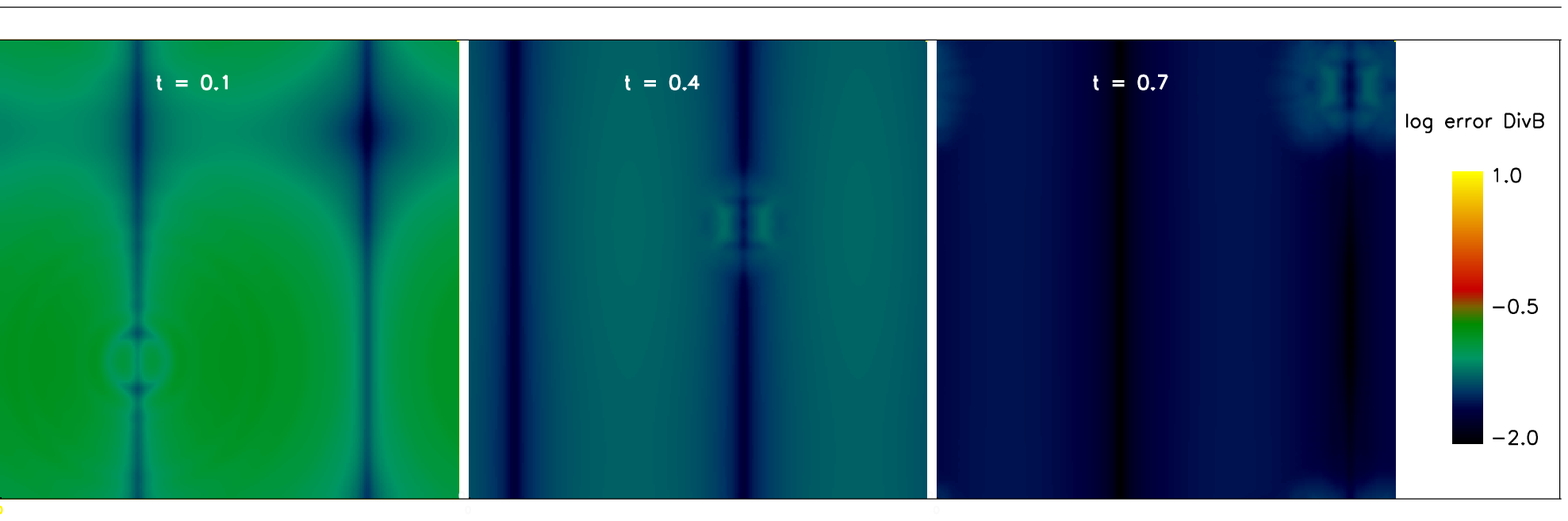}
 \caption
 {
 Blob of divergence within the magnetic field of a fluid moving at uniform velocity, as described in Appendix \ref{appsub:div}.  
 {\it Top row:} No cleaning applied.  The magnetic field is simply advected.
 {\it Middle row:} Undamped cleaning.  Only the hyperbolic term is applied to the divergence cleaning term.
 {\it Bottom row:} Damped cleaning.  Both the hyperbolic and parabolic terms are applied. 
 Damped cleaning quickly removes the divergence.
 }
\label{monopole}
\end{figure*}

\subsection{Additional Divergence Cleaning}
\label{appsub:add}

We perform further divergence cleaning before applying the magnetic field determined from {\sc gadget-2} to an {\sc orion2} grid.  
We map the magnetic field of individual particles $ \vecB_{i,\rm sph}$ onto the {\sc orion2} grid in the same fashion as the other SPH quantities.  Each grid cell center $(i,j,k)$ is assigned a value $\vecB_{(i,j,k)}$.  To satisfy the divergence condition, we perform a scalar divergence cleaning similar to that described in \cite{kimetal1999} and \cite{bals04}.  In this method  $\vecB_{(i,j,k)}$ is defined such that
\begin{equation}
\vecB = \vecB_{\rm uncor} + \grad\phi \mbox{}
\end{equation}
where $\vecB_{\rm uncor}$ is the uncorrected magnetic field, before divergence cleaning. The scalar function $\phi$ is defined as 
\begin{equation}
\grad^2\phi= -\div\vecB_{\rm uncor} \mbox{,}
\end{equation}
so that
$\vecB_{(i,j,k)}$ is divergence free.

\begin{figure}
\includegraphics[width=.48\textwidth]{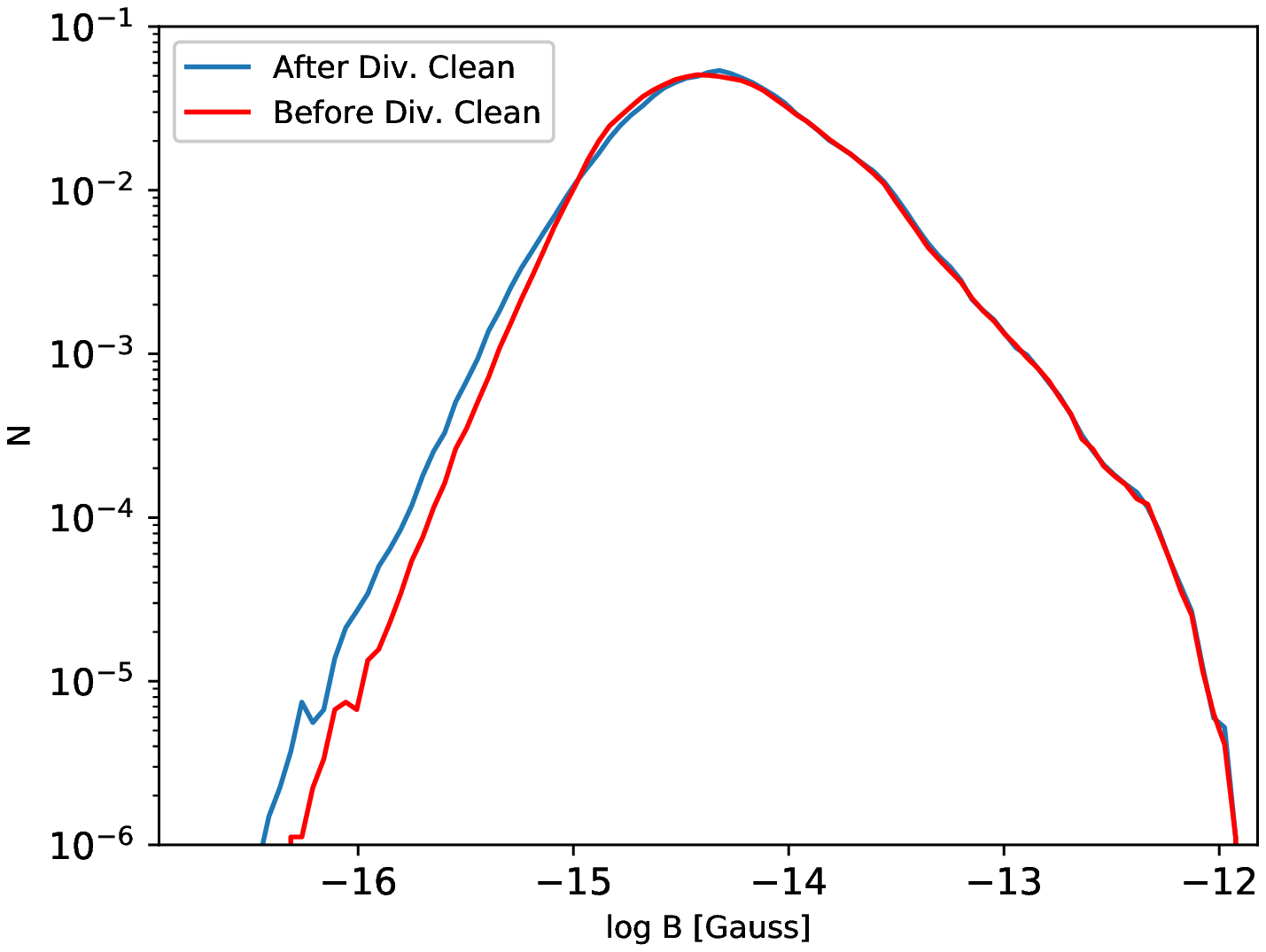}
 \caption
 {
 Divergence cleaning: Probability distribution function (pdf) of original magnetic field magnitude $B$ (blue line) as evolved during the {\sc gadget-2} simulation and of divergence-cleaned component $B_{\rm clean}$ (red line).  Due to effective divergence cleaning throughout the SPH simulation, there was little additional change between  $B$ and $B_{\rm clean}$ when initializing the Orion2 simulation.
 }
\label{Bpdf}
\end{figure}
We determine $\phi$ using the Jacobi iterative method and assuming periodic boundary conditions.  
There was minimal overall change in the value of $\vecB$ after divergence cleaning (see Fig. \ref{Bpdf}).

\section{Mapping From SPH to AMR}
\label{app:mapping}

To initialize the {\sc orion2} simulations we map data from the cosmological SPH {\sc gadget-2} simulation onto a uniform {\sc orion2} grid 
(cf. \citealt{richardsonetal2013}).  Each SPH particle is described by a mass $m_i$, a smoothing length $h_a$, and a smoothing kernel $W(r,h_a)$:
\begin{equation} 
W(r,h_a)=\frac{8}{\pi h_a^3} 
\left\{
\begin{array}{ll}
1-6\left(\frac{r}{h_a}\right)^2 + 6\left(\frac{r}{h_a}\right)^3, &
0\le\frac{r}{h_a}\le\frac{1}{2} ,\\
2\left(1-\frac{r}{h_a}\right)^3, & \frac{1}{2}<\frac{r}{h_a}\le 1 ,\\
0 , & \frac{r}{h_a}>1 ,
\end{array}
\right.
\label{eq:kernel}
\end{equation}
(\citealt{springeletal2001}), where $r$ is the distance to the particle $a$ from a given point in the computational box.  The contribution of a particle $a$ to the density at a given point is
\begin{equation}
\rho_a(r) = m_a \, W(r,h_a).
\end{equation}
We translate an SPH particle density to a corresponding density within an AMR grid cell $j$ by first considering the mass contribution of particle $a$ to the cell:
\begin{equation}
m_{aj} = \int\limits_{\rm cell} m_a W(r_{aj};h_a) dV = m_a f_{aj}
\end{equation}
where $r_{aj}$ is the distance from the SPH particle to a point within the grid cell.  The fraction $f_{aj}$ of the kernel-weighted volume of particle $a$ that overlaps with grid cell $j$ is taken as
\begin{equation}
f_{aj} = \int\limits_{\rm cell} W(r_{aj};h_a) dV \simeq l_{\rm grid}^3 \left< W \right>
\end{equation}
where $\left< W \right>$ is an estimate of the  
average value of the
kernel and defined as
\begin{equation}
 \left< W \right> = \frac{1}{N}  \sum\limits_{a=1}^N W(r_{aj};h_a) \mbox{.}
\end{equation}
If an SPH particle is entirely enclosed with a grid cell, then we set $f_{aj} = 1$.  Otherwise, for each partially overlapping particle-cell pair, we evaluate $\left< W \right>$ by summing over $N=2500$ random points within the cell. We then check the standard deviation 
of the mean, $\sigma$, 
of the 2500 values.  If $\sigma < 0.001$, then $\left< W \right>$ is used in the final evaluation of $f_{ab}$.  Otherwise we take a separate sample of 2500 points.
Once $f_{aj}$ is determined for all particles overlapping a cell, the density of cell $j$ is then
\begin{equation}
 \rho_j = \sum\limits_{a} f_{aj} m_a /  l_{\rm grid}^3 \mbox{.}
\end{equation}

The other physical properties of the grid cell that must be initialized are the internal energy and the chemical abundances as well as the $x$, $y$, and $z$ components of velocity.  For each particle we first subtract the 
the center-of-mass velocity of all particles within the 1-pc `cut-out' box to be mapped onto the uniform {\sc orion2} grid.  The value $X_j$ of the grid cell velocity component, internal energy, etc., is then given by the corresponding $X_a$ values of the SPH particles:
\begin{equation}
X_j =   \frac{\sum\limits_{a}   f_{aj} X_a   m_a }  { l_{\rm grid}^3 \rho_j}\mbox{.}
\end{equation}
Our mapping procedure conserves the total mass, internal energy, and linear momentum when summed over the entire box. However, as discussed by \citet{richardsonetal2013}, the mapping procedure does not conserve the total angular momentum because during the mapping, SPH particles that are offset from the center of the grid are effectively ``moved" to the center of the grid while their velocities remain unchanged.
As shown in Fig. \ref{amom}, this effect is minimal when particles are shifted small distances on high-resolution grids.  
At the peak densities above $\sim8\times10^7$ cm$^{-3}$, representing approximately only the central 2$^3$ grid cells, the {\sc orion2} angular momentum is as much as twice as high as that for {\sc gadget-2}.
However, at densities between $10^6$ cm$^{-3}$ and $\sim5\times10^7$ cm$^{-3}$, the difference in angular momentum between the {\sc gadget-2} simulation and the mapped {\sc orion2} grid is generally of order ten percent or less.  
This difference increases again
at densities less than 10$^6$ cm$^{-3}$, but these regions undergo minimal evolution in the relatively short timescale of the {\sc orion2} simulation.

\begin{figure}
\includegraphics[width=.48\textwidth]{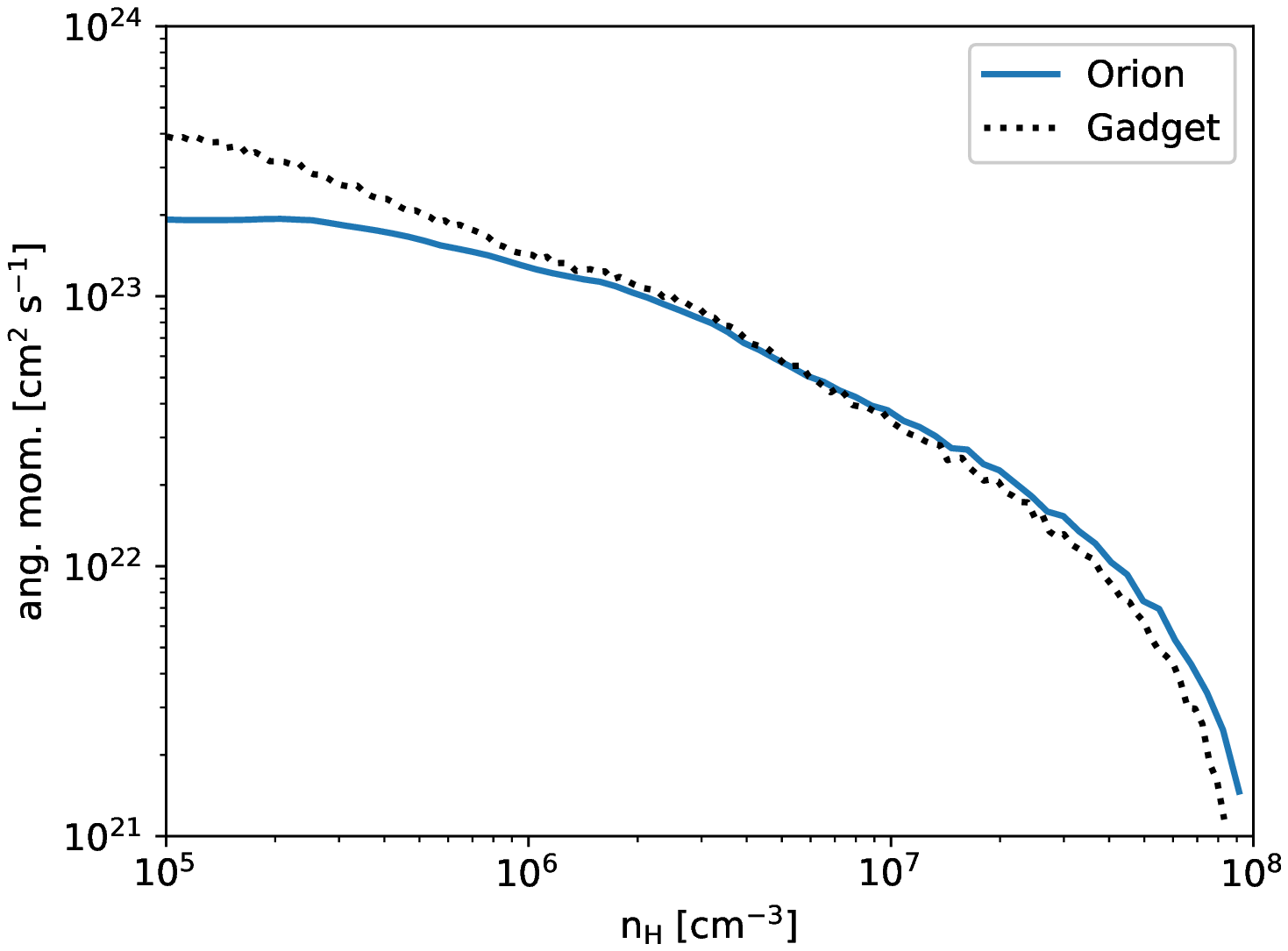}
 \caption{Specific angular momentum versus density in at the end of the {\sc gadget-2} run (dotted line), and after the {\sc gadget-2} SPH particles are mapped onto the {\sc orion2} grid (solid blue line).  Specific angular momentum is taken mass-weighted average within 200 density bins.  At the high densities relevant to the {\sc orion2} simulation timescale, the {\sc gagdet-2} data was mapped onto the grid with minimal error.}
\label{amom}
\end{figure}

To map the magnetic field onto the {\sc orion2} grid, we apply the above method to each field component separately.  We then apply the divergence cleaning to the grid as described in Section A4.

\section{Refinement and Sinks in ORION2}
\label{app:refine}

Cells on level $n$ that are flagged for refinement are sub-divided into 8 smaller cells of length $\Delta x_{n+1} = \Delta x_n / 2$. Cells are flagged when they have a sufficiently large density or velocity gradient:
\beqa
\nabla \rho &>& 0.8 \rho /  \Delta x \mbox{,}\\
\nabla v &>& 0.8 v /  \Delta x \mbox{.}
\eeqa
Cells are also refined to higher levels in order to ensure that they are 
smaller than $J_{\max}\lj$--i.e., 
if they have $\rho_{\rm cell} > \rho_{\rm max}$, where $\rho_{\rm max}$ is the density at which $\Delta x = J_{\rm max}\lj(\rho_{\rm max})$,
\begin{equation}
\rho_{\rm max} = \frac{\pi J_{\rm max}^2 c_s^2} { G \Delta x^2} \mbox{.}
\label{eq:rhomax}
\end{equation}
For grid refinement we set $J_{\rm max} = 1/64$, as recommended by \cite{fede11b} and \citet{turketal2012} to resolve turbulent magnetic field growth.  However, for sink creation we set $J_{\rm max} = 1/4$.

The sink formation and accretion within {\sc orion2} are based upon the methods introduced in \cite{krumholzetal2004}.
If a cell within the highest refinement level reaches a density greater than the above-defined Jeans density, then a sink is placed in that cell.  
In the MHD run, the density criterion is modified to account for the fragmentation-suppressing effect of magnetic pressure:
\begin{equation}
\rho_{\rm max} = \frac{\pi J_{\rm max}^2 c_s^2} { G \Delta x^2}\left( 1 + \frac{0.74}{\beta} \right) \mbox{,}
\end{equation}
where $\beta = 8\pi\rho c_s^2 / B^2$ characterizes the magnetic pressure relative to the thermal pressure
(\citealt{myersetal2013}).
The initial mass of the sink is determined as
\begin{equation}
m_{\rm sink} = [\rho - \rho_{\rm max}(J_{\rm max}=0.25)]\Delta x^3 \mbox{.}
\end{equation}
The density of the remaining gas in the cell is then $\rho_{\rm max}(J_{\rm max}=0.25$), and a proportional amount of momentum and energy is transferred to the sink as well \citep{krumholzetal2004}.  
As discussed in Section \ref{sec:effect}, the magnetic field is left unchanged.

A sink particle accretes mass from an accretion region with a radius of $4 \Delta x$, where $\Delta x \sim 3$ au is the length of the most refined grid cell in our simulation.  The accretion rate is set by the kernel-weighted average density of cells within this region along with the sound speed of the gas in the host cell and the relative velocity of the sink to the gas in its host cell (see \citealt{krumholzetal2004}).
To conserve angular momentum during accretion onto a sink particle, we use the following procedure: Once the mass to be accreted onto the sink is determined, this mass is divided among the cells in the accretion proportional to each cell's kernel weight.  The mass contribution from each cell is reduced depending on the fraction of the cell's mass which would not reach the sink due to having too much angular momentum (see \citealt{krumholzetal2004} for details).

When there are multiple sinks within a single accretion region ($\simeq 12$~au), these sinks are merged.  While this is a crude approximation, doing significantly better would require a sophisticated sub-grid model. \citet{greifetal2011b} merged gravitationally bound sinks within a radius of 100~au; removal of the gravitational binding criterion had only a modest effect on the number of protostars separated by more than 100~au.  
Primordial protostars are often inflated due to a high rate of gas accretion, with radii of the order of 1 au \citep{omukai&palla2003,tan04, hosokawaetal2010}. The large radii of Pop III protostars increases the chance that close encounters will result in mergers, particularly as gas dynamical friction removes angular momentum from their orbit. \citet{greifetal2011b} found that only about 1/5 of the protostars formed in their simulations avoided physical contact with another protostar, which likely would have resulted in a merger. In a simulation with very high resolution (0.25~R$_\odot$) so that sinks were not needed, \citet{greifetal2012} found that about 2/3 of the stars that formed underwent mergers; however, the high resolution of this simulation meant that it stopped after only about 10 yr after the formation of the first protostar,
so this is probably an underestimate of the merger fraction. 
Simulations by \cite{hirano&bromm2017}, which also did not employ sink particles, found that the rate of mergers depended upon the viscosity in the disk of the primary protostar and was high at high resolution ($m_\sph=0.01\;\msunm$). The results of these simulations imply efficient mergers and are thus qualitatively consistent with our merger criterion.


\section{Chemistry and Cooling in ORION2}
\label{app:chem}


Each cell's temperature and chemical abundances are evolved using the same chemothermal network as described in detail by \cite{greifetal2009} and used in \cite{stacyetal2012}.   The code follows the abundance evolution of  
H, H$^{+}$, H$^{-}$, H$_{2}$, H$_{2}^{+}$, He, He$^{+}$, He$^{++}$, and e$^{-}$, as well as the three deuterium species D, D$^{+}$, and HD.  
Every cell also has an adiabatic index $\gamma$ that is updated according its temperature and chemical abundances.

All relevant cooling mechanisms, including H$_2$ collisions with  H and He as well as other H$_2$ molecules, are included.  The thermal network also includes cooling through  
H$_2$ collisions with protons and electrons, H and He collisional excitation and ionization, recombination, bremsstrahlung, and inverse Compton scattering.  

Further H$_2$ processes are included to properly model gas evolution to high densities.  
In particular, the chemistry and thermal network includes three-body H$_2$ formation and the concomitant H$_2$ formation heating, which become important at  $n \ga 10^8$ cm$^{-3}$.  As noted in Section \ref{sec:caveat},
we use the rates presented in \cite{forrey2013}, 
which are intermediate between the higher rates published by \cite{flower&harris2007} and the lower rates published by \cite{abeletal2002}.
When $n \ga 10^9$ cm$^{-3}$, cooling through H$_2$ ro-vibrational lines becomes less effective as these lines grow optically thick.  
We estimate the optical depth  using an escape probability formalism together with the Sobolev approximation (see \citealt{yoshidaetal2006,greifetal2011b} for further details).

\begin{figure*}
\includegraphics[width=.95\textwidth]{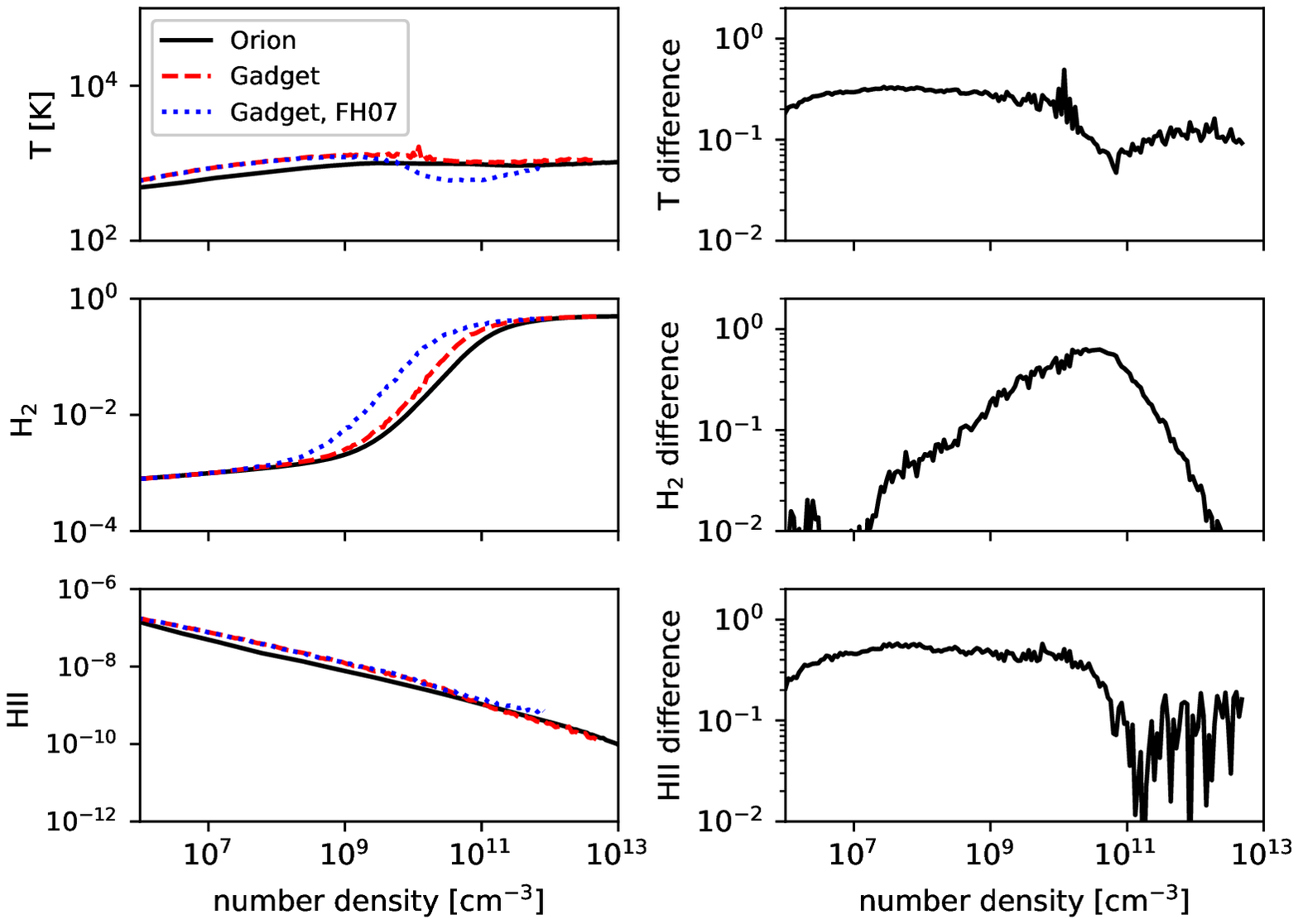}
 \caption
 {
 Comparison of primordial gas evolution from 10$^7$ to 10$^{12}$ cm$^{-3}$ 
 as evolved in 
a {\sc gadget-2} hydrodynamic run from \citet{stacyetal2012} versus the {\sc orion2} hydrodynamic run.  
 Black lines denote the evolution in  {\sc orion2}, while dashed red lines show evolution in {\sc gadget-2}, both using the same H$_2$ rates
 presented in Forrey et al 2013.  For comparison, dotted blue lines show the {\sc gadget-2} run, but instead using the rates of Flower \& Harris (2007),
 showing that the uncertainty in the chemical rates is substantially greater than the difference between the two codes.
 {\it Top Panels:}  Temperature versus number density and relative difference between the {\sc orion2} and {\sc gadget-2} results.
 {\it Middle Panels:} H$_2$ abundance versus number density and relative difference between the {\sc orion2} and {\sc gadget-2} results.
 {\it Bottom Panels:} H{\sc II} abundance versus number density along with relative difference between the {\sc orion2} and {\sc gadget-2} results.
 }
\label{chem_comp}
\end{figure*}

To test the accuracy the chemothermal network incorporated into {\sc orion2}, we map a small 1 parsec test box from a minihalo taken from \cite{stacyetal2012} onto an {\sc orion2} grid and compare the gas evolution in both codes as the density peak evolves from 10$^7$ to 10$^{12}$ cm$^{-3}$ (Fig. \ref{chem_comp}) .  As expected, in both cases the gas remains roughly isothermal at $\sim$ 1000 K and transitions from atomic fully molecular.  This evolution is furthermore consistent with the results of previous cosmological simulations (e.g. \citealt{yoshidaetal2006,greifetal2011b,stacy&bromm2013}).  The relative error between {\sc gadget-2} and {\sc orion2} was typically a few tens percent.  This is less than the relative error between the original {\sc gadget-2} run and a repeated run where when we instead used the three-body H$_2$ rates of \cite{flower&harris2007}.

\section{Growth Rate of the Kinematic Dynamo}
\label{app:growth}

As discussed in Section \ref{sec:predict}, the growth rate of the field in a kinematic dynamo can be expressed as (eq. \ref{eq:gamma})
\beq
\Gamma=C_\Gamma\Gamma_\nu=C_\Gamma \left(\frac{v_L}{L}\right) Re^{1/2},
\label{eq:Gamma}
\eeq
where $C_\Gamma$ is a numerical constant. For large values of the magnetic Prandtl number, 
$C_\Gamma=(1,\,3/8)$ when resistivity is negligible or important, respectively \citep{kuls92,sche02a}. 
On the other hand, numerical simulations have found $C_\Gamma\ll 1$, presumably because of the greater importance of dissipation at the moderate values of the magnetic Prandtl number in simulations, $P_m\sim 1$ \citep{les07}. \citet{haug04} verified the $Re^{1/2}$ dependence of the growth rate in a set of simulations with $P_m=1$, so that $Re=R_m$. They defined the Reynolds number in terms of the wavenumber of the driving eddies, $Re_k=v_{\rm rms}/k_f\nu=Re/2\pi$.
A necessary condition for the operation of a kinematic dynamo is that $R_m$ be large enough that the growth of the field is more rapid than its dissipation, and they showed that this requires $R_m>2\pi\times 35 P_m^{-1/2}=220 P_m^{-1/2}$ for $3\ga P_m\ga 0.1$. Paper I showed that this criterion agrees with the condition found by \citet{fede11b} that the minimum resolution required for the operation of a dynamo in a collapsing cloud is between 16 and 32 cells per Jeans length.
After converting $Re_k$ to $Re$ and replacing $R_m$ by $Re=R_m/P_m$, their results for $P_m=1$ can be fit to within about 10\% by
\beq
C_\Gamma=8.3\times 10^{-3}\left(1-\frac{220}{Re}\right)^{1/2}~~~~(P_m=1)
\eeq
for $Re\ga 450$.

In a set of ideal MHD simulations with $128^3$ grid cells, \citet{fede11a} explored the properties of small-scale dynamos as a function of the Mach number of the turbulence. Comparing the ideal MHD results with those from several non-ideal MHD simulations, they found that the ideal MHD simulations were consistent with $Re=1500$ and $P_m=2$. In Paper I we showed that the Reynolds number of a grid-based simulation is $Re=2\caln_g^{4/3}$, which is 1290 for $\caln_g=128$ and is in good agreement with \citet{fede11a}. They found $\Gamma\simeq v_L/L$ for subsonic, solenoidally driven turbulence, which corresponds to $C_\Gamma\simeq Re^{-1/2}=0.026$. Compressively driven turbulence was much less effective at driving a dynamo. The growth rate found by \citet{fede11a} is several times larger than that found by \citet{haug04}, possibly due to the difference in driving patterns and/or the fact that \citet{haug04}'s simulation included explicit viscosity and resistivity whereas \citet{fede11a}'s simulation was an ideal MHD simulation with numerical viscosity and resistivity. Since \citet{fede11a} considered only one value of $Re$, they could not determine how the growth rate depends on $Re$.

\citet{fede11b} studied the growth rate of kinematic dynamos in gravitationally collapsing clouds. They showed that the effective outer scale in such clouds is approximately equal to the Jeans length, $\lj$. As noted above, the Reynolds number for a grid-based code is $Re\propto \caln_g^{4/3}$, where now $\caln_g$ is the number of grid cells per Jeans length. It follows that $Re$ is constant for an adaptive-mesh refinement simulation in which the number of cells per Jeans length is kept approximately constant. As a result, equation (\ref{eq:intg1}) becomes
\beq
\int\Gamma dt = \left(\frac{3}{32}\right)^{1/2}C_\Gamma Re^{1/2}\avg{\calm}\int\frac{dt}{\tff},
\label{eq:intg2}
\eeq
where $\avg{\calm}$ is the time-averaged value of the Mach number. 
\citet{fede11b} define $\Omega$ as the coefficient of the integral, so it follows that
\beq
C_\Gamma=\left(\frac{32}{3}\right)^{1/2}\frac{\Omega}{\avg{\calm}Re^{1/2}}.
\eeq
They found $\Omega\simeq 0.1 J_{\max}^{-0.3}$ for Jeans numbers $J_{\max}=1/32-1/128$; for $J_{\max}=1/16$, the growth rate is substantially reduced, so the the critical magnetic Reynolds number corresponds to a resolution in the range $1/16>J_{\max}>1/32$, as noted above. In the range of times over which they measured $\Omega$, the average Mach number was $\avg{\calm}\simeq 0.5$. Since the Reynolds number in the simulation is $Re=2\caln_g^{4/3}=2/J_{\max}^{4/3}$, their results imply
\beq
C_\Gamma\simeq 0.46 J_{\max}^{0.37},
\eeq
which is 0.10 at their recommended value $J_{\max}=1/64$. In contrast to \citet{haug04}'s results for a turbulent box, there is no evidence for a gradual reduction in $C_\Gamma$ as $Re$ decreases; instead there is a dramatic reduction between $J_{\max}=1/32$ and $J_{\max}=1/16$.

In sum, Federrath and collaborators have found $C_\Gamma\simeq 0.026$ for a turbulent box with $\caln_g=128$ (corresponding to $Re=1500$), and $C_\Gamma\simeq 0.1$ for a cloud undergoing gravitational collapse at a resolution of 64 cells per Jeans length. We attribute this reduction in $C_\Gamma$ from the theoretically expected value to the moderate value of $P_m$ in simulations. It appears that this reduction is less severe in a collapsing cloud.
In Section \ref{sec:initc} we characterized the rate of collapse in terms of
the ratio of the time taken to collapse to a star to the central free-fall time,
$\tcoll/\tff$. Analysis of the results of \citet{fede11b} shows that over the time interval in which they determined $\Omega$, this ratio was $\tcoll/\tff\simeq 1.5$ in their simulation. By contrast, our simulation from cosmological initial conditions had a slower collapse, with $\tcoll/\tff\simeq 3$ during the {\sc gadget-2} stage of the collapse and $\tcoll/\tff\simeq 2$ during the {\sc orion2} stage. Since the {\sc gadget-2} stage is intermediate between static conditions considered by \citet{fede11a} and the rapid collapse considered by \citet{fede11b}, we anticipate that the value of $C_\Gamma$ is intermediate between the values for a static medium and a rapidly collapsing one, and this is borne out by the results discussed in the text.

\label{lastpage}
\end{document}